\documentclass[twocolumn]{aastex631}

\hypersetup{linkcolor=red,citecolor=blue,filecolor=magenta,urlcolor=cyan}

 \accepted{June 17, 2022}
\shorttitle{Neptune's resonance: 2:1}
\shortauthors{Thirouin \& Sheppard}

\begin{document}

\title{Lightcurves and Rotations of Trans-Neptunian Objects \\ in the 2:1 Mean Motion Resonance with Neptune}

\correspondingauthor{Audrey Thirouin}
\email{thirouin@lowell.edu}

\author[0000-0002-1506-4248]{Audrey Thirouin}
\affiliation{Lowell Observatory, 1400 W Mars Hill Road, Flagstaff, AZ 86001, USA}

\author[0000-0003-3145-8682]{Scott S. Sheppard}
\affiliation{Earth and Planets Laboratory, Carnegie Institution for Science,\\ 5241 Broad Branch Rd. NW, Washington, District of Columbia, 20015, USA.}



\begin{abstract}
 
 We report the rotational lightcurves of 21 trans-Neptunian objects (TNOs) in Neptune's 2:1 mean motion resonance obtained with the 6.5~m \textit{Magellan-Baade telescope} and the 4.3~m \textit{Lowell Discovery Telescope}. The main survey's goal is to find objects displaying a large lightcurve amplitude which is indicative of contact binaries or highly elongated objects. In our sample, two 2:1 resonant TNOs showed a significant short-term lightcurve amplitude: 2002~VD$_{130}$ and (531074) 2012~DX$_{98}$. The full lightcurve of 2012~DX$_{98}$ infers a periodicity of 20.80$\pm$0.06~h and amplitude of 0.56$\pm$0.03~mag whereas 2002~VD$_{130}$ rotates in 9.85$\pm$0.07~h with a 0.31$\pm$0.04~mag lightcurve amplitude. Based on lightcurve morphology, we classify (531074) 2012~DX$_{98}$ as a likely contact binary, but 2002~VD$_{130}$ as a likely single elongated object. Based on our sample and the lightcurves reported in the literature, we estimate the lower percentage of nearly equal-sized contact binaries at only 7-14~$\%$ in the 2:1 resonance, which is comparable to the low fraction reported for the dynamically Cold Classical trans-Neptunian objects. This low contact binary fraction in the 2:1 Neptune resonance is consistent with the lower estimate of the recent numerical modeling. We report the Sloan g', r', i' surface colors of 2002~VD$_{130}$ which is an ultra-red TNO whereas 2012~DX$_{98}$ is a very red object based on published surface colors. \\  

\end{abstract}



\section{Introduction}
\label{sec:intro}

Located at about 47.7~AU, the 2:1 mean motion resonance with Neptune is the second known most populated resonance after Neptune's 3:2 mean motion resonance at $\sim$39.4~AU \citep{Volk2016, Bannister2018, Chen2019}. The 2:1 Neptune resonance is just beyond the main classical Kuiper Belt and is likely made up of objects that formed in the main classical belt as well as objects scattered outward from the giant planet region before being trapped into the resonance \citep{Sheppard2012}. The dynamically classical population$\footnote{For the purpose of this work, our definition of the Cold Classical population is the same as in \citet{ThirouinSheppard2019a}.}$ is trapped between the 3:2 and 2:1 resonances and is generally divided between the dynamically Hot and Cold classical. Typically, the Cold classicals have an inclination i$\leq$4-5$^\circ$, but some works infer that the inclination threshold should be at about 12$^\circ$ \citep{Brown2001, Elliot2005, Peixinho2008, Gladman2008}. Also, \citet{Petit2011} suggested that the Cold Classical population is composed of at least two sub-groups, the stirred and the kernel.  

As of February 2022, the \textit{Deep Ecliptic Survey}\footnote{The Deep Ecliptic Survey (DES) object classification is available at \url{https://www.boulder.swri.edu/~buie/kbo/desclass.html}} has classified 106 trans-Neptunian objects (TNOs) confined in the 2:1 resonance. Some objects are classified as likely 2:1 TNOs but some other classifications (e.g., Centaur or Scattered Disk Object) are also possible based on the currently available astrometry. These objects are 2006~SG$_{415}$\footnote{The partial lightcurve of 2006~SG$_{415}$ is presented in this paper. Based on the DES classification, 2006~SG$_{415}$ is likely a 2:1 resonant TNO, but it can also be a Scattered Disk Object (SDO). Therefore, care will be taken to include or exclude this object during the presentation and discussion of our results.},
2009~MG$_{10}$, 2013~TD$_{228}$, 2014~SX$_{349}$, 2014~YZ$_{91}$, (554102) 2012~KW$_{51}$, 2016~SJ$_{56}$, 2017~FD$_{163}$, and 2017~FQ$_{161}$. 

For decades, lightcurves have been used to estimate the rotational period and the lightcurve amplitude, as well as to derive the rotational properties (shape, binarity, surface features, and others) of small bodies across the Solar System (e.g., \citet{Pravec2000, Sheppard2008, Thirouin2016, ThirouinSheppard2019a}). Most TNO lightcurve surveys use small 1 to 2~m class telescopes and thus are limited to bright objects with, typically, a visual magnitude (V) brighter than $\sim$21~mag \citep{Thirouin2010, Sheppard2008}. Therefore, there is a bias in the literature toward brighter and thus larger objects, which skews our current understanding of the rotational properties of the TNOs as a population. Recently, several surveys dedicated to the rotational lightcurves of small TNOs were designed using 4 to 8~m class telescopes to observe TNOs as faint as V$\sim$25~mag \citep{Alexandersen2019, ThirouinSheppard2019a, ThirouinSheppard2018}. These new surveys aim to improve our overall understanding of the TNO rotational properties by pushing the facilities to their limit of detectability, but more work has to be done. Observing fainter objects is required, but it is also necessary to increase the number of objects with rotational lightcurves in most of the TNO sub-populations. 

Little is known about the rotational characteristics of the 2:1 resonant TNOs. Only four objects (see Section~\ref{sec:literaturedis} for more details) -- (26308) 1998~SM$_{165}$, (119979) 2002~WC$_{19}$, (469505) 2003~FE$_{128}$, and (312645) 2010~EP$_{65}$-- have published rotational lightcurve studies \citep{Romanishin2001, SheppardJewitt2002, Kern2006, Spencer2006, Sheppard2007, Benecchi2013, Thirouin2013}. Three of them have resolved companions while 2010~EP$_{65}$ is the only one with no satellite detected based on \textit{Hubble Space Telescope} images. The satellites of 1998~SM$_{165}$, 2002~WC$_{19}$, and 2003~FE$_{128}$ were discovered after their lightcurve studies. Because the lightcurve sample of 2:1 resonant TNOs is extremely limited, biased towards resolved binary systems, and biased towards bright objects, we aim for this paper to observe faint single 2:1 resonant objects to improve our understanding of this sub-population. Below, we will describe our survey strategy and target selection. We will also summarize the rotational properties of the 2:1 TNOs and estimate the contact binary percentage in this resonance.

\section{Survey Description and Facilities} 
\label{sec:obs}

\begin{figure}
 \includegraphics[width=8cm,angle=0]{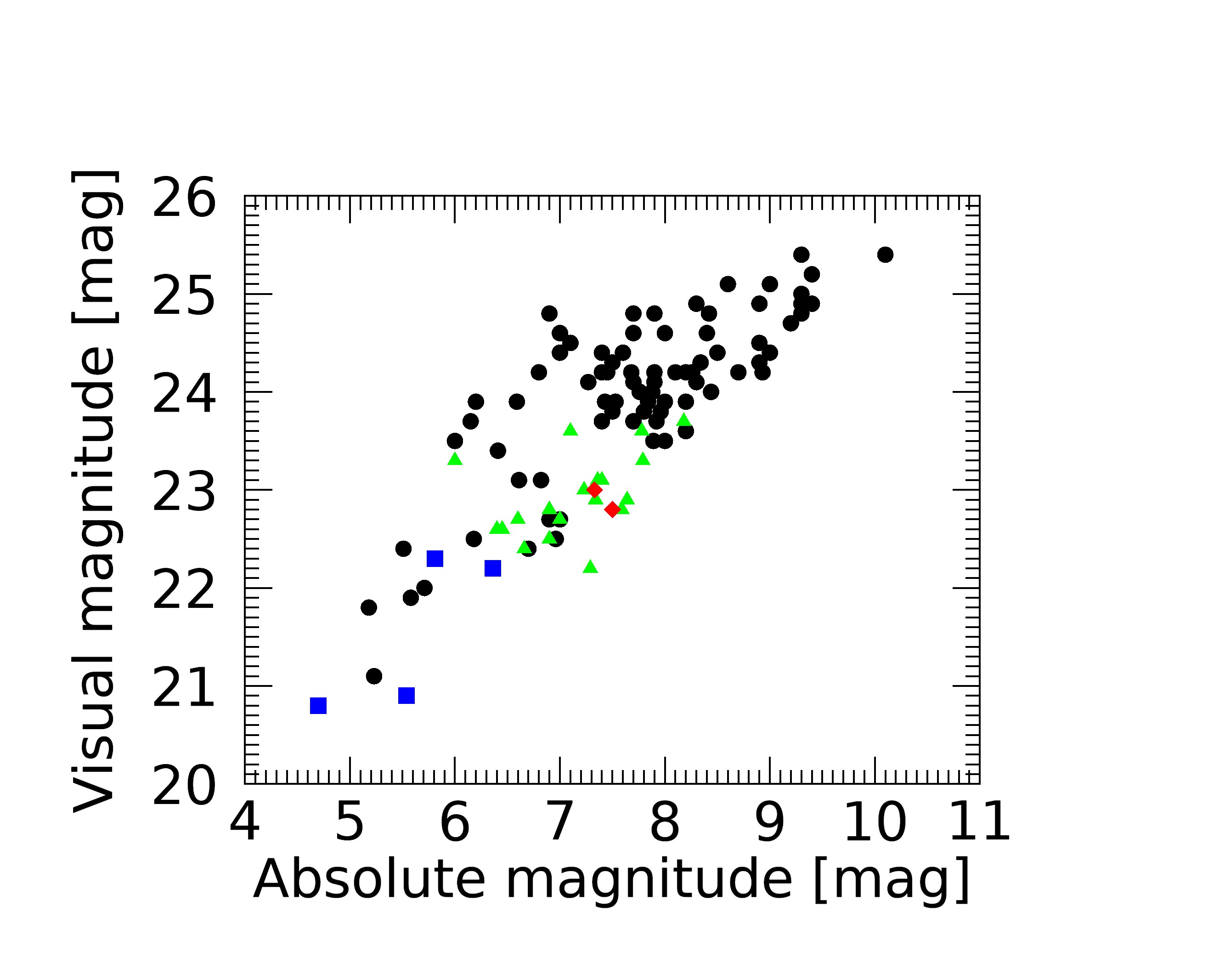} 
  \includegraphics[width=8cm,angle=0]{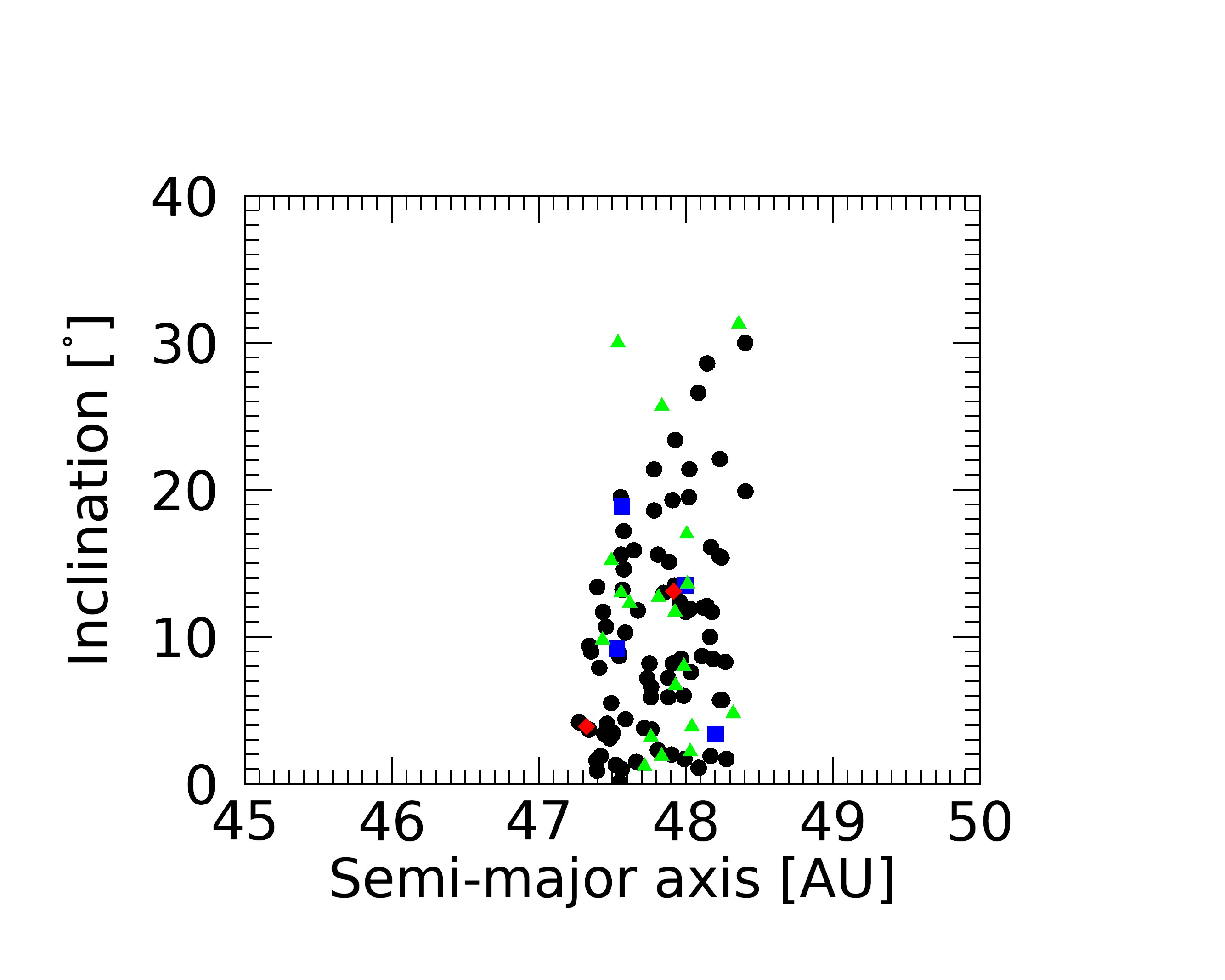} 
    \includegraphics[width=8cm,angle=0]{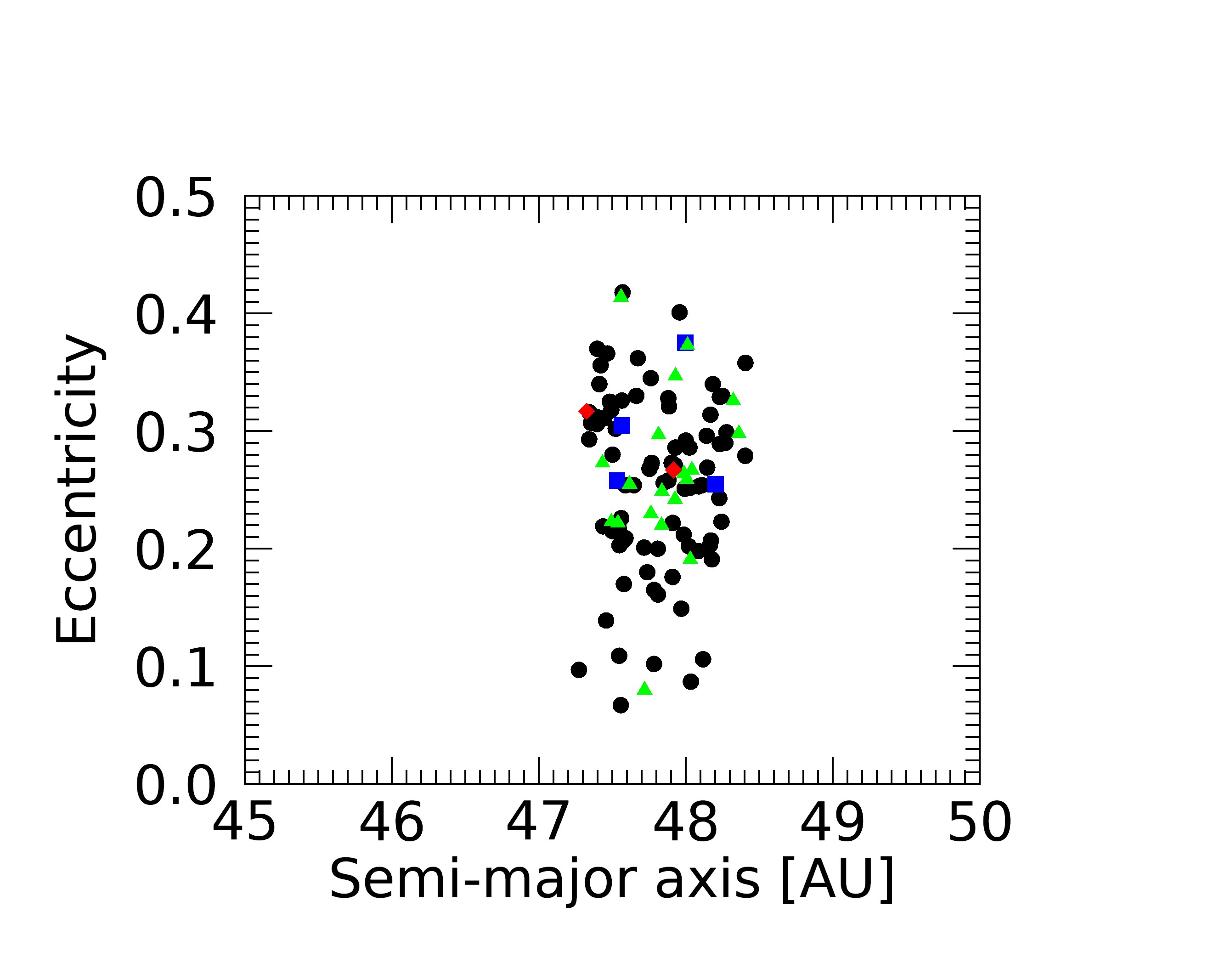} 
\caption{The 2:1 resonant TNOs are classified into 4 groups: (1) objects with a lightcurve from the literature (blue squares, see Section~\ref{sec:literaturedis} for more details), (2) objects with a flat to moderate lightcurve amplitude from our survey (green triangles), (3) objects with a large lightcurve amplitude from this survey (red diamond), and (4) objects never observed for lightcurve study (black circles). Our survey is combined with published lightcurves of 2:1 TNOs to cover a range of inclination, eccentricity and absolute magnitude. \textit{\underline{Notes:}} The nine likely 2:1 TNOs mentioned in the Introduction are plotted. Orbital elements, absolute and visual magnitudes are from the Minor Planet Center (February 2022). }
\label{fig:Survey}
\end{figure}

Our 2:1 Neptune resonance lightcurve survey strategy is inspired by the dynamically Cold Classical survey published in \citet{ThirouinSheppard2019a}. The strategy is to image a substantial set of TNOs for partial lightcurves to constrain their rotational periods and amplitudes as well as to discover interesting objects with a large amplitude (typically, larger than 0.4~mag) which can be indicative of contact binaries or highly elongated objects. 

Our first target selection criterion is the visual magnitude. Bright objects (V$\lesssim$22~mag) are already covered by the literature and because we are using 4 and 6~m class telescopes, we select objects with a V between $\sim$22 and 23.5-24~mag (more details about facilities below). Preferentially, TNOs without a known resolved companion are chosen, but several of our targets have never been observed with \textit{Hubble Space Telescope} thus their resolved binarity status is unknown (see Table~\ref{tab:Summary_photo}). Targets are also selected to cover a large range of eccentricities and inclinations (semi-major axis is also considered but the range is limited due to the definition of the 2:1 resonant population), as well as absolute magnitudes (i.e., sizes). Our selected targets are plotted in Figure~\ref{fig:Survey}.

Our survey makes use of two facilities. The \textit{Magellan-Baade} telescope at Las Campanas observatory is equipped with IMACS which is a wide-field imager with 8 CCDs giving a 27.4$\arcmin$ diameter field and a pixel scale of 0.20$\arcsec$/pixel. Our runs at the \textit{Lowell Discovery Telescope} (LDT), formerly known as \textit{Lowell's Discovery Channel Telescope} (DCT), uses the Large Monolithic Imager which is a 6144$\times$6160 pixels CCD with a field of view of 12.5$\arcmin$$\times$12.5$\arcmin$ and a pixel scale of 0.12$\arcsec$/pixel. Exposure times range from 200 to 600~s and are adjusted based on the weather conditions and the facility. All observations are obtained with a broad-band filters (VR and WB4800-7800 filters at \textit{LDT} and \textit{Magellan-Baade}, respectively) to maximize the target's signal-to-noise ratio. In one instance, we used the Sloan g'r'i' filters for surface color determination. 

All images are calibrated with bias and dome or twilight flats before proceeding with aperture photometry. Once the photometry is on hand, we search for periodicity using the Lomb and the Phase Dispersion Minimization (PDM) techniques \citep{Lomb1976, Stellingwerf1978}. This series of steps is standard and has been already described in greater detail in \citet{Thirouin2010}.

\section{From flat to large amplitude lightcurves} 
\label{sec:res}

 \subsection{What is a lightcurve?}
 
A lightcurve (i.e. brightness variation as a function of time) of a small body is determined by the periodic variation of the body brightness due to its rotation. The two main parameters derived from a lightcurve are: (1) the time separation of repeated brightness peaks in the lightcurve gives the object's \textit{rotational period} (P) and (2) the full (or peak-to-peak) \textit{lightcurve amplitude} ($\Delta m$). Rotational period, lightcurve amplitude, and lightcurve morphology can be used to infer some physical and rotational properties of the body: shape, surface heterogeneity/homogeneity, internal structure, density, and binarity \citep{Sheppard2008}. A lightcurve is due to (1) albedo variation(s) on the object surface, (2) non-spherical shape, and/or (3) binarity \citep{Sheppard2008}. Assuming hydrostatic equilibrium, a small body with a spherical shape is called MacLaurin spheroid whereas an elongated triaxial ellipsoidal object is a Jacobi object \citep{Chandrasekhar1969}. As illustrated in \citet{Thirouin2014}, a MacLaurin object will have a single-peaked lightcurve whereas a Jacobi or contact binary will display a double-peaked lightcurve. Typically, a spheroidal body with albedo spot(s) on its surface will have a low amplitude lightcurve such as $\Delta$m$\lesssim$0.15-0.2~mag. A triaxial ellipsoidal object will have a sinusoidal lightcurve with a moderate lightcurve amplitude of $\sim$0.15-0.2 mag$\lesssim$$\Delta$m$\lesssim$0.4~mag. The lightcurve of a nearly equal-sized contact binary observed equator-on will have a $\Delta$m$\geq$0.9~mag and the maximum/minimum of brightness is an inverted U-shape/V-shape from shadowing effects \citep{Lacerda2007, Lacerda2011, Lacerda2014, Harris2020}. If a contact binary is imaged when the line of sight is further off the equator, the lightcurve will have lower amplitude, and the V-/U-shapes will be less prominent \citep{Lacerda2011}. Therefore, \citet{ThirouinSheppard2019a} uses: an object with an amplitude greater than 0.9~mag, and the V-/U-shapes is a confirmed contact binary but a likely contact binary will have large amplitude except that under a 0.9~mag cutoff and the U-/V-shapes are less prominent \citep{Descamps2015, Lacerda2011, Leone1984}. Following \citet{ThirouinSheppard2019a}, for an object with  $\Delta$m$>$0.4~mag, we will discuss if it is a likely contact binary or a single triaxial object. We note that ideally two lightcurves obtained at significantly different epochs are needed for lightcurve modeling purposes to confirm the morphology of the object/system \citep{Lacerda2014, Lacerda2011}. \\

Following, we classify the 21 confirmed 2:1 resonant TNOs (plus 2006~SG$_{415}$) lightcurves$\footnote{Photometry and partial/flat lightcurves are available in Appendix~A and Appendix~B.}$ obtained for this work in categories based on their amplitude: (1) a flat lightcurve displays no significant variability, (2) a low amplitude lightcurve has a $\Delta m$$<$0.2~mag, (3) a moderate amplitude lightcurve with 0.2~mag$<$$\Delta m$$<$0.4~mag, and (4) a large amplitude lightcurve with $\Delta m$$>$0.4~mag.  

 \subsection{Large Amplitude Lightcurves}
 
\begin{figure}
 \includegraphics[width=9.5cm,angle=180]{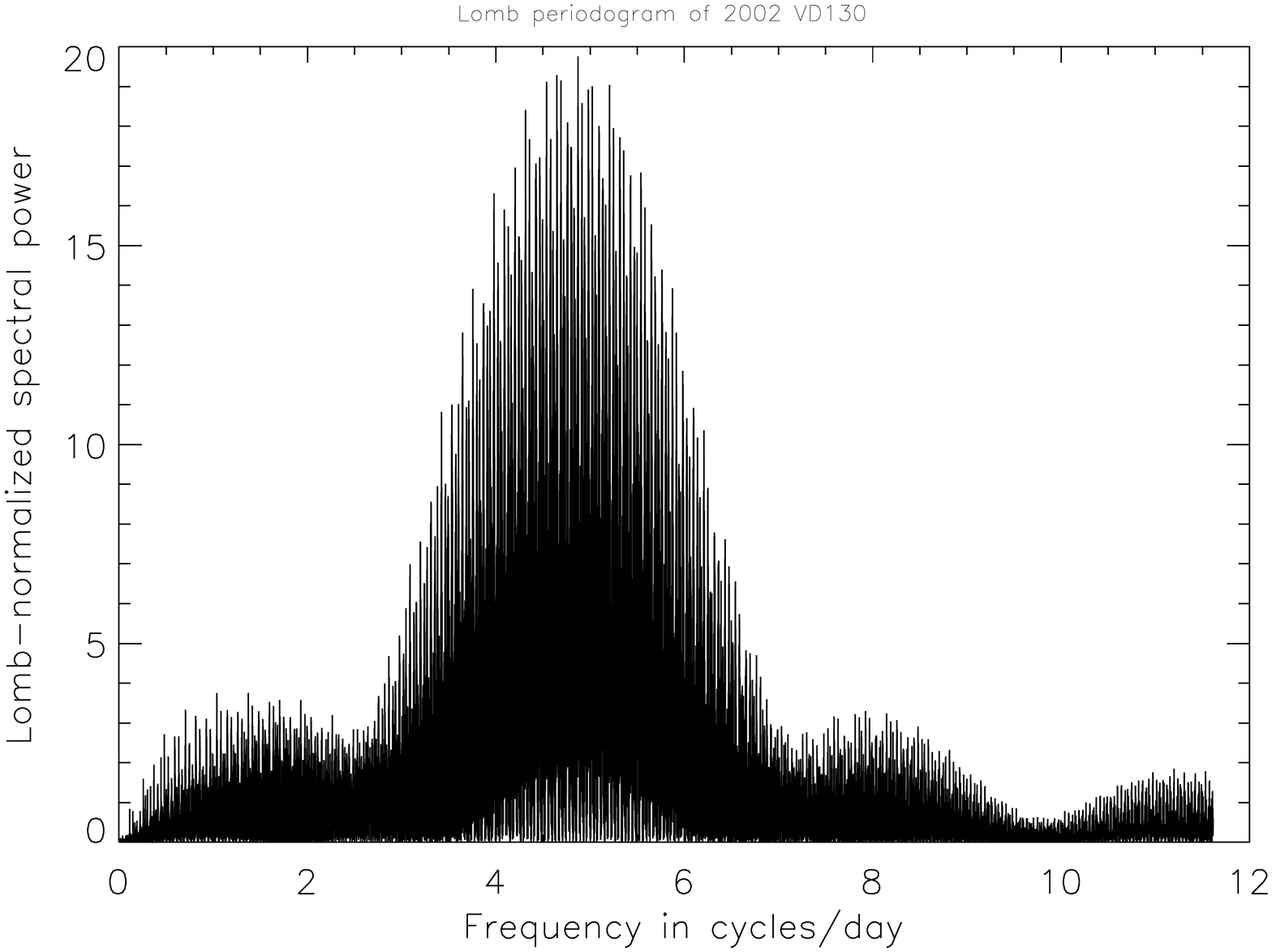}
 \includegraphics[width=9.5cm, angle=0]{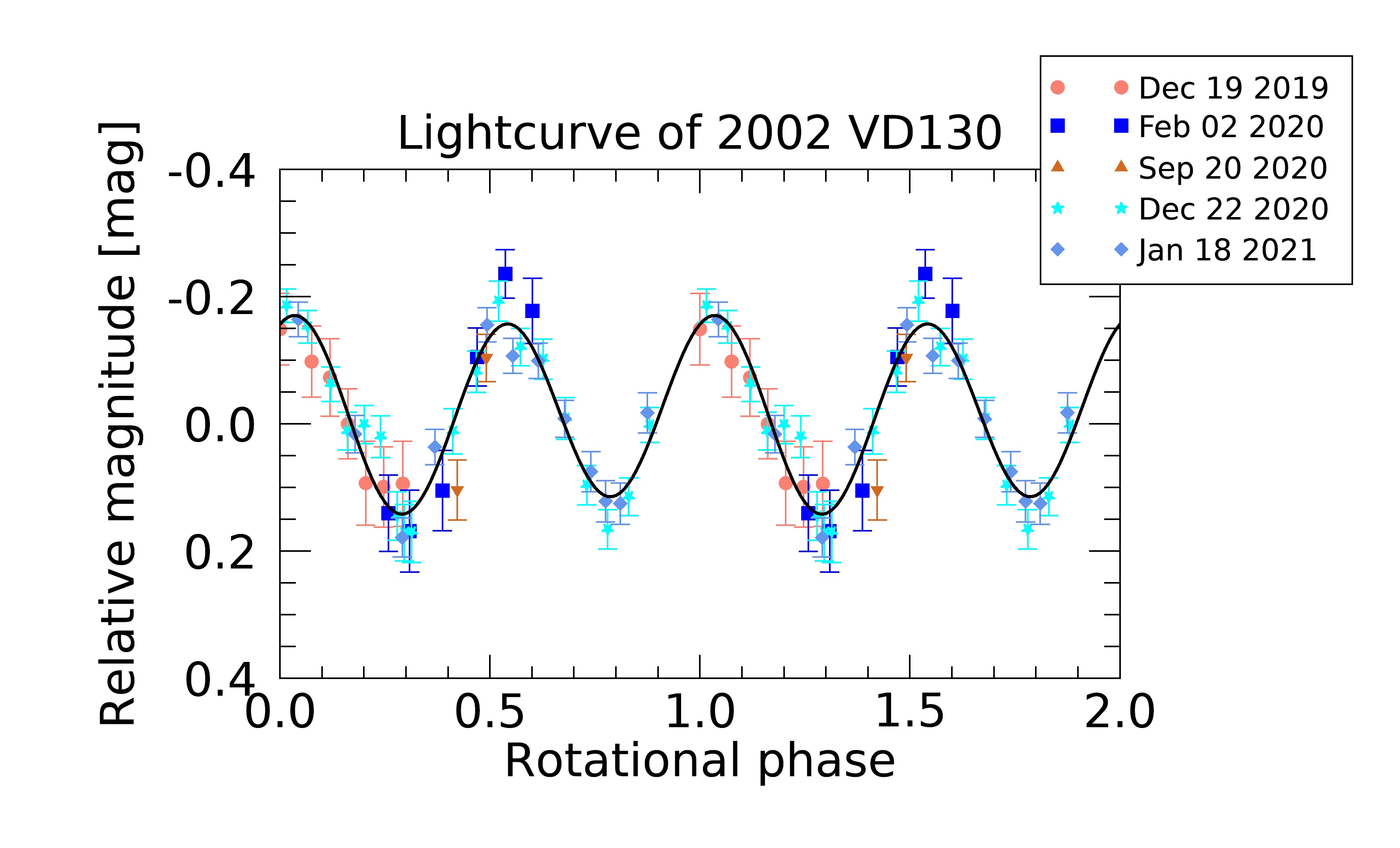}
\caption{The Lomb periodogram (upper plot) favors a rotational frequency at 4.87~cycles/day. Due to the large amplitude and asymmetric lightcurve, the double-peaked rotational lightcurve with a periodicity of 9.85~h is favored (lower plot). A second-order Fourier fit is overplotted (black line). }
\label{fig:VD130}
\end{figure}

 \paragraph{2002~VD$_{130}$} This object was observed on several occasions with the \textit{LDT} from 2019 to 2021 (Figure~\ref{fig:VD130}). The Lomb periodogram inferred a rotational period of 4.87~cycles/day (or 4.93~h), but there are several nearby aliases with a high confidence level as well. Due to the large amplitude and asymmetric lightcurve with the first minimum being deeper than the second one, the double-peaked rotational period is favored. The rotational period of 2002~VD$_{130}$ is about 9.85$\pm$0.07~h and the full lightcurve amplitude is 0.31$\pm$0.04~mag from the second-order Fourier fit. The lightcurve of 2002~VD$_{130}$ displays a large amplitude, but there is no sign of the characteristics V-shape and U-shape of a contact binary. The sinusoidal morphology of this lightcurve would currently suggest that 2002~VD$_{130}$ is an elongated object and not a clear contact binary candidate. Future observations at a significantly later epoch will help determine the true nature of this object, but for now, we do not classify 2002~VD$_{130}$ as a candidate contact binary TNO.

Following the formalism described in \citet{Chandrasekhar1987}, one can derive the axis ratio and lower limit to the density of an ellipsoidal object in hydrostatic equilibrium for a given rotational period. If 2002~VD$_{130}$ is an elongated Jacobi body with axes such as a$>$b$>$c and is rotating along its c-axis, its density is $\rho$$\geq$0.42~g cm$^{-3}$ and the axis ratios are a/b=1.46, and c/a=0.47 assuming that this object was observed under an equatorial view. If we consider that the lightcurve of 2002~VD$_{130}$ is single-peaked, then its rotational period is 4.93~h (half the period of the double-peaked lightcurve), and in this case, the lower limit to the density would be 1.69~g cm$^{-3}$. \citet{Grundy2012} reported the densities of several binary/multiple systems which were extracted from their mutual orbitals and mass determinations. Based on Figure~7 in \citet{Grundy2012}, there is a clear trend inferring that small objects have density lower than 1~g cm$^{-3}$ whereas the large objects have densities above 1~g cm$^{-3}$ limit. 2002~VD$_{130}$ has a diameter between $\sim$100 and $\sim$200~km (assuming an albedo of 0.20 or 0.04, respectively), therefore its density is likely below or around 1~g cm$^{-3}$. Therefore, we can rule out the density estimate derived from the single-peaked lightcurve, and also clearly favor the double-peaked lightcurve.

\begin{figure}
 \includegraphics[width=9.5cm,angle=180]{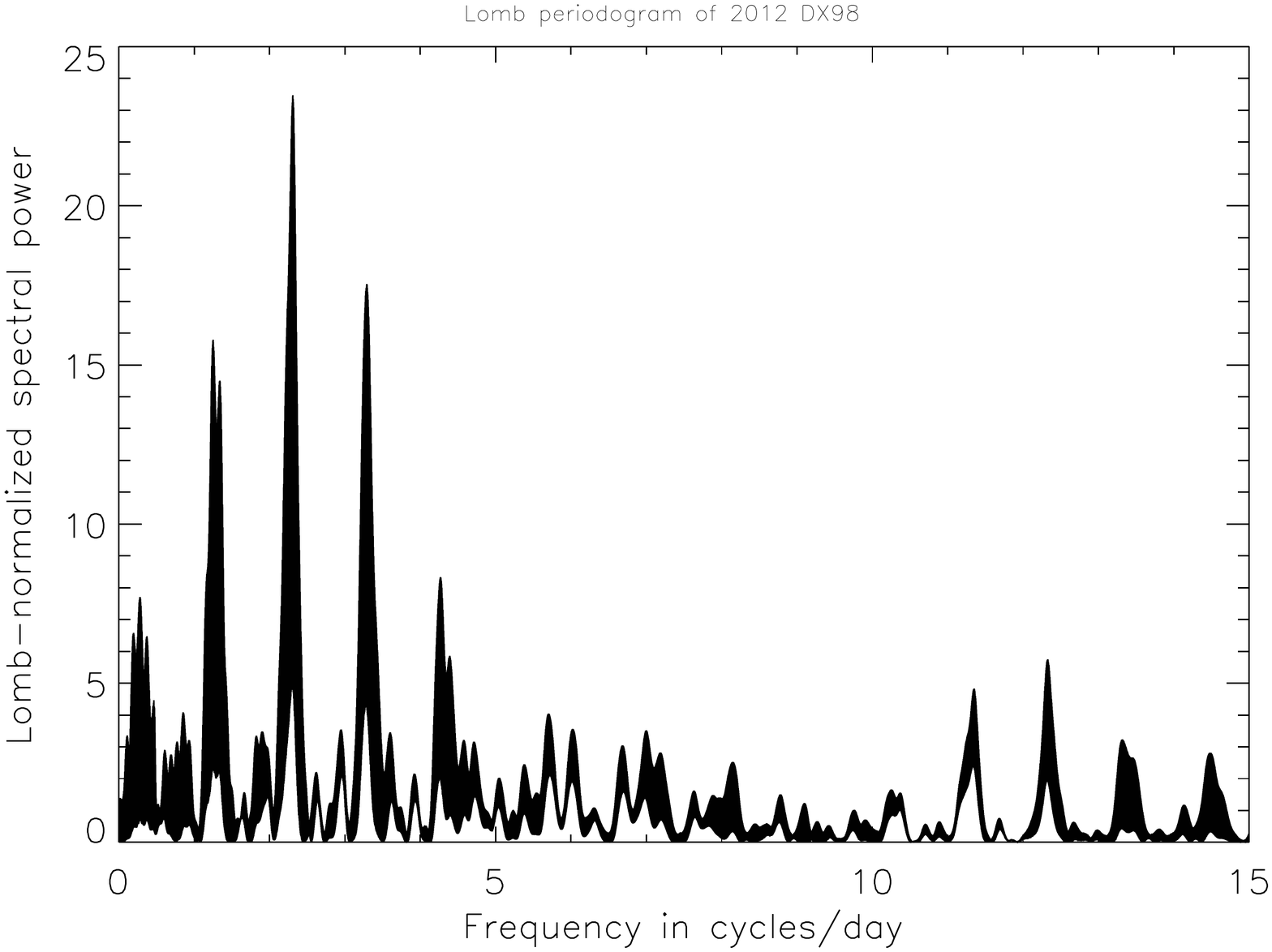}
 \includegraphics[width=9.5cm, angle=0]{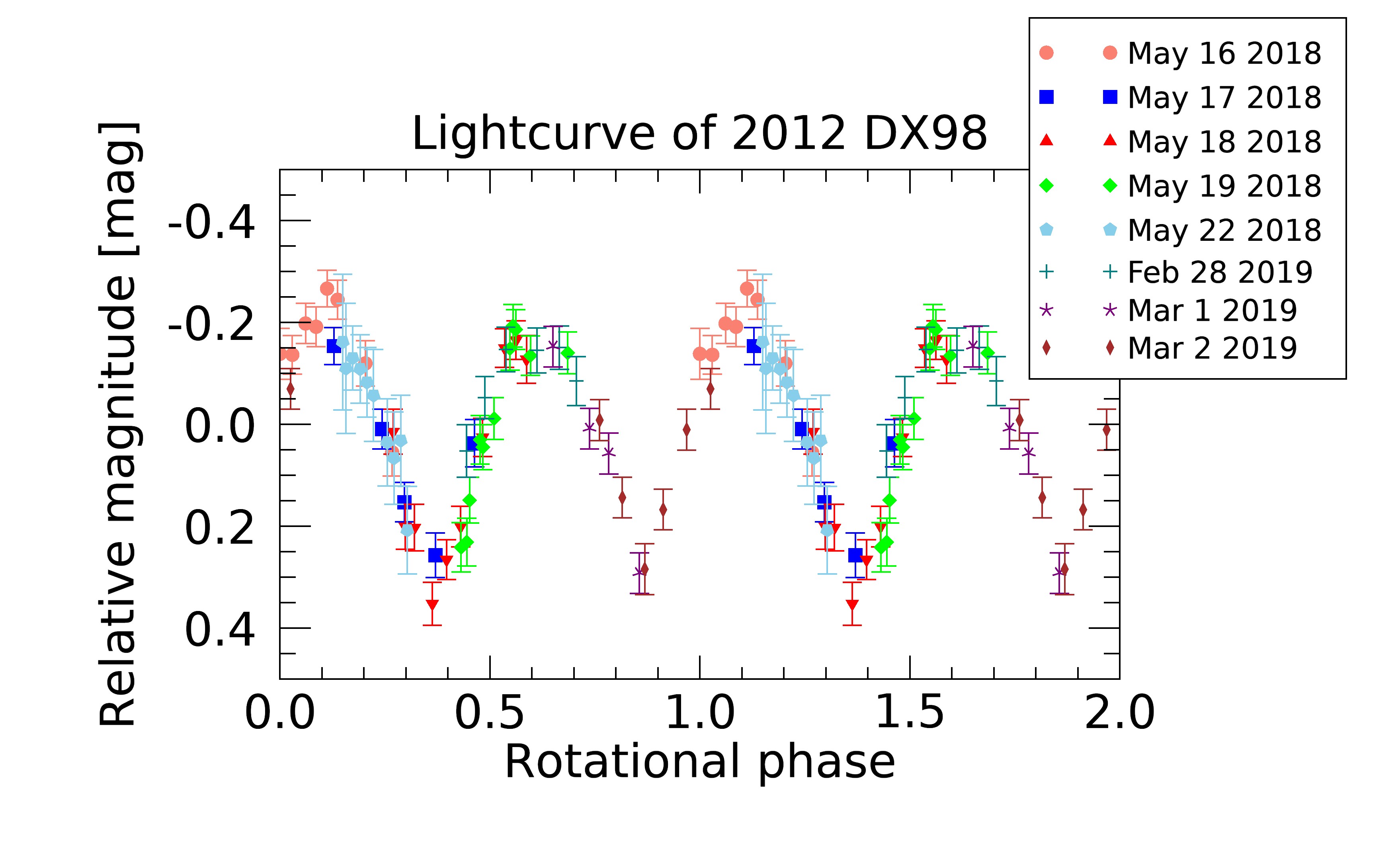}
 \caption{The main peak favored by the Lomb periodogram is at 2.31~cycles/day (10.40~h). Due to the large amplitude and assymetric lightcurve, we choose the double-peaked period of 2$\times$10.40~h=20.80~h.  }
\label{fig:DX98}
\end{figure}

 \paragraph{(531074) 2012~DX$_{98}$} This object was observed seven times with the \textit{Magellan-Baade} telescope and during one night with the \textit{LDT} between May 2018 and March 2019. The Lomb periodogram in Figure~\ref{fig:DX98} favored a single-peaked period of 2.31~cycles/day (i.e., 10.40~h), but the double-peaked rotational period of 20.80~h, and amplitude of 0.56~mag is preferred (Figure~\ref{fig:DX98}). The lightcurve morphology with the V- and U-shapes is characteristic of a contact binary from the shadowing effects of the two components. However, the lightcurve amplitude is below the threshold generally used to classify an object as a nearly equal-sized contact binary \citep{SheppardJewitt2004}. So, following the terminology used in \citet{ThirouinSheppard2019a}, we consider 2012~DX$_{98}$ as a likely contact binary as future observations will determine if the amplitude gets larger as the object becomes more viewed equator-on. 

If 2012~DX$_{98}$ is a contact binary, its mass ratio is between q$_{max}$=0.53 with $\rho_{max}$=5~g cm$^{-3}$ and q$_{min}$=0.47 with $\rho_{min}$=1~g cm$^{-3}$. Because q$_{min}$ and q$_{max}$ are comparable, we consider the case where q=0.5 and $\rho$=1~g cm$^{-3}$ to estimate that the axis ratios for the primary are b$_{p}$/a$_{p}$=0.97, c$_{p}$/a$_{p}$=0.94, and for the secondary b$_{s}$/a$_{s}$=0.93, c$_{s}$/a$_{s}$=0.91, while the separation between the two components is D=0.48 corresponding 252/113~km (with an albedo of 0.04/0.20). 

Even if the contact binary configuration is favored, we also consider the case of a single elongated object to estimate that the density should be larger than 0.10~g cm$^{-3}$ for a body with a/b=1.67 and c/a=0.43 \citep{Chandrasekhar1987}.

 \subsection{Moderate Amplitude Lightcurves}

   \paragraph{2004~HP$_{79}$} In about 3~h of observations, 2004~HP$_{79}$ has a variability of 0.29~mag. We do not have enough data to derive a rotational period. 
 
  \paragraph{(470083) 2006~SG$_{369}$} This object was observed over two non-consecutive nights in October 2019 over 4.5~h and 5~h, respectively. The variability is not consistent over the two observing blocks as we reported a variability of 0.38~mag and 0.08~mag. After a detailed inspection of our dataset, there is no obvious background contamination able to explain such different amplitudes. Therefore, we considered that both runs tested different phases of the lightcurve, possibly indicating a very long period for this object. For the following statistical analysis, we will use a mean amplitude of 0.23~mag.    
    
   \paragraph{(55025) 2014~WT$_{509}$} Based on one isolated night at the Magellan telescope, we report a moderate lightcurve amplitude of 0.26~mag over 5~h of observations. We re-observed this object at the \textit{LDT}, but the weather, as well as technical difficulties, resulted in low-quality data. Therefore, for our study, we will only consider the results from the \textit{Magellan-Baade} lightcurve. 
   
 \subsection{Flat and Low Amplitude Lightcurves}
 
   \paragraph{(137295) 1999~RB$_{216}$} In about 6~h, this object presents an amplitude of $\sim$0.12~mag. We imaged a minimum and part of the maximum in one night, therefore we constrain the rotational period to be about 6/12~h assuming a single-/double-peaked option. 
 
  \paragraph{2000~QL$_{251}$} This object is the only known resolved binary observed during our survey. The variability is low, about 0.15~mag over 6~h. 
  
     \paragraph{(524179) 2001~FQ$_{185}$} We report a consecutive maximum and minimum during our observations allowing us to constrain the rotational period to $\sim$6.8~h while the amplitude is $\sim$0.06~mag. 
  
 \paragraph{2001~UP$_{18}$} Based on about 5~h of observations under variable weather conditions, we report an amplitude likely lower than or near 0.2~mag for this object. 

    \paragraph{2003~UP$_{292}$}  We only have a few images for the above reasons, and can only conclude that the object's variability was low over about 1~h.  

     \paragraph{2012~KW$_{51}$} In approximately 5~h, 2012~KW$_{51}$ displayed a variability of $\sim$0.12~mag. 
     
     \paragraph{2013~GW$_{136}$} This body was observed over 2 consecutive nights at the \textit{LDT}. Because we only obtained 2 usable images on the second night, the period and amplitude constraints are based on the first night. The period of 2013~GW$_{136}$ is longer than 4~h and the amplitude is likely greater than 0.17~mag.  
   
    \paragraph{2014~GE$_{154}$}  Based on $\sim$4~h of observations with the \textit{Magellan-Baade} telescope over one night, 2014~GE$_{154}$ variability is only $\sim$0.1~mag. This dataset alone is insufficient to derive the periodicity of this object. 
    
   \paragraph{(534626) 2014~UT$_{224}$} In three isolated nights, we observed this object under variable conditions, and thus we only report a handful of images suggesting an amplitude larger than 0.1~mag over 3.5~h. 
   
    \paragraph{2017~DN$_{121}$} With only one night of data for 2017~DN$_{121}$, we can only infer that the variability of this object is low, around 0.1~mag in 5~h. 
   
   \paragraph{2002~PU$_{170}$, 2004~TV$_{357}$, 2006~SG$_{415}$, 2011~EY$_{90}$, 2012~WE$_{37}$, 2012~XR$_{157}$, and 2013~TG$_{172}$} The objects 2002~PU$_{170}$, 2006~SG$_{415}$, 2011~EY$_{90}$, 2012~WE$_{37}$, 2012~XR$_{157}$ and 2013~TG$_{172}$ were observed over one or two observing nights and they all displayed a very low variability. The only known 2:1 resonant TNO with neutral surface colors according to \citet{Sheppard2012}, 2004~TV$_{357}$, was scheduled for observations over 3 non-consecutive nights with the \textit{LDT} in 2019 and 2020. Over this amount of time, the lightcurve amplitude was very low.

\startlongtable
\begin{deluxetable*}{lccccccc|cc|cccc|c}
\tabletypesize{\scriptsize}
\tablecaption{\label{tab:Summary_photo} Observing log with the date of observations (UT-obs), number of images (N), heliocentric and geocentric distances (r$_h$ and $\Delta$),\\
and phase angle ($\alpha$) for our runs. \\
The last column presents any hints for resolved wide binary based on \textit{Hubble Space Telescope (HST)} observations: \\
(1) objects with a satellite detected are indicated with a \textit{yes}, \\
(2) objects with no detected moon with a \textit{no},\\
(3) a \textit{?} means that an object has not been observed with the \textit{HST} and thus we do not know if they have a resolved \\
companion. \\
  We also summarize our results regarding rotational period and lightcurve amplitude.  }
\tablewidth{0pt}
\tablehead{ TNO  & UT-obs  & N  & r$_h$ &  $\Delta$ & $\alpha$   & Filter & Telescope & Period & Amplitude & H$_{MPC}$ & a &e & i& Binary$^{*}$ \\
        &           &   &  [AU]  &  [AU]  &  [$^{\circ}$]   & &  & [h] & [mag] & [mag]& [AU] & & [$^\circ$]&yes/no/? }
\startdata
 \hline
(137295) 1999~RB$_{216}$ & 09/20/2020 &  9  & 33.091 & 33.851  &  1.1 & VR & LDT  & $>$6  & $>$0.15  & 7.3  & 47.814 & 0.297  & 12.7 &  no  \\ 
 \hline
 2000~QL$_{251}$ & 10/06/2019 & 8  &  39.782 & 40.773 & 0.2  & VR & LDT  & $>$6 & $>$0.15 &  6.8 & 47.935&0.220 & 3.7& yes  \\ 
 \hline
(524179) 2001~FQ$_{185}$ &  05/16/2018   & 9 &  37.002  &36.033&  0.5 & WB  & Magellan  & $\sim$6.8 & $\sim$0.06 & 6.9 &47.762 &  0.230	& 3.2& no \\
 \hline
2001~UP$_{18}$ & 09/24/2020 &  10  &   50.198 & 51.011  &  0.7 & VR & LDT  &  $>$5 & $>$0.2  & 6.0  & 47.719 & 0.080  & 1.2 &  ?  \\ 
  & 10/17/2020 &  7  &  50.056 & 51.027  &  0.3 & VR & LDT  &  ... &... & ...  & ... &...  &...&  ...  \\ 
 \hline
 2002~PU$_{170}$ & 09/20/2020 &  7  &   42.469 & 43.460  &  0.2 & VR & LDT  &  - & $\sim$0.1  & 7.1  & 47.835 & 0.220  & 1.9 &  ?  \\ 
 \hline
2002~VD$_{130}$ $^{a}$ &     12/19/2019    &  8 &   31.461 & 32.419  &  0.4 & VR & LDT  & 9.85 & 0.31$\pm$0.04 & 7.5  &47.324 & 0.317  & 3.9 & no  \\ 
    &   02/02/2020   &  6 &   31.602&  32.424& 1.0  & VR & LDT  & ... & ... & ...  &  ... &  ...& ...& ...  \\ 
    &   09/20/2020   &  6 &   32.704 &  32.454 & 1.7  & gri,VR & LDT  & ...& ...  & ...  &  ...&  ...& ...& ...  \\ 
   &  12/22/2020   & 19 &  31.505 & 32.468   & 0.4   & VR & LDT  & ...& ...  & ...  &  ...&  ...& ...& ...  \\ 
  &  01/18/2021   &  12  &   31.530 & 32.472   & 0.5   & VR & LDT  & ...& ...  & ...  &  ...&  ...& ...& ...  \\ 
     \hline     
2003~UP$_{292}$ &  12/06/2019  &3    &    27.936 & 28.900 &  0.4 & VR & LDT  & $>$1 & $>$0.1 & 7.3  &47.559 &0.414 & 13.0& ? \\ 
    &    12/19/2019     &  4  &  27.949 & 28.907 & 0.4 & VR & LDT  & ... & ... & ...  & ...& ... &... & ... \\ 
     \hline
  2004~HP$_{79}$ &   05/22/2018      & 5 &  38.877  &37.884&  0.3 & VR &  LDT & $>$3 & $>$0.29 & 6.6&48.030&0.191&2.2&  ?\\
    &   05/20/2020      & 5 &  37.852 & 38.858&  0.2 & VR &  LDT & ... & ... & ... & ... & ... & ... & ... \\
     \hline
    2004~TV$_{357}$     &   10/03/2019 & 6 &   34.397 & 34.806 & 1.5 & VR & LDT  & - & $\sim$0.1 &  6.9 &47.433&0.273	& 9.8& no \\  
    &    10/06/2019 & 7 &   34.351  &34.805  & 1.5   & VR & LDT  & ... &   ...& ... & ...& ...	&  & ... \\  
    &  02/14/2020  & 5 &  34.441 & 34.779  & 1.5  & VR  & LDT  &  ...  &  ... &  ... & ...&  ...	&  &  ... \\
     \hline
  (470083) 2006~SG$_{369}$ &  10/03/2019 & 9 & 30.802  & 31.375  & 1.5  & VR & LDT  & $>$4.5 & $>$0.38 &  7.6 &48.011&0.373& 13.6&  no \\ 
  &  10/06/2019 & 7 & 30.763 & 31.376  &1.5  & VR & LDT  & $>$5  & $>$0.08 & ...  &... &... &... & ... \\ 
   \hline
  2011~EY$_{90}$ & 05/20/2020   &  5  &  35.315 & 36.143   &  0.9 & VR & LDT  &  - & $\sim$0.1   &   7.0 & 47.986 &0.264& 8.0 & ? \\
   \hline
(531074) 2012~DX$_{98}$ $^{a}$&     05/16/2018    &  8 &  35.119  &34.306&  1.0 & WB & Magellan  & 20.80 & 0.56$\pm$0.03 & 7.3 &47.916&0.267&13.1& ? \\   
 &  05/17/2018  & 5 &   35.119 &34.316&  1.0 & WB &  Magellan & ...  &  ... & ...  &...&...&...& ... \\
& 05/18/2018  & 10 &  35.119  &34.327& 1.0  &  WB& Magellan  & ...  &  ... & ...  &...&...&...& ... \\
 &  05/19/2018  & 11 & 35.119   &34.337&  1.1 & WB & Magellan  & ...  &  ... & ...  &...&...&...& ... \\
& 05/22/2018  &10  & 35.119   &34.370& 1.1  & VR &LDT   & ...  &  ... & ...  &...&...&...& ... \\
&  02/28/2019     &   6   & 34.486  & 35.123 &  1.1  & WB & Magellan&  ... & ...& ...  &...&...&...& ...\\
& 03/01/2019       &  6   & 34.375  & 35.123 &  1.1  & WB & Magellan&  ... & ...& ... &...&...&... & ...\\
&    03/02/2019    &   6  &  34.363 & 35.123 &  1.0  & WB & Magellan&  ... & ...& ...  &...&...&...& ...\\
 \hline
(554102) 2012~KW$_{51}$ & 05/19/2020   &  8  &  38.738&  39.715  &  0.4 & VR & LDT  &  $>$5 &$>$0.12   &   6.5 &47.926 &0.242& 11.7 & ? \\
  \hline
 2012~WE$_{37}$ & 09/20/2020 &  5  & 35.533 & 36.175  &  1.2 & VR & LDT  & -  & $\sim$0.1  & 7.8  & 47.837 & 0.249  & 25.7 &  ?  \\ 
 \hline
2012~XR$_{157}$ &   11/30/2019   &  4  &   39.495&  40.414 &  0.5 & WB & Magellan  &  - &$\sim$0.1  &   6.4&47.538 & 0.222& 30.0& ? \\
&  12/01/2019    & 7   &    39.492 & 40.413  & 0.5  & WB & Magellan  & ... & ... &  ...& ...& ...& ...&...\\
&     02/14/2020    &  3 & 40.008&  40.377 & 1.3  & VR & LDT  & ... & ... & ...  &... & ...& ...&... \\   
 \hline
(577578) 2013~GW$_{136}$ &   05/19/2020   &  6  &  30.850 & 31.857  &  0.2 & VR & LDT  &  $>$4 &$>$0.17   &  7.6 & 47.929 & 0.347 & 6.7 & ? \\
    &   05/20/2020   &  3  &  30.852 &  31.857  &  0.2 & VR & LDT  & ... &...  & ... & ... & ...& ... & ... \\
 \hline
 2013~TG$_{172}$ & 09/20/2020 & 8  &   32.591 & 33.506  &  0.7 & VR & LDT  & -  &  $\sim$0.1 & 8.2  & 48.322 &0.326  & 4.8 &  ?  \\ 
 \hline     
2014~GE$_{54}$ &     05/18/2018    &  6 &   38.715 &37.824 &  0.7 & WB & Magellan  & $>$4 & $>$0.1 &6.6&48.006&0.259&17.0& ?\\
 \hline
(534626) 2014~UT$_{224}$ &  11/19/2019   &  3  &   34.662&  35.622 & 0.4  & VR & LDT  & $>$3.5 & $>$0.1 &  6.7 & 48.041 & 0.267 & 3.9& ? \\ 
&  12/19/2019   &  6  &   34.687 & 35.629 &  0.5 & VR & LDT  & ... &...  & ...  &... &... & & ... \\ 
    &  02/14/2020   &  3  &  35.385&  35.642 &  1.5 & VR & LDT  & ... & ... &...   & ...& ...&... & ... \\ 
     \hline
(535025) 2014~WT$_{509}$ &  12/02/2019    & 7   &   34.969 & 35.935 & 0.3  & WB & Magellan  & $>$5 & $>$0.26 & 7.2  &47.617 & 0.255 & 12.3& ? \\   
&  12/19/2019    & 7   &   35.024 & 35.939  & 0.6  & VR & LDT  &  ... &  ... & ...  & ...&... &... & ... \\ 
&  09/24/2020    &  6   &   35.658 & 36.007  & 1.5  & VR & LDT  &  ... &  ... & ...  & ...&... &... & ... \\ 
&  10/17/2020    &  5   &  35.341 & 36.012  & 1.2  & VR & LDT  &  ... &  ... & ...  & ...&... &... & ... \\ 
 \hline
2017~DN$_{121}$ &   02/14/2020    &  10  & 36.335 & 37.312  &  0.2 & VR & LDT  & $>$5 & $>$0.1 & 7.4  &47.493  &0.223 &15.2 & ? \\ 
 \hline  
 \hline    
 2006~SG$_{415}$$^b$ & 10/06/2019   & 8 &   32.981 & 33.976 & 1.2  & VR  & LDT  &  -  & $\sim$0.1  &  7.8 & 48.360 &0.298 	&31.3  &  ? \\
      \hline
\enddata
\tablenotetext{a}{2012~DX$_{98}$ is classified as a likely contact binary. The current lightcurve of 2002~VD$_{130}$ favors an elongated single object.  }
\tablenotetext{b}{2006~SG$_{415}$ is likely a 2:1 resonant TNO, but additional astrometry can favor/discard a Scattered Disk object orbit. 
 \tablenotetext{*}{The discovery of the binarity of 2001~QL$_{251}$ was reported in \citet{Noll2006}. Several \textit{HST} programs were awarded to search for satellites and/or derive colors: \#11113 (PI: Noll), 11644 (PI: Brown), and 12234 (PI: Fraser), but no moons were detected.  } 
}
\end{deluxetable*}
 

\section{Rotational properties}
\label{sec:dis}

\subsection{Lightcurves from the literature}
 \label{sec:literaturedis}
Only four bright 2:1 resonant TNOs have significant time-resolved photometric information in the literature (Table~\ref{Summary_resonant} and Figure~\ref{fig:Survey}). Three of them are known resolved binary systems, and thus the published lightcurves are the system's lightcurves because the components are unresolvable with ground-based observations. 

\citet{Sheppard2007} observed 2002~WC$_{19}$ over 3 nights in December 2003 with observing blocks of about 6~h, 3~h, and 5.5~h and reported a nearly flat lightcurve. Based on data obtained over three nights in January 2004, \citet{Thirouin2013} also concluded that 2002~WC$_{19}$ has a nearly flat lightcurve with an amplitude less than 0.10~mag. \citet{Benecchi2013} reported several potential rotational periods for 2010~EP$_{65}$ but it seems that the best fit is obtained for a double-peaked lightcurve with a rotational period of 14.97~h and an amplitude of 0.17~mag. \citet{Kern2006} presented 18 images obtained over one observing night of 2003~FE$_{128}$. A single-peaked lightcurve with a rotational period of 5.85~h and amplitude of 0.50~mag was derived. Unfortunately, less than half of the single-peaked lightcurve was covered during the observing time. Based on Figure~20 of \citet{Kern2006}, the amplitude of the dataset is only about 0.25~mag as that is the maximum amplitude actually observed. The 0.50~mag inferred by \citet{Kern2006} is based on the lightcurve fit but as the maximum of the curve is missing it is unclear if the fit is realistic. We use an amplitude of 0.25~mag for this object. Also, because of the moderate to potentially large amplitude the double-peaked lightcurve is maybe a better option compared to the single-peaked one \citep{Thirouin2014}. For this work, we used the double-peaked lightcurve with a period of 11.70~h. Additional data to confirm the amplitude and secure the rotational period are warranted. Several lightcurves of 1998~SM$_{165}$ have been published. The first ligtcurve was obtained by \citet{Romanishin2001} and they favored a period of 7.966~h and amplitude of 0.56~mag based on four nights of data in 1999 and 2000. However, \citet{Spencer2006} inferred a period of 8.40~h using 3 nights from December 2005. The amplitude is consistent with \citet{Romanishin2001} result and a period of 8.40~h appears to adequately fit the \citet{Romanishin2001} and the \citet{Spencer2006} datasets (J. Spencer, private communication). \citet{SheppardJewitt2002} obtained a partial lightcurve suggesting an amplitude of at least 0.45~mag and a period of at least 7.1~h. Here, we will use the period and amplitude estimated by \citet{Spencer2006}. 

Finally, several similarities between 1998~SM$_{165}$ and 2006~SG$_{369}$ are highlighted. Both objects have similar orbital elements\footnote{Orbital parameters computed by the Minor Planet Center for 1998~SM${165}$ and 2006~SG$_{369}$ are at \url{https://minorplanetcenter.net/db_search/show_object?object_id=470083} and \url{https://minorplanetcenter.net/db_search/show_object?object_id=26308}} and they might both display large lightcurve amplitudes (see this Section and Section~\ref{sec:res}). Also, they have similar surface colors with g'-r'=0.91$\pm$0.04~mag and g'-i'=1.31$\pm$0.03~mag for 2006~SG$_{369}$ according to \citet{Sheppard2012} and the colors$\footnote{The BVRI colors reported in \citet{Delsanti2001} were converted to g'r'i' using the equations from \citet{Smith2002}.}$ of 1998~SM$_{165}$ are g'-r'=0.90$\pm$0.05~mag and g'-i'=1.34$\pm$0.07~mag from \citet{Delsanti2001}. Such similar orbital elements and surface colors may indicate that these objects are a pair \citep{Vokrouhlicky2008, Abedin2021}. Numerical modeling is warranted to confirm such a find.

\subsection{Lightcurves from our survey}

Our survey observed twenty-one 2:1 resonant TNOs (22 TNOs if 2006~SG$_{415}$ is included). The TNO 2000~QL$_{251}$ is a binary system \citep{Noll2006}, but the satellite is not resolved in our images and thus we report the combined lightcurve of the primary and secondary. Four TNOs were observed with \textit{HST} and have no detected moon: 2001~FQ$_{185}$, 2002~VD$_{130}$, 2004~TV$_{357}$ and 2006~SG$_{369}$. The other TNOs have never been observed for binarity using \textit{HST} and so their binarity status is unclear. 

Most of the lightcurves reported in this work are partial as our main goal is to identify high amplitude lightcurves and thus potential contact binaries for future more intensive observations. Only the lightcurve of 2001~FQ$_{185}$ is nearly complete as this object has a rotational period consistent with the observing time spent on this object. Our survey found one likely contact binary, 2012~DX$_{98}$, and one single elongated object, 2002~VD$_{130}$.  

\startlongtable
\begin{deluxetable*}{lccccc}
\tablecaption{\label{Summary_resonant} Published lightcurves of four 2:1 resonant TNOs with information regarding rotational period, lightcurve amplitude and absolute magnitude are summarized.  }
\tablewidth{0pt}
\tablehead{
  Object & Single-peaked  & Double-peaked & $\Delta m$  & H$_{MPC}$ & Ref.$^{a}$ \\ 
              & P [h] & P [h]& [mag] & [mag] &   }
\startdata
 (26308) 1998~SM$_{165}$$^{b}$ & - & 7.966 & 0.56 & 5.8 & R01\\
                              & - &  7.1&0.45  & ... & S02\\
                              & - &  8.40$\pm$0.05 & 0.56 & ... & S06\\ 
 (119979) 2002~WC$_{19}$$^{b}$ & - & - & $<$0.05 & 4.7 & S07 \\
                              & - &  -& $<$0.10  & ... & T13 \\
  (469505) 2003~FE$_{128}$$^{b}$ & 5.85$\pm$0.15 & - & 0.50$\pm$0.14 & 6.4 & K06 \\
  (312645) 2010~EP$_{65}$$^{c}$  & 7.48 & 14.97 & 0.17$\pm$0.03 & 5.5 & B13 \\
  \hline
\enddata
 \tablenotetext{a}{References list: \\ R01: \citet{Romanishin2001}, S02: \citet{SheppardJewitt2002}, K06: \citet{Kern2006}, S06: \citet{Spencer2006}, S07: \citet{Sheppard2007}, B13: \citet{Benecchi2013}, T13: \citet{Thirouin2013}. } 
\tablenotetext{b}{Known resolved wide binaries: \citet{Brown2002}, \citet{Noll2007}, \url{http://www2.lowell.edu/users/grundy/tnbs/469505_2003_FE128.html}. 2010~EP$_{65}$ was observed by \textit{HST} program 12468 (PI: K.S. Noll) and no satellite was discovered.  }
  \tablenotetext{c}{Several aliases are reported by \citet{Benecchi2013}. }
 \end{deluxetable*}

\subsection{Amplitude and period distributions}

 \begin{figure*}
 \includegraphics[width=9.5cm,angle=0]{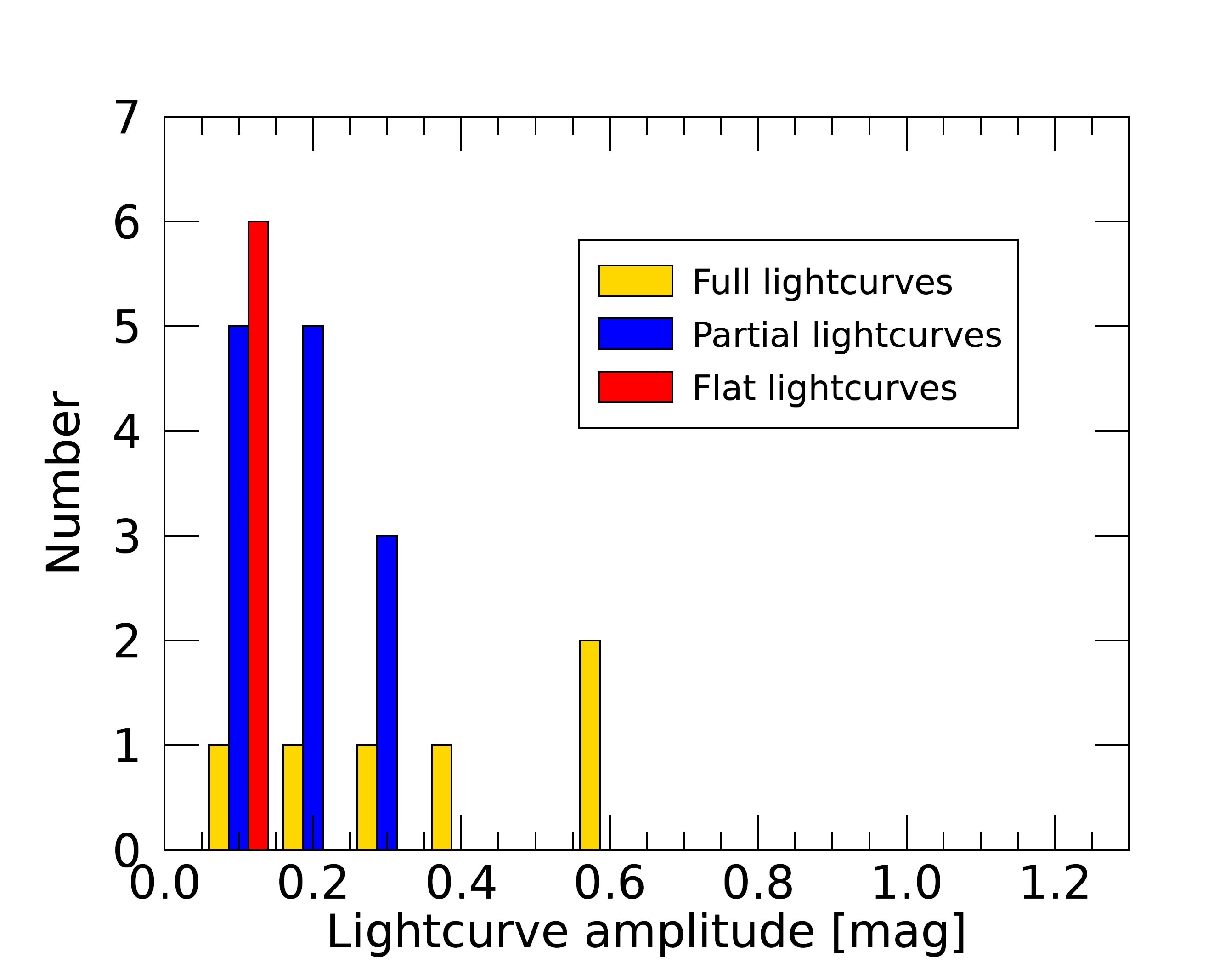}
  \includegraphics[width=9.5cm,angle=0]{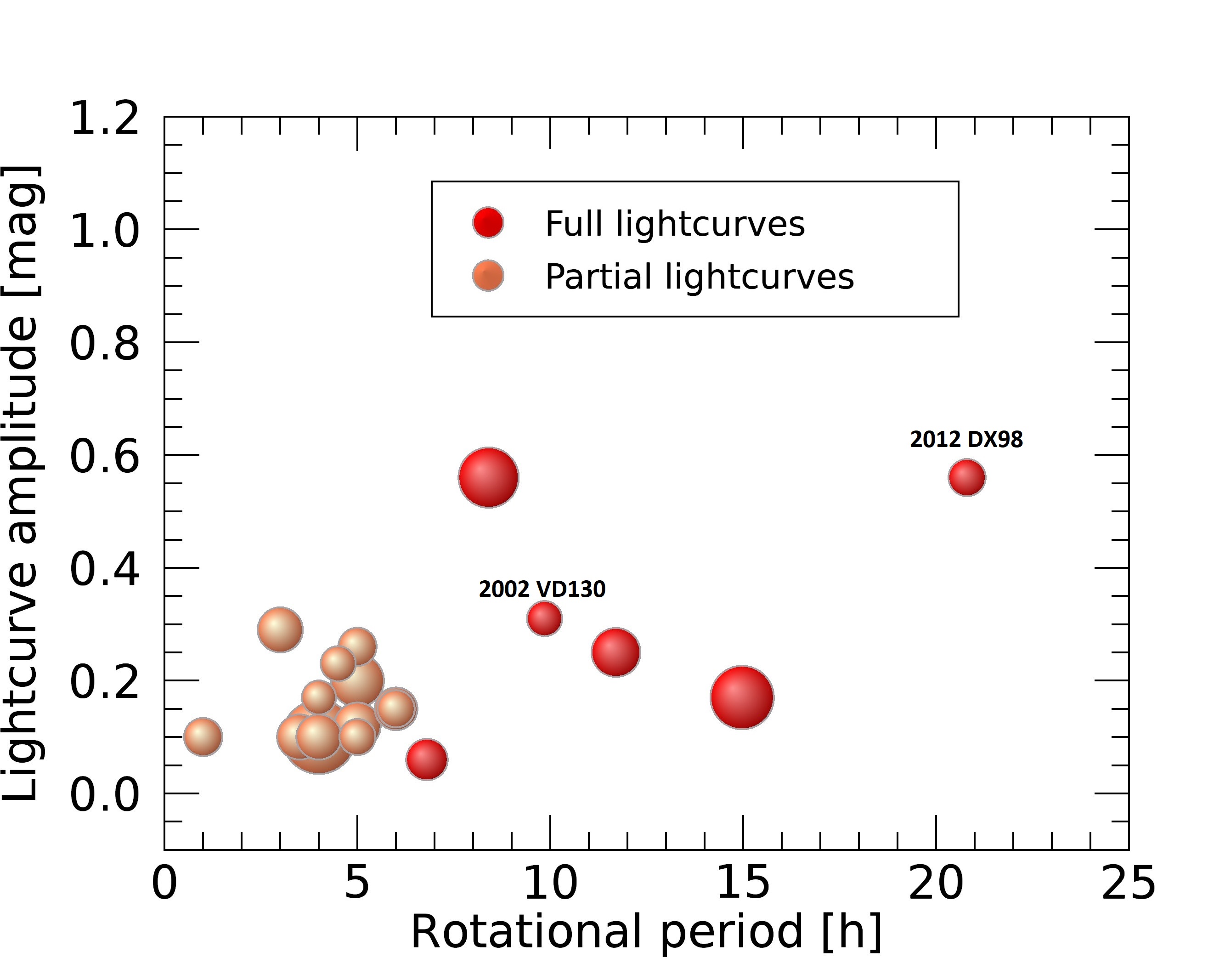}
 \caption{We summarize the number of partial, flat and full lightcurves in the 2:1 resonance (left plot). On the right, the bubble size varies with the object's absolute magnitude. There is no obvious trend between lightcurve variability and absolute magnitude. Binaries display a larger amplitude than the single objects. The two large amplitude TNOs found by our survey are indicated.}  
\label{fig:histo}
\end{figure*}

The literature and our survey report 6 full lightcurves, 12 partial lightcurves, and 7 flat lightcurves (excluding the flat lightcurve of 2006~SG$_{415}$). Two 2:1 TNOs have the large amplitudes, the likely contact binary 2012~DX$_{98}$ and the known resolved binary 1998~SM$_{165}$ (Figure~\ref{fig:histo}). The average amplitude is 0.32~mag for the full lightcurves and 0.16~mag for the partial lightcurves. The mean rotational period of the binary systems is about 13.6~h. There is not enough full lightcurve of single objects to estimate the median period, but it seems that the binaries tend to rotate slower than the single objects in other dynamical populations \citep{ThirouinSheppard2019a, ThirouinSheppard2018, Thirouin2014}.   

\subsection{Resolved and unresolved binaries}
\label{sec:fractions}
 
From this work, 2012~DX$_{98}$ is likely a nearly equal-sized contact binary whereas the lightcurve of 2002~VD$_{130}$ is best interpreted as a single triaxial object. But, the case of 1998~SM$_{165}$ has to be discussed in more detail. 1998~SM$_{165}$ has a large lightcurve amplitude of 0.56~mag (i.e., the same amplitude as 2012~DX$_{98}$), but the lightcurve does not present the typical U- and V-shapes of a contact binary \citep{Romanishin2001, Spencer2006}. The lightcurve interpretation proposed by \citet{Romanishin2001} is that 1998~SM$_{165}$ is an irregularly-shaped object. Since the lightcurve publication, a small satellite has been discovered orbiting 1998~SM$_{165}$ \citep{Grundy2011}. To summarize, based on the large lightcurve amplitude of 1998~SM$_{165}$, it can be a contact binary, but the lightcurve is not displaying the usual V- and U-shapes, and thus the lightcurve interpretation is still open to debate. Below, we will consider 1998~SM$_{165}$ as a contact binary and as an elongated object. 
 
 \citet{SheppardJewitt2004} estimated the fraction of contact binaries in the trans-Neptunian population using the following approach. The lightcurve amplitude of a small body with axis such as a$>$b and b=c varies with the angle of the object's pole relative to the perpendicular of the line sight ($\theta$): 
\begin{equation}
\Delta_m = 2.5 \log \left(\frac{1+\tan \theta}{(b/a)+\tan \theta}\right)
\label{eq:frac1}
\end{equation}
whereas the lightcurve amplitude of a triaxial ellipsoid varies as: 
\begin{equation}
\Delta_m = 2.5 \log \left(\frac{a}{b}\right) - 1.25 \log \bigg[ \left( \left(\frac{a}{b}\right)^2 -1 \right) \sin^2 \theta +1\bigg]
\label{eq:frac2}
\end{equation}
Using Equation~\ref{eq:frac1} and considering an object with a/b=3 (corresponding to a contact binary) has a lightcurve amplitude of 0.5~mag, we compute that $\theta$ has to be $\sim$36$^\circ$. The probability that Earth would lie within 36$^\circ$ of the equator of randomly oriented objects is P($\theta$$\leq$36$^\circ$)$\sim$0.59. Using the same axis ratio and amplitude cut-off, Equation~\ref{eq:frac2} infers $\theta$$\sim$34.5$^\circ$, and P($\theta$$\leq$34.5$^\circ$)$\sim$0.57. The amplitude cut-off of 0.5~mag was chosen based on the lightcurve amplitudes of 2012~DX$_{98}$ and 1998~SM$_{165}$.  

Excluding 2006~SG$_{415}$ from our sample (because its dynamical classification is uncertain), we have a total of 25 objects observed for lightcurve studies. Assuming that 2012~DX$_{98}$ is the only contact binary in the 2:1 resonance, the fraction of contact binaries is f($\Delta$m$\geq$0.5mag)$\sim$1/(25$\times$P($\theta$))$\sim$7$\%$ (same result with Equation~1 and 2). Including 1998~SM$_{165}$ as a contact binary, we estimate that f($\Delta$m$\geq$0.5mag)$\sim$2/(25$\times$P($\theta$))$\sim$14$\%$ using the previous equations. Therefore, we report a nearly equal-sized contact binary fraction at $\sim$7-14\% for the 2:1 resonance with Neptune. This is only a lower limit because of possible projection effects. Some nearly equal-sized contact binaries would still only show low amplitude lightcurves when observed near pole-on.

\citet{Noll2020} reported that the fraction of 2:1 resolved binaries with i$<$12$^\circ$ and an absolute magnitude between 5 and 8~mag is 27$_{-9}^{+16}$\% which is higher than the 5$_{-2}^{+6}$\% estimate for the 3:2 resonance, but similar to the fraction of resolved binaries in the dynamically Cold Classicals at 29$^{+7}_{-6}$\%. \citet{Noll2020} also pointed out that the resolved binaries tend to be at higher inclinations in the 3:2 resonance. The use of the widely separated equal-sized systems in the trans-Neptunian belt to constrain Neptune's migration was proposed by \citet{MurrayClay2011}. If the \textit{migration-induced capture} scenario is favored, some of the testable outcomes are (1) the 2:1 and 3:2 mean motion resonance with Neptune should have a low-inclination Cold Classical component, and (2) the low-inclination group in the 2:1 resonance should have a higher binary fraction than the high-inclination counterpart whereas the 3:2 low-inclination component should have a fraction $\sim$20-30\% lower than the Cold Classicals. \citet{Sheppard2012} demonstrated that both resonances have a Cold Classical component at low inclination based on color measurements, and the wide binary fractions reported by \citet{Noll2020} are also in agreement with a migration-induced capture scenario. Contact binaries are not considered by \citet{MurrayClay2011}, but they are also at low to moderate inclinations ($<$20$^\circ$) in the 3:2 and the 2:1 resonances \citep{ThirouinSheppard2018}. However, the higher fraction of contact binaries is in the 3:2, and not in the 2:1 as for the wide binaries.  

\citet{ThirouinSheppard2018} reported a nearly equal-sized contact binary fraction of $\sim$40-50$\%$ (corrected from projection effects) for the 3:2 mean motion resonance and \citet{ThirouinSheppard2019a} inferred a fraction up to $\sim$10-25$\%$ (corrected from projection effects) for the dynamically Cold Classical. Therefore, the estimate for the 2:1 resonance is even lower than for the Cold Classical population, but as our 2:1 sample is smaller than the Cold Classical one, we will consider that the estimates for these two sub-populations are similarly low compared to the 3:2 resonance. \citet{Nesvorny2019} predicted a contact binary fraction between 10$\%$ and 30$\%$ for the excited TNO populations (as the 2:1 resonance), therefore our fraction is consistent with the lower range of the modeling estimate. \citet{Nesvorny2019} model only considers the collapse of a wide binary as the genesis of a contact binary which is not the only proposed mechanism to create these systems \citep{Nesvorny2019, Porter2012, Nesvorny2010, Weidenschilling2002, Goldreich2002}. Also, the modeling is not focused on each TNO sub-populations, but rather on the Cold Classicals versus the excited populations. Therefore, there is a clear need for additional modeling efforts to understand the formation of contact binaries and their fractions across the trans-Neptunian belt. 

The different contact binary fractions can also be due to the past history and interactions with Neptune of these three sub-populations (e.g., \citet{Morbidelli2020, Volk2019, Nesvorny2019, Nesvorny2015, Parker2010}). The Cold Classical population had very limited interactions with Neptune whereas the 3:2 was sculpted by these interactions. The 2:1 resonant objects were also likely affected by Neptune interactions in order to get them into resonance, but they present a low ratio of contact binaries, which is unlike the 3:2 resonance but similar to the Cold Classicals. This may suggest the 2:1’s Neptune interactions were not as strong as the 3:2 resonance’s Neptune interactions, as the 2:1 also has a large number of wide binaries, again similar to the less dynamically stirred Cold Classicals and unlike the 3:2 resonance population. More simulations on how the contact binaries form through Neptune's interactions and/or migration is warranted.  

\subsection{Colors of large amplitude objects in the 2:1 resonance}

Three objects, 1998~SM$_{165}$, 2002~VD$_{130}$, and 2012~DX$_{98}$, have large lightcurve amplitudes from 0.31 to 0.56~mag. 

Based on surface color measurements, 1998~SM$_{165}$ is an ultra-red object \citep{Delsanti2001, Peixinho2015}. \citet{ThirouinSheppard2019b} evaluated that the Sloan g'r'i' surface colors of 2012~DX$_{98}$ are g'-r'=0.85$\pm$0.06~mag and g'-i'=1.25$\pm$0.06~mag. Based on its very red colors which are typical of the dynamically Cold Classical population, \citet{ThirouinSheppard2019b} suggested that 2012~DX$_{98}$ is potentially an escaped Cold Classical.   

On 2020 September 20UT, the g'r'i' colors of 2002~VD$_{130}$ were derived using the \textit{LDT}. Its colors are g'-r'=1.03$\pm$0.06~mag and g'-i'=1.35$\pm$0.06~mag, which correspond to an ultra-red object. Therefore, all 3 objects with a large lightcurve amplitude have similar surface colors possibly linking them to the dynamically Cold Classicals \citep{ThirouinSheppard2019b}.

\section{Summary and Conclusions}

The main results of this work are: 
\begin{itemize}
    \item We report short-term variability of twenty-one 2:1 resonant TNOs. With our survey and the literature, we compile eighteen TNOs showing a low to high lightcurve amplitude. 
    \item We propose the first lightcurve of 2012~DX$_{98}$ which rotates in $\sim$21~h and has a large amplitude of 0.56~mag. Based on its lightcurve morphology, 2012~DX$_{98}$ is likely a contact binary whose mass ratio is about 0.5. 
    \item The lightcurve of 2002~VD$_{130}$ with an amplitude of $\sim$0.3~mag and periodicity of 9.85~h seems to be due to a single elongated object. 
    \item The contact binary fraction in the 2:1 resonance is very low, at $\sim$7-14$\%$ which is similar to the fraction in the Cold Classical population. Modeling from \citet{Nesvorny2019} suggested that excited TNO populations should have a fraction of contact binaries of 10-30$\%$. Therefore, estimates based on observations are consistent with the lower hand of the modeling results.  
    \item The surface g'r'i' colors of 2002~VD$_{130}$ and 2012~DX$_{98}$ are very red/ultra-red which is consistent with an origin in the dynamically Cold Classical population \citep{ThirouinSheppard2019b}.
\end{itemize}

\begin{acknowledgments}
We thank two anonymous reviewers for their careful reading of this work. This paper includes data gathered with the 6.5~m Magellan-Baade Telescope located at Las Campanas Observatory, Chile. This research is based on data obtained at the Lowell Discovery Telescope (LDT, previously known as the Discovery Channel Telescope, DCT). Lowell Observatory is a private, non-profit institution dedicated to astrophysical research and public appreciation of astronomy and operates the LDT in partnership with Boston University, the University of Maryland, the University of Toledo, Northern Arizona University, and Yale University. Partial support of the LDT was provided by Discovery Communications. LMI was built by Lowell Observatory using funds from the National Science Foundation (AST-1005313). The authors acknowledge the LDT and Magellan staffs. 
The authors also acknowledge support from the National Science Foundation (NSF), grant No. AST-1734484 awarded to the ``Comprehensive Study of the Most Pristine Objects Known in the Outer Solar System'' and grant No. AST-2109207 awarded to the ``Resonant Contact Binaries in the Trans-Neptunian Belt''.  
\end{acknowledgments}

%

\facilities{Lowell Discovery Telescope (LDT), Magellan-Baade Telescope}





\bibliography{biblio}{}
\bibliographystyle{aasjournal}



\appendix

 \section{Appendix A}
 
 The photometry of all targets observed in this paper is available below. No light-time correction applied.  
 \startlongtable
\begin{deluxetable}{lccc}
 \tablecaption{\label{Tab:Summary_photo2}   }
\tablewidth{0pt}
\tablehead{
TNO  & Julian Date & Relative Magnitude &  Error  \\
        &          &   [mag]           &  [mag]\\
}
\startdata 
1999~RB$_{216}$   &   &   &   \\
&	2459112.74797	&	-0.07	&	0.04	\\
&	2459112.76947	&	0.01	&	0.04	\\
&	2459112.79098	&	0.05	&	0.04	\\
&	2459112.81929	&	0.08	&	0.04	\\
&	2459112.84876	&	0.01	&	0.03	\\
&	2459112.87824	&	-0.07	&	0.03	\\
&	2459112.90750	&	-0.05	&	0.03	\\
&	2459112.93563	&	-0.07	&	0.03	\\
&	2459113.00314	&	0.00	&	0.04	\\
\hline
2000~QL$_{251}$   &   &   &   \\
&	2458762.73490	&	-0.04	&	0.04	\\
&	2458762.76565	&	0.03	&	0.04	\\
&	2458762.83763	&	0.03	&	0.03	\\
&	2458762.87612	&	0.00	&	0.03	\\
&	2458762.94669	&	-0.04	&	0.04	\\
&	2458762.97795	&	-0.12	&	0.05	\\
\hline
2001~FQ$_{185}$   &   &   &   \\
&	2458254.51509	&	0.01	&	0.05	\\
&	2458254.53911	&	-0.01	&	0.05	\\
&	2458254.58800	&	-0.02	&	0.05	\\
&	2458254.61038	&	-0.02	&	0.05	\\
&	2458254.63125	&	-0.02	&	0.05	\\
&	2458254.68931	&	0.01	&	0.06	\\
&	2458254.74416	&	0.04	&	0.06	\\
&	2458254.78098	&	0.02	&	0.06	\\
&	2458254.81173	&	0.00	&	0.06	\\
\hline 
2001~UP$_{18}$   &   &   &   \\
&	2459116.76200	&	0.00	&	0.10	\\
&	2459116.78163	&	0.03	&	0.11	\\
&	2459116.81072	&	0.03	&	0.10	\\
&	2459116.83948	&	0.19	&	0.10	\\
&	2459116.86734	&	0.05	&	0.10	\\
&	2459116.89474	&	-0.02	&	0.09	\\
&	2459116.92245	&	-0.07	&	0.10	\\
&	2459116.95027	&	-0.17	&	0.10	\\
&	2459116.97821	&	-0.03	&	0.10	\\
&	2459139.71629	&	-0.08	&	0.07	\\
&	2459139.83822	&	-0.06	&	0.07	\\
&	2459139.86728	&	0.00	&	0.07	\\
&	2459139.89683	&	0.01	&	0.08	\\
&	2459139.94693	&	0.03	&	0.08	\\
&	2459139.97563	&	0.00	&	0.08	\\
&	2459140.00518	&	0.10	&	0.10	\\
\hline 
2002~PU$_{170}$   &   &   &   \\
&	2459112.65877	&	-0.03	&	0.08	\\
&	2459112.68620	&	-0.03	&	0.07	\\
&	2459112.71814	&	0.00	&	0.06	\\
&	2459112.77651	&	0.02	&	0.06	\\
&	2459112.83442	&	0.02	&	0.06	\\
&	2459112.86365	&	0.01	&	0.07	\\
&	2459112.89305	&	0.00	&	0.07	\\
\hline
2002~VD$_{130}$ &&&\\
&	2458836.89721	&	-0.15	&	0.06	\\
&	2458836.92818	&	-0.10	&	0.06	\\
&	2458836.94618	&	-0.07	&	0.06	\\
&	2458836.96362	&	0.00	&	0.06	\\
&	2458836.98115	&	0.09	&	0.07	\\
&	2458836.99845	&	0.10	&	0.06	\\
&	2458837.01730	&	0.09	&	0.07	\\
&	2458881.74653	&	0.14	&	0.06	\\
&	2458881.76728	&	0.17	&	0.06	\\
&	2458881.79936	&	0.11	&	0.06	\\
&	2458881.83335	&	-0.11	&	0.05	\\
&	2458881.86067	&	-0.24	&	0.04	\\
&	2458881.88736	&	-0.18	&	0.05	\\
&	2459112.92134	&	0.10	&	0.05	\\
&	2459112.94969	&	-0.10	&	0.04	\\
&	2459205.67903	&	0.01	&	0.04	\\
&	2459205.70198	&	-0.08	&	0.03	\\
&	2459205.72354	&	-0.19	&	0.03	\\
&	2459205.74507	&	-0.12	&	0.03	\\
&	2459205.76718	&	-0.10	&	0.03	\\
&	2459205.78903	&	-0.01	&	0.03	\\
&	2459205.80974	&	0.10	&	0.03	\\
&	2459205.83027	&	0.17	&	0.03	\\
&	2459205.85085	&	0.11	&	0.03	\\
&	2459205.87140	&	0.00	&	0.03	\\
&	2459205.92695	&	-0.19	&	0.03	\\
&	2459205.94763	&	-0.15	&	0.03	\\
&	2459205.97008	&	-0.06	&	0.03	\\
&	2459205.98636	&	0.01	&	0.03	\\
&	2459206.00259	&	0.00	&	0.03	\\
&	2459206.01887	&	0.02	&	0.03	\\
&	2459206.03494	&	0.14	&	0.04	\\
&	2459206.04207	&	0.17	&	0.04	\\
&	2459206.04917	&	0.17	&	0.05	\\
&	2459232.61988	&	-0.16	&	0.03	\\
&	2459232.67510	&	0.02	&	0.03	\\
&	2459232.72134	&	0.18	&	0.03	\\
&	2459232.75327	&	0.04	&	0.03	\\
&	2459232.80425	&	-0.16	&	0.03	\\
&	2459232.82945	&	-0.11	&	0.03	\\
&	2459232.85436	&	-0.10	&	0.03	\\
&	2459232.88012	&	-0.01	&	0.03	\\
&	2459232.90582	&	0.08	&	0.03	\\
&	2459232.92018	&	0.12	&	0.03	\\
&	2459232.93456	&	0.13	&	0.03	\\
&	2459232.96105	&	-0.02	&	0.03	\\
\hline 
2003~UP$_{292}$ &&&\\
&	2458823.63995	&	-0.04	&	0.09	\\
&	2458823.65549	&	0.00	&	0.05	\\
&	2458823.67951	&	0.04	&	0.05	\\
&	2458836.61365	&	0.01	&	0.07	\\
&	2458836.63000	&	0.00	&	0.06	\\
&	2458836.63778	&	-0.10	&	0.06	\\
\hline 
2004~HP$_{79}$ &&&\\
&	2458260.72835	&	0.13	&	0.07	\\
&	2458260.75631	&	0.17	&	0.07	\\
&	2458260.79835	&	-0.02	&	0.05	\\
&	2458260.82598	&	0.00	&	0.05	\\
&	2458260.85343	&	-0.03	&	0.04	\\
&	2458989.77376	&	-0.14	&	0.06	\\
&	2458989.79988	&	-0.07	&	0.06	\\
&	2458989.82586	&	0.07	&	0.07	\\
&	2458989.85142	&	0.15	&	0.09	\\
\hline 
2004~TV$_{357}$ &&&\\
&	2458759.82999	&	-0.05	&	0.03	\\
&	2458759.85793	&	-0.02	&	0.03	\\
&	2458759.88218	&	0.01	&	0.03	\\
&	2458759.90660	&	0.03	&	0.03	\\
&	2458759.93094	&	0.00	&	0.03	\\
&	2458759.95523	&	0.00	&	0.03	\\
&	2458759.97936	&	0.00	&	0.03	\\
&	2458760.00344	&	0.03	&	0.03	\\
&	2458762.85756	&	0.02	&	0.02	\\
&	2458762.89594	&	-0.01	&	0.02	\\
&	2458762.93457	&	0.04	&	0.03	\\
&	2458762.96664	&	0.04	&	0.02	\\
&	2458762.99734	&	0.00	&	0.02	\\
&	2458762.81880	&	-0.01	&	0.02	\\
&	2458893.59921	&	0.04	&	0.03	\\
&	2458893.67778	&	0.00	&	0.02	\\
&	2458893.71174	&	-0.01	&	0.02	\\
&	2458893.75144	&	-0.02	&	0.02	\\
&	2458893.78930	&	0.00	&	0.04	\\
\hline 
2006~SG$_{369}$ &&&\\
&	2458759.82338	&	-0.15	&	0.04	\\
&	2458759.85116	&	-0.22	&	0.04	\\
&	2458759.87644	&	-0.10	&	0.04	\\
&	2458759.90056	&	-0.12	&	0.04	\\
&	2458759.92491	&	0.00	&	0.04	\\
&	2458759.94936	&	0.02	&	0.04	\\
&	2458759.97341	&	0.16	&	0.05	\\
&	2458759.99745	&	0.10	&	0.04	\\
&	2458760.01537	&	0.11	&	0.06	\\
&	2458762.78258	&	0.02	&	0.04	\\
&	2458762.81531	&	0.06	&	0.03	\\
&	2458762.85389	&	0.04	&	0.03	\\
&	2458762.89244	&	0.00	&	0.03	\\
&	2458762.93102	&	0.00	&	0.03	\\
&	2458762.96300	&	-0.03	&	0.03	\\
&	2458762.99376	&	-0.03	&	0.03	\\
\hline 
2006~SG$_{415}$ &&&\\
&	2458762.64956	&	0.08	&	0.06	\\
&	2458762.66966	&	0.11	&	0.07	\\
&	2458762.83007	&	-0.11	&	0.03	\\
&	2458762.86883	&	0.02	&	0.04	\\
&	2458762.90713	&	-0.08	&	0.04	\\
&	2458762.63417	&	0.00	&	0.06	\\
\hline  
2011~EY$_{90}$ &&&\\
&	2458989.70527	&	0.09	&	0.08	\\
&	2458989.72632	&	0.00	&	0.06	\\
&	2458989.75363	&	0.07	&	0.06	\\
&	2458989.78008	&	-0.01	&	0.06	\\
&	2458989.80622	&	0.00	&	0.07	\\
\hline  
2012~DX$_{98}$ &&&\\
&	2458254.50217	&	-0.14	&	0.05	\\
&	2458254.52781	&	-0.14	&	0.04	\\
&	2458254.55497	&	-0.20	&	0.04	\\
&	2458254.57684	&	-0.19	&	0.04	\\
&	2458254.59943	&	-0.27	&	0.04	\\
&	2458254.62117	&	-0.24	&	0.04	\\
&	2458254.67869	&	-0.12	&	0.04	\\
&	2458254.73377	&	0.05	&	0.05	\\
&	2458255.48029	&	-0.15	&	0.04	\\
&	2458255.57953	&	0.01	&	0.04	\\
&	2458255.62615	&	0.15	&	0.04	\\
&	2458255.68974	&	0.26	&	0.04	\\
&	2458255.77080	&	0.04	&	0.05	\\
&	2458256.46938	&	0.01	&	0.04	\\
&	2458256.49453	&	0.20	&	0.04	\\
&	2458256.51369	&	0.20	&	0.05	\\
&	2458256.55033	&	0.35	&	0.04	\\
&	2458256.57975	&	0.27	&	0.04	\\
&	2458256.60886	&	0.20	&	0.04	\\
&	2458256.65415	&	0.03	&	0.04	\\
&	2458256.69962	&	-0.15	&	0.04	\\
&	2458256.72320	&	-0.17	&	0.04	\\
&	2458256.74463	&	-0.13	&	0.05	\\
&	2458257.47652	&	0.24	&	0.05	\\
&	2458257.48830	&	0.23	&	0.05	\\
&	2458257.49420	&	0.15	&	0.05	\\
&	2458257.51570	&	0.03	&	0.05	\\
&	2458257.52153	&	0.04	&	0.04	\\
&	2458257.54486	&	-0.01	&	0.04	\\
&	2458257.57767	&	-0.15	&	0.04	\\
&	2458257.58351	&	-0.19	&	0.04	\\
&	2458257.58940	&	-0.19	&	0.04	\\
&	2458257.61967	&	-0.14	&	0.04	\\
&	2458257.69655	&	-0.14	&	0.04	\\
&	2458260.69950	&	-0.16	&	0.13	\\
&	2458260.70639	&	-0.11	&	0.13	\\
&	2458260.72017	&	-0.13	&	0.06	\\
&	2458260.73542	&	-0.11	&	0.07	\\
&	2458260.74945	&	-0.08	&	0.07	\\
&	2458260.76318	&	-0.06	&	0.09	\\
&	2458260.79135	&	0.04	&	0.09	\\
&	2458260.80522	&	0.07	&	0.09	\\
&	2458260.81917	&	0.03	&	0.09	\\
&	2458260.83292	&	0.21	&	0.09	\\
&	2458542.65777	&	0.05	&	0.04	\\
&	2458542.69574	&	-0.05	&	0.03	\\
&	2458542.73925	&	-0.15	&	0.03	\\
&	2458542.80342	&	-0.15	&	0.03	\\
&	2458542.84969	&	-0.15	&	0.03	\\
&	2458542.88473	&	-0.09	&	0.03	\\
&	2458543.70326	&	-0.15	&	0.03	\\
&	2458543.77856	&	0.01	&	0.03	\\
&	2458543.81837	&	0.06	&	0.03	\\
&	2458543.88160	&	0.29	&	0.04	\\
&	2458544.66612	&	-0.01	&	0.03	\\
&	2458544.71304	&	0.14	&	0.03	\\
&	2458544.75924	&	0.28	&	0.03	\\
&	2458544.79724	&	0.17	&	0.03	\\
&	2458544.84580	&	0.01	&	0.03	\\
&	2458544.89477	&	-0.07	&	0.04	\\
\hline  
2012~KW$_{51}$ &&&\\
&	2458988.67237	&	0.01	&	0.04	\\
&	2458988.69880	&	0.00	&	0.03	\\
&	2458988.73188	&	-0.04	&	0.03	\\
&	2458988.75694	&	-0.01	&	0.03	\\
&	2458988.79030	&	0.00	&	0.03	\\
&	2458988.82268	&	-0.01	&	0.03	\\
&	2458988.85507	&	0.04	&	0.03	\\
&	2458988.88425	&	0.09	&	0.04	\\
\hline  
2012~WE$_{37}$ &&&\\
&	2459112.82663	&	0.00	&	0.06	\\
&	2459112.85552	&	-0.02	&	0.06	\\
&	2459112.88510	&	0.01	&	0.06	\\
&	2459112.91424	&	-0.04	&	0.05	\\
&	2459112.94239	&	0.03	&	0.05	\\
\hline  
2012~XR$_{157}$ &&&\\
&	2458817.78954	&	0.02	&	0.04	\\
&	2458817.81494	&	-0.02	&	0.03	\\
&	2458817.84533	&	0.02	&	0.03	\\
&	2458817.85522	&	-0.04	&	0.04	\\
&	2458818.59544	&	0.00	&	0.04	\\
&	2458818.63797	&	0.00	&	0.04	\\
&	2458818.67970	&	-0.07	&	0.03	\\
&	2458893.68362	&	0.08	&	0.03	\\
&	2458893.71761	&	0.00	&	0.03	\\
&	2458893.75723	&	-0.02	&	0.03	\\
\hline  
2013~GW$_{136}$ &&&\\
&	2458988.70526	&	-0.06	&	0.04	\\
&	2458988.73810	&	-0.05	&	0.04	\\
&	2458988.79652	&	0.00	&	0.04	\\
&	2458988.82889	&	0.06	&	0.04	\\
&	2458988.86515	&	0.11	&	0.05	\\
&	2458989.83175	&	-0.04	&	0.05	\\
&	2458989.85723	&	0.04	&	0.05	\\
\hline  
2013~TG$_{172}$ &&&\\
&	2459112.72583	&	-0.01	&	0.06	\\
&	2459112.74100	&	0.05	&	0.06	\\
&	2459112.78414	&	0.00	&	0.05	\\
&	2459112.81243	&	0.06	&	0.05	\\
&	2459112.84196	&	0.02	&	0.05	\\
&	2459112.90053	&	-0.03	&	0.05	\\
&	2459112.92879	&	0.00	&	0.05	\\
&	2459112.95725	&	-0.01	&	0.06	\\
\hline 
2014~GE$_{54}$ &&&\\
&	2458256.54462	&	0.04	&	0.05	\\
&	2458256.57386	&	0.01	&	0.05	\\
&	2458256.60303	&	0.01	&	0.04	\\
&	2458256.64606	&	-0.01	&	0.05	\\
&	2458256.66951	&	-0.04	&	0.05	\\
&	2458256.71785	&	-0.05	&	0.06	\\
\hline  
2014~UT$_{224}$ &&&\\
&	2458806.86524	&	-0.03	&	0.03	\\
&	2458806.89122	&	0.00	&	0.04	\\
&	2458806.91727	&	0.05	&	0.05	\\
&	2458836.77668	&	0.00	&	0.03	\\
&	2458836.81671	&	-0.03	&	0.03	\\
&	2458836.84798	&	0.01	&	0.03	\\
&	2458836.87970	&	-0.11	&	0.04	\\
&	2458836.91051	&	-0.06	&	0.06	\\
&	2458893.69049	&	-0.05	&	0.04	\\
&	2458893.72372	&	0.00	&	0.04	\\
&	2458893.76271	&	0.04	&	0.05	\\
\hline  
2014~WT$_{509}$ &&&\\
&	2458819.58651	&	-0.21	&	0.04	\\
&	2458819.62462	&	-0.16	&	0.04	\\
&	2458819.70264	&	-0.04	&	0.04	\\
&	2458819.73390	&	0.04	&	0.04	\\
&	2458819.76861	&	0.04	&	0.04	\\
&	2458819.80816	&	0.06	&	0.04	\\
&	2458836.76352	&	-0.06	&	0.05	\\
&	2458836.80983	&	-0.06	&	0.05	\\
&	2458836.84161	&	0.13	&	0.08	\\
&	2458836.87335	&	0.00	&	0.08	\\
&	2458836.90418	&	0.12	&	0.12	\\
&	2459116.84736	&	-0.06	&	0.06	\\
&	2459116.87499	&	-0.07	&	0.06	\\
&	2459116.90246	&	-0.01	&	0.06	\\
&	2459116.93023	&	0.01	&	0.05	\\
&	2459116.95822	&	0.04	&	0.06	\\
&	2459116.98597	&	0.16	&	0.06	\\
&	2459139.85249	&	-0.11	&	0.10	\\
&	2459139.88183	&	-0.07	&	0.08	\\
&	2459139.91759	&	0.04	&	0.05	\\
&	2459139.96195	&	0.05	&	0.06	\\
&	2459139.99140	&	0.09	&	0.04	\\
\hline  
2017~DN$_{121}$ &&&\\
&	2458893.72750	&	0.04	&	0.02	\\
&	2458893.76637	&	0.03	&	0.02	\\
&	2458893.79362	&	0.06	&	0.03	\\
&	2458893.82124	&	0.00	&	0.03	\\
&	2458893.84135	&	-0.02	&	0.03	\\
&	2458893.86146	&	-0.04	&	0.03	\\
&	2458893.90200	&	0.04	&	0.05	\\
&	2458893.69531	&	-0.01	&	0.03	\\
\hline 
\enddata
\end{deluxetable}

\clearpage

 \section*{Appendix B} 

Sparse lightcurves discussed in this paper are in Appendix~B. 
\begin{figure*}
\includegraphics[width=9.5cm, angle=0]{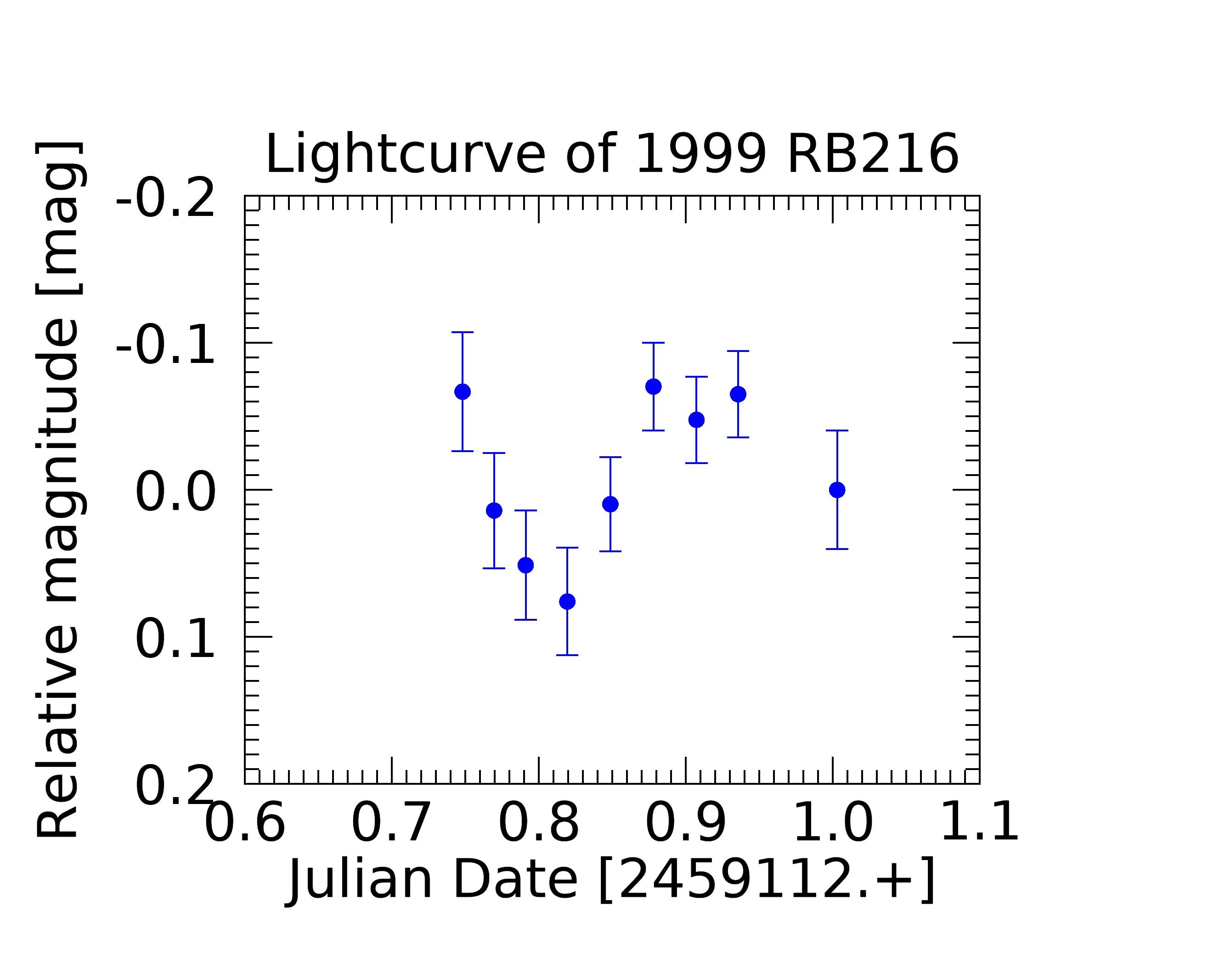}
\includegraphics[width=9.5cm, angle=0]{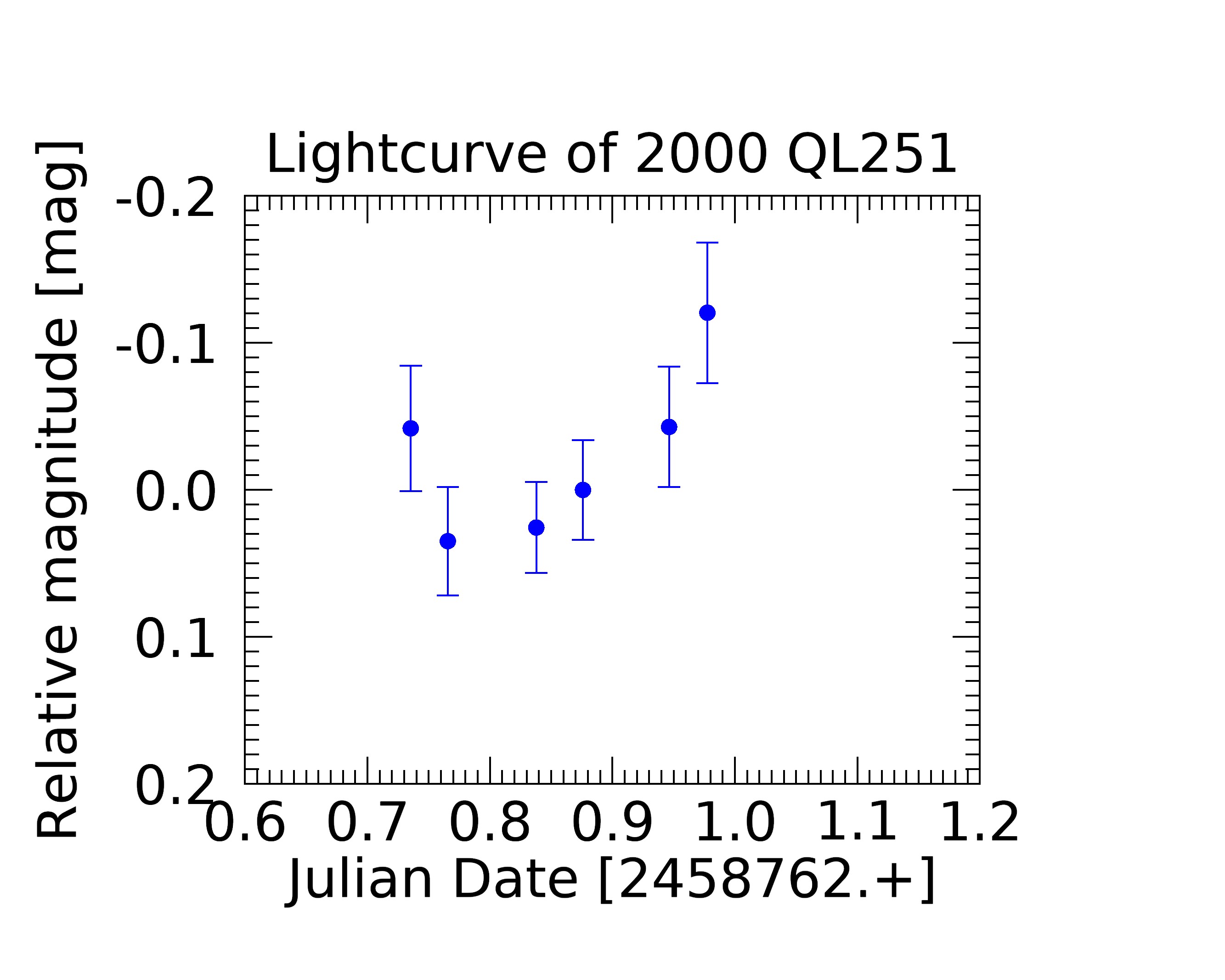}
\includegraphics[width=9.5cm, angle=0]{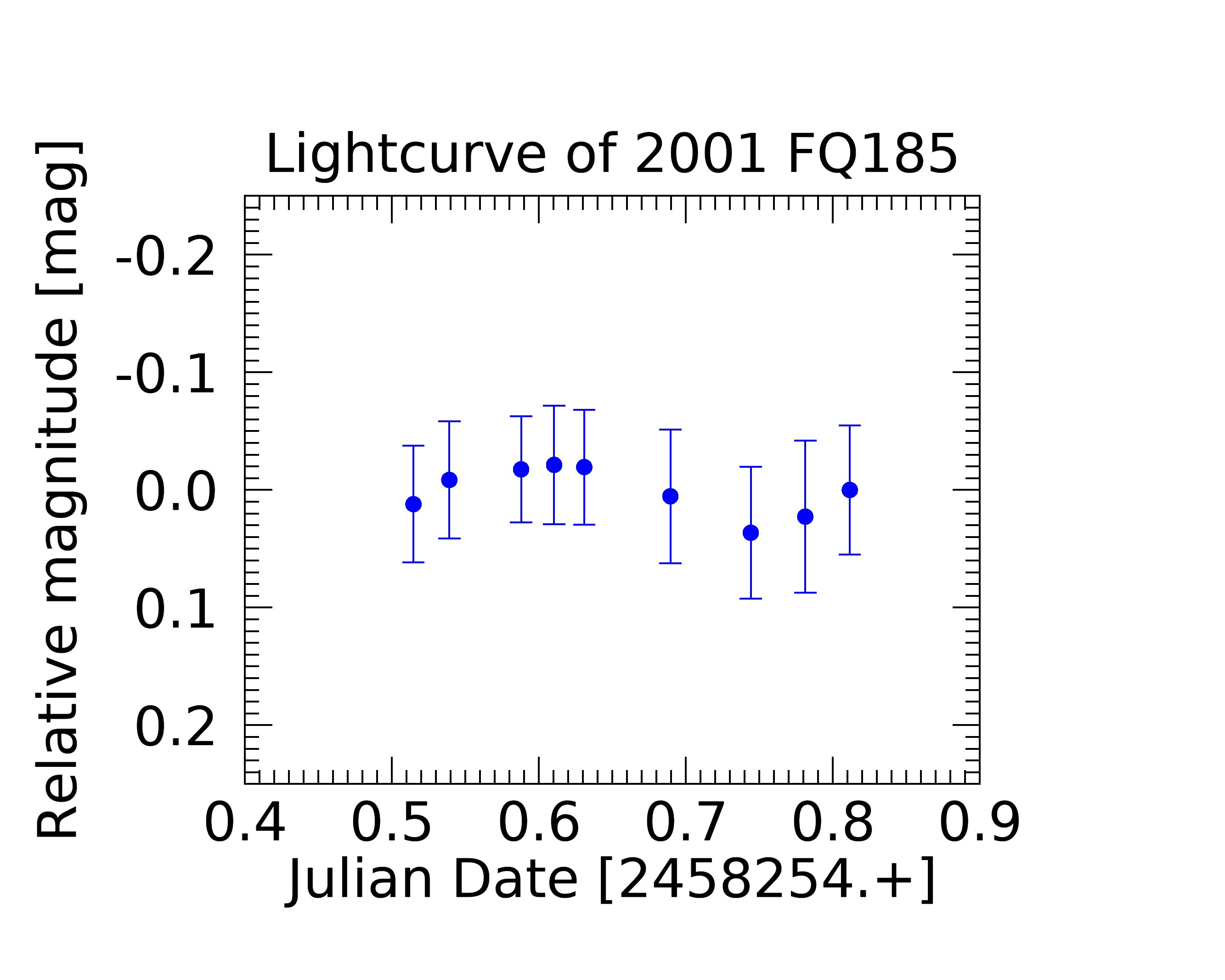}
\includegraphics[width=9.5cm, angle=0]{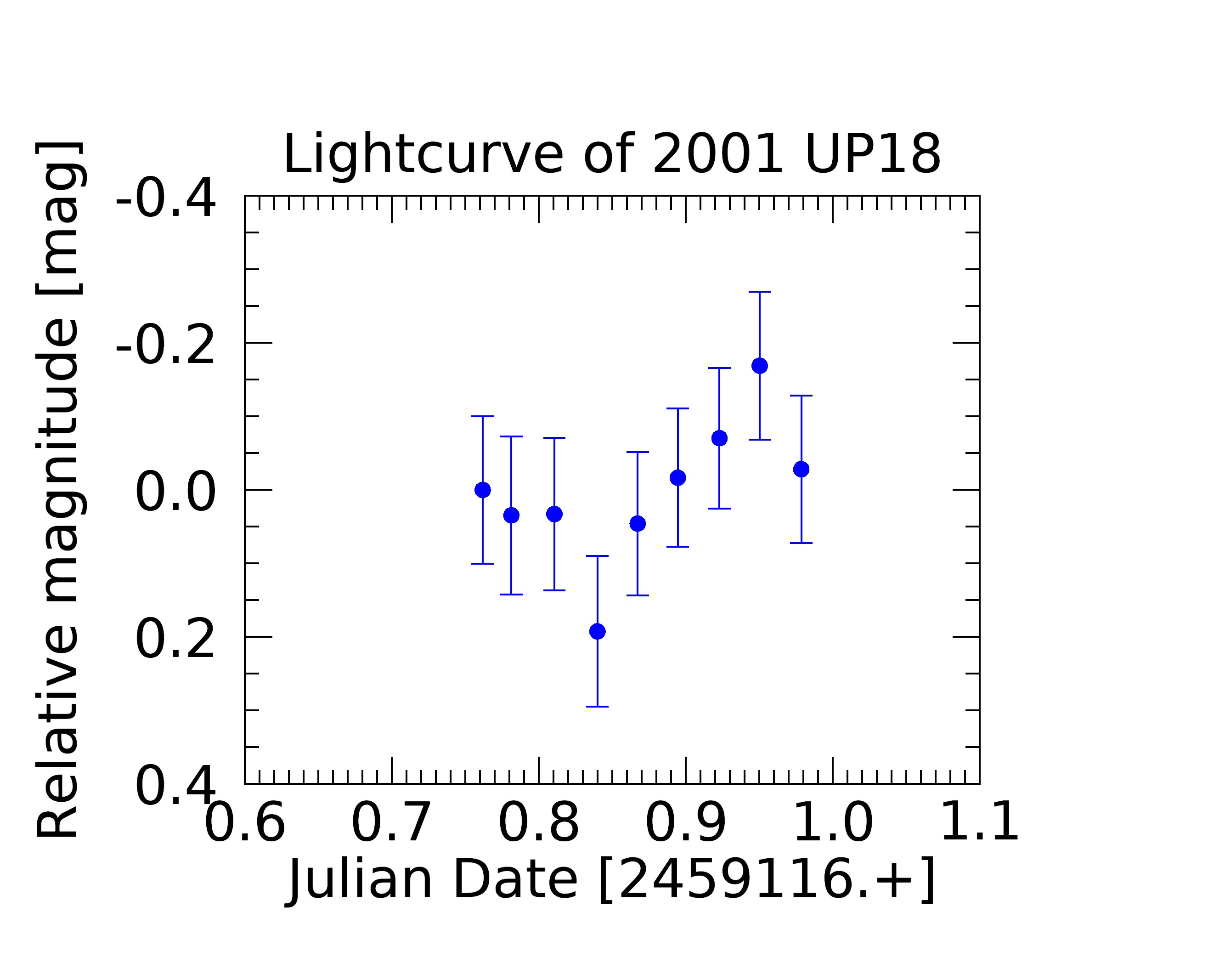}
\includegraphics[width=9.5cm, angle=0]{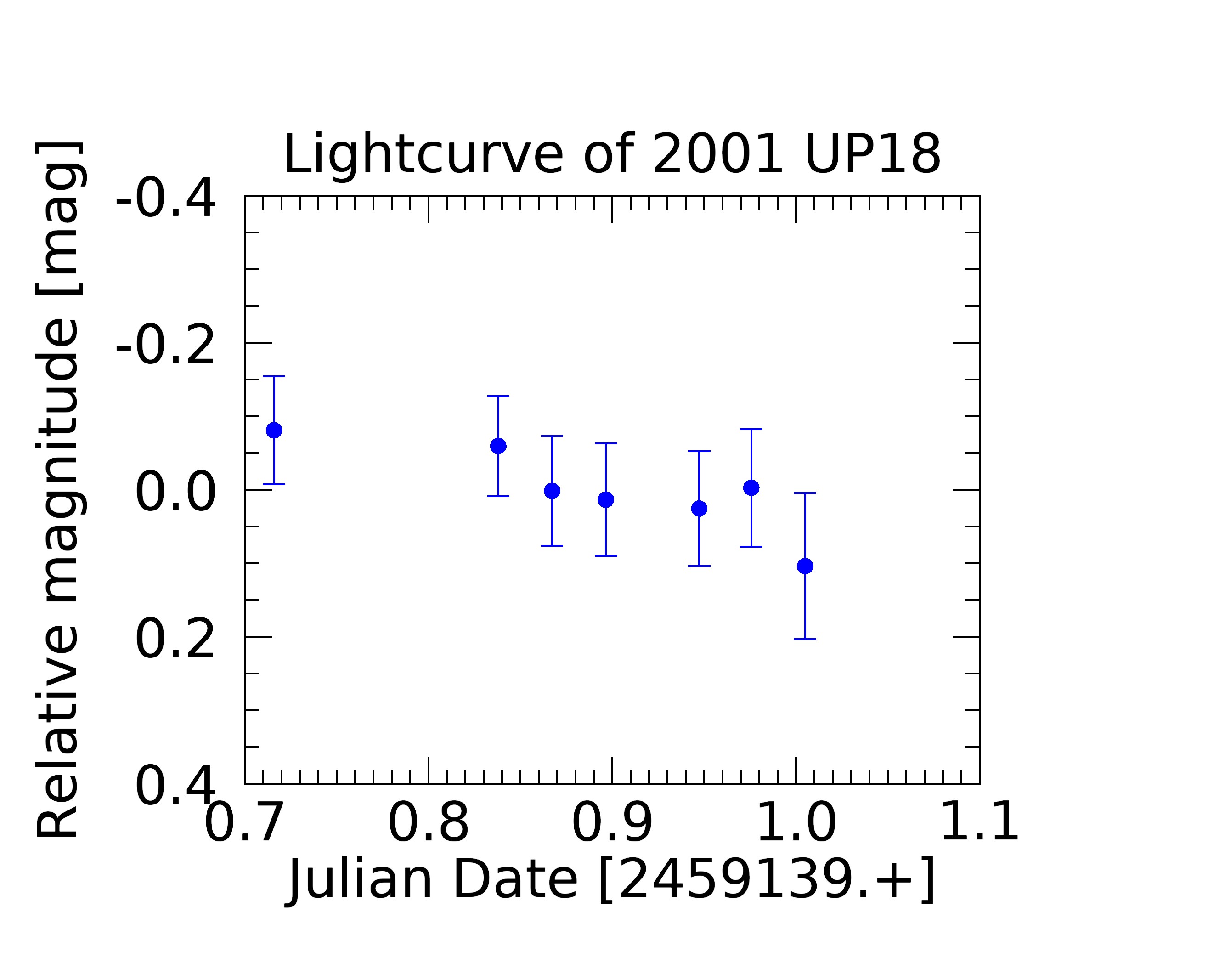}
\includegraphics[width=9.5cm, angle=0]{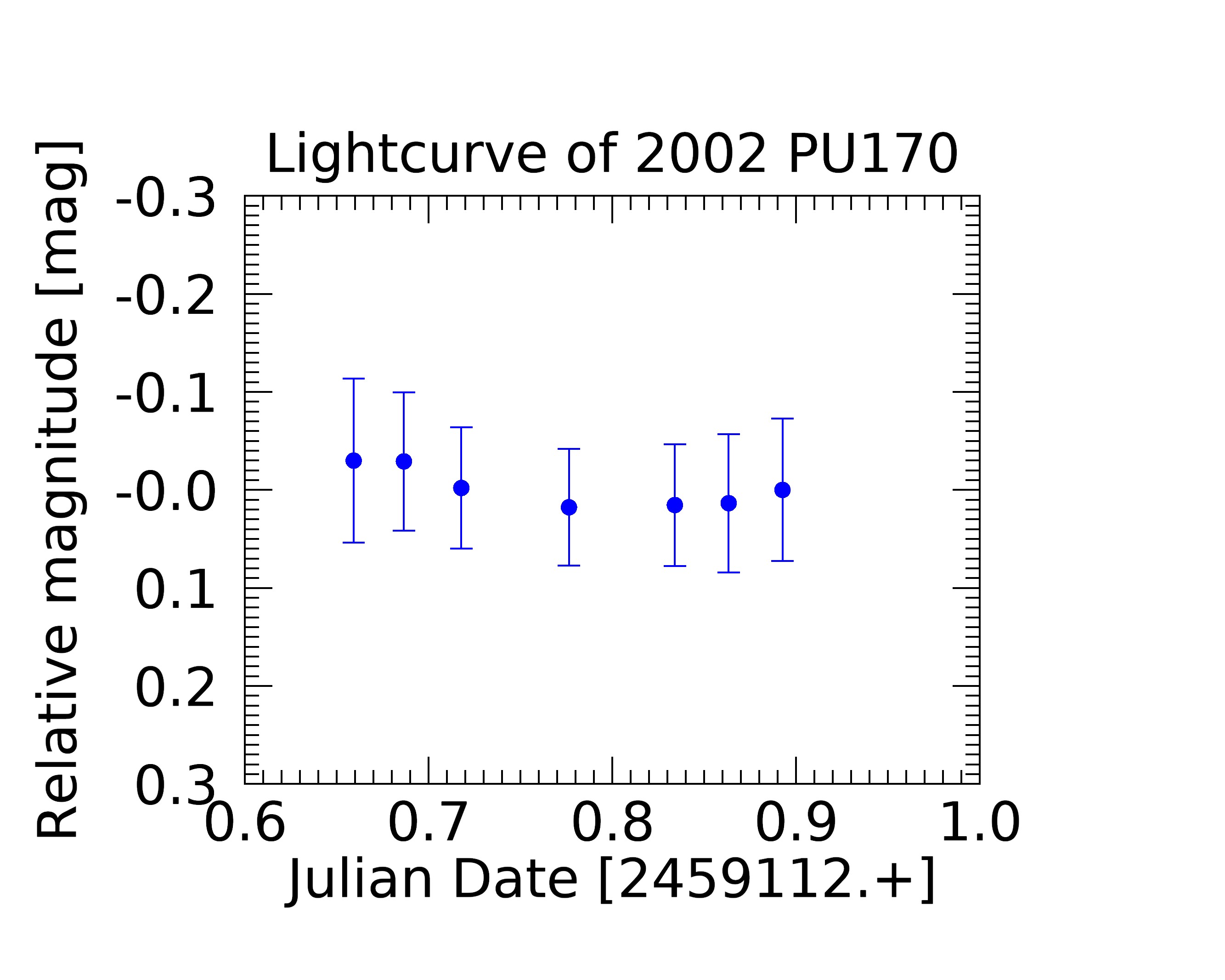}
\caption{\textit{Objects in the 2:1 mean motion resonance with Neptune }   }
\label{fig:LC21}
\end{figure*}

\begin{figure*}
\includegraphics[width=9.5cm, angle=0]{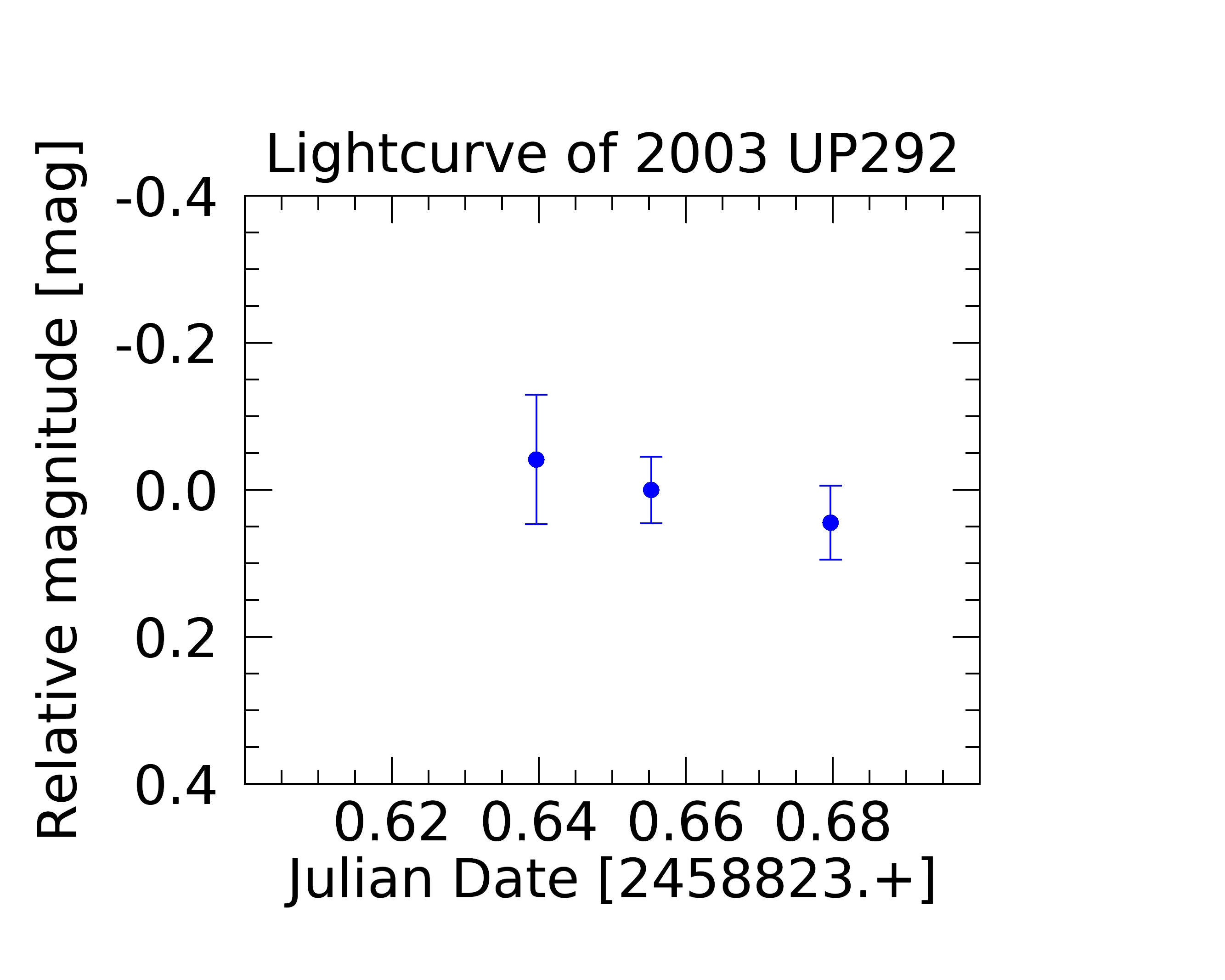}
\includegraphics[width=9.5cm, angle=0]{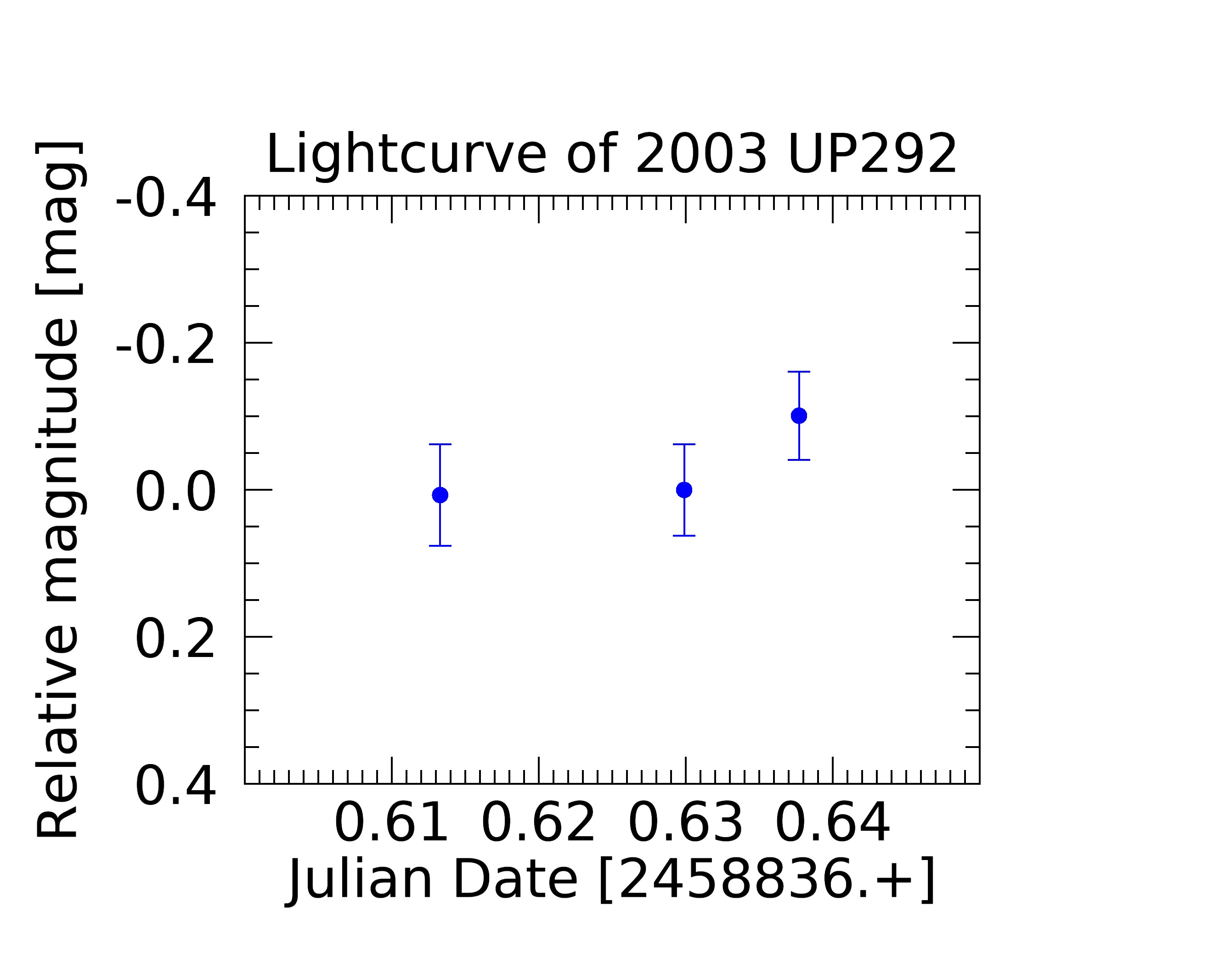}
\includegraphics[width=9.5cm, angle=0]{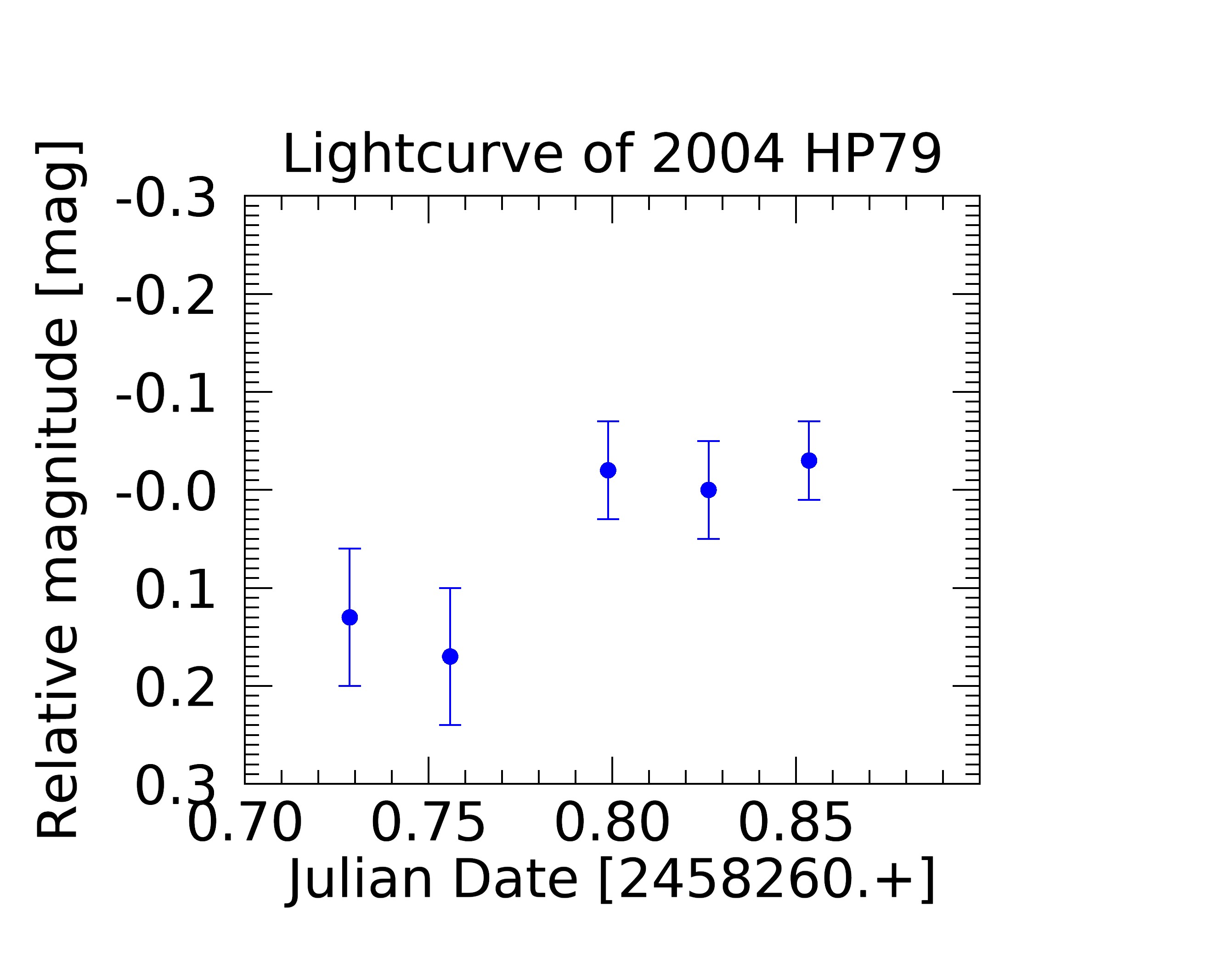}
\includegraphics[width=9.5cm, angle=0]{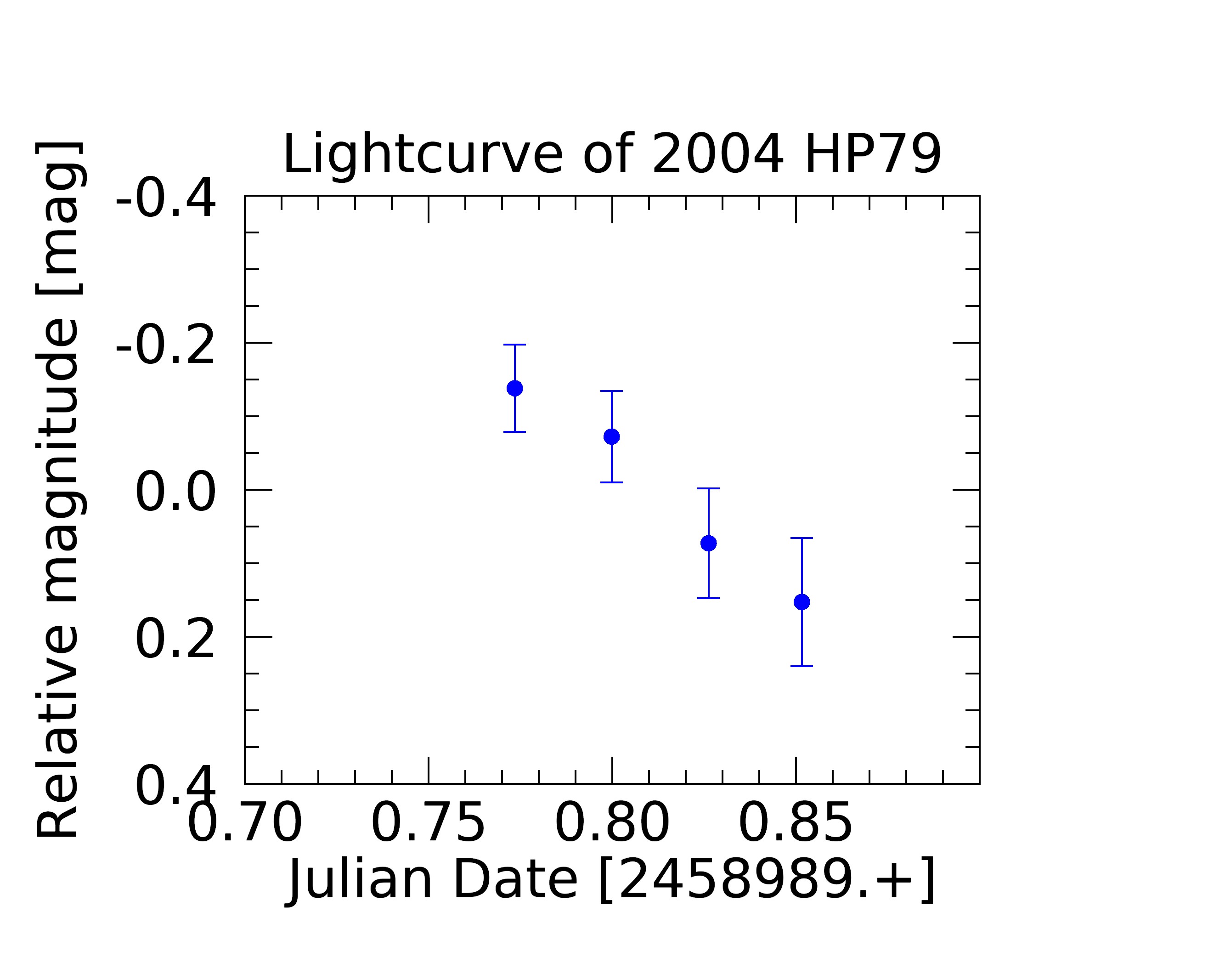}
 \includegraphics[width=9.5cm, angle=0]{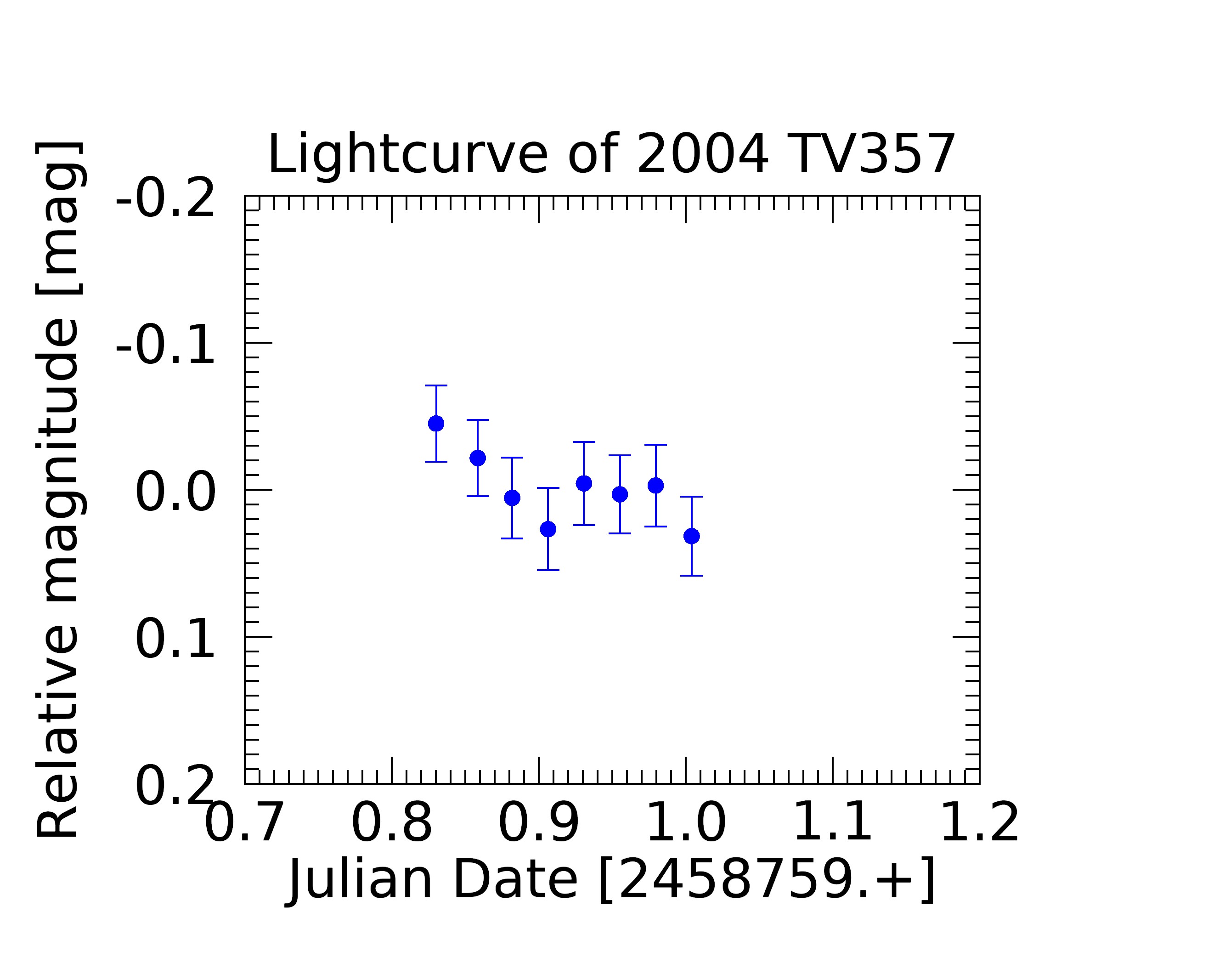}
 \includegraphics[width=9.5cm, angle=0]{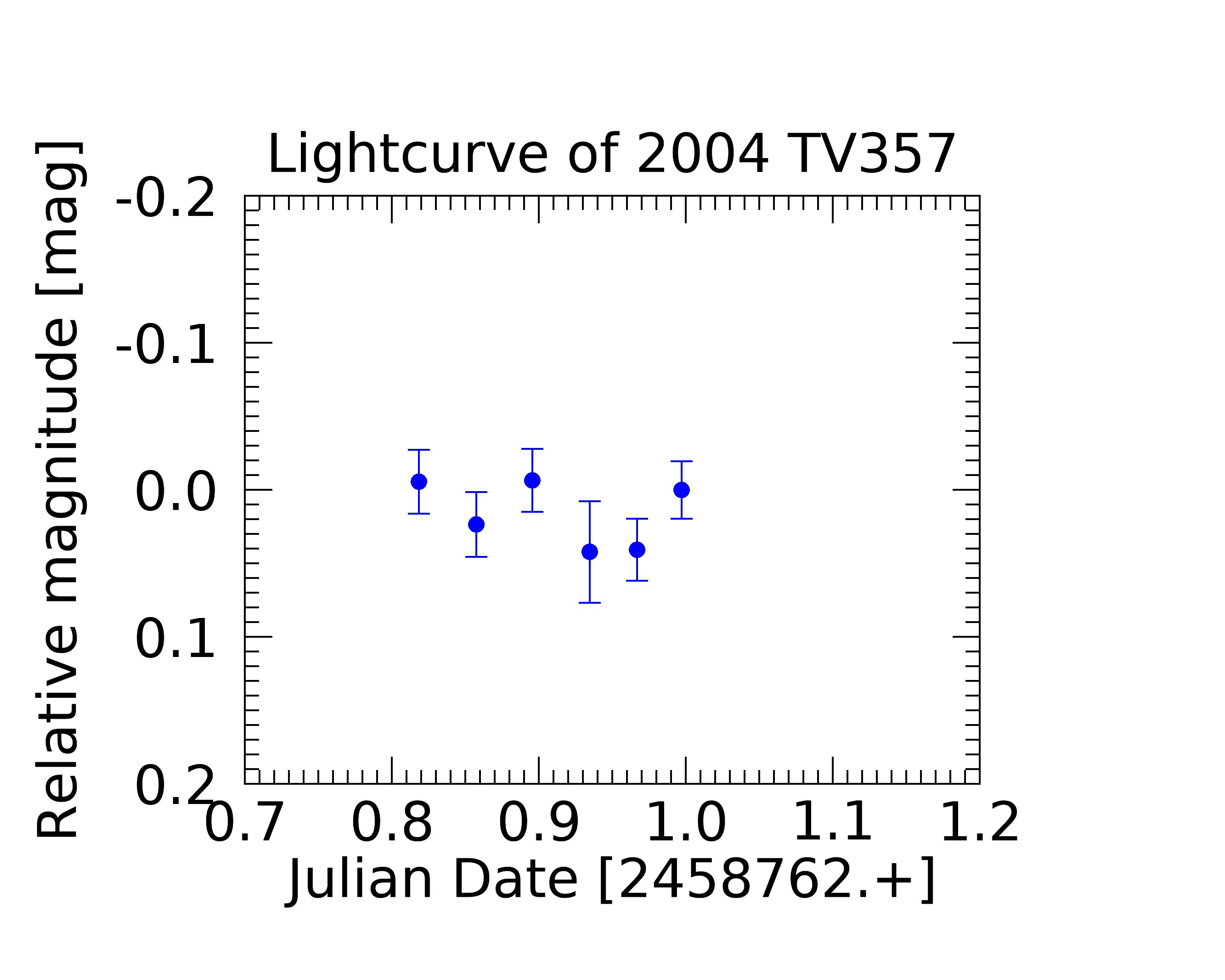}
\caption{\textit{Objects in the 2:1 mean motion resonance with Neptune }   }
\label{fig:LC21}
\end{figure*}

\begin{figure*}
  \includegraphics[width=9.5cm, angle=0]{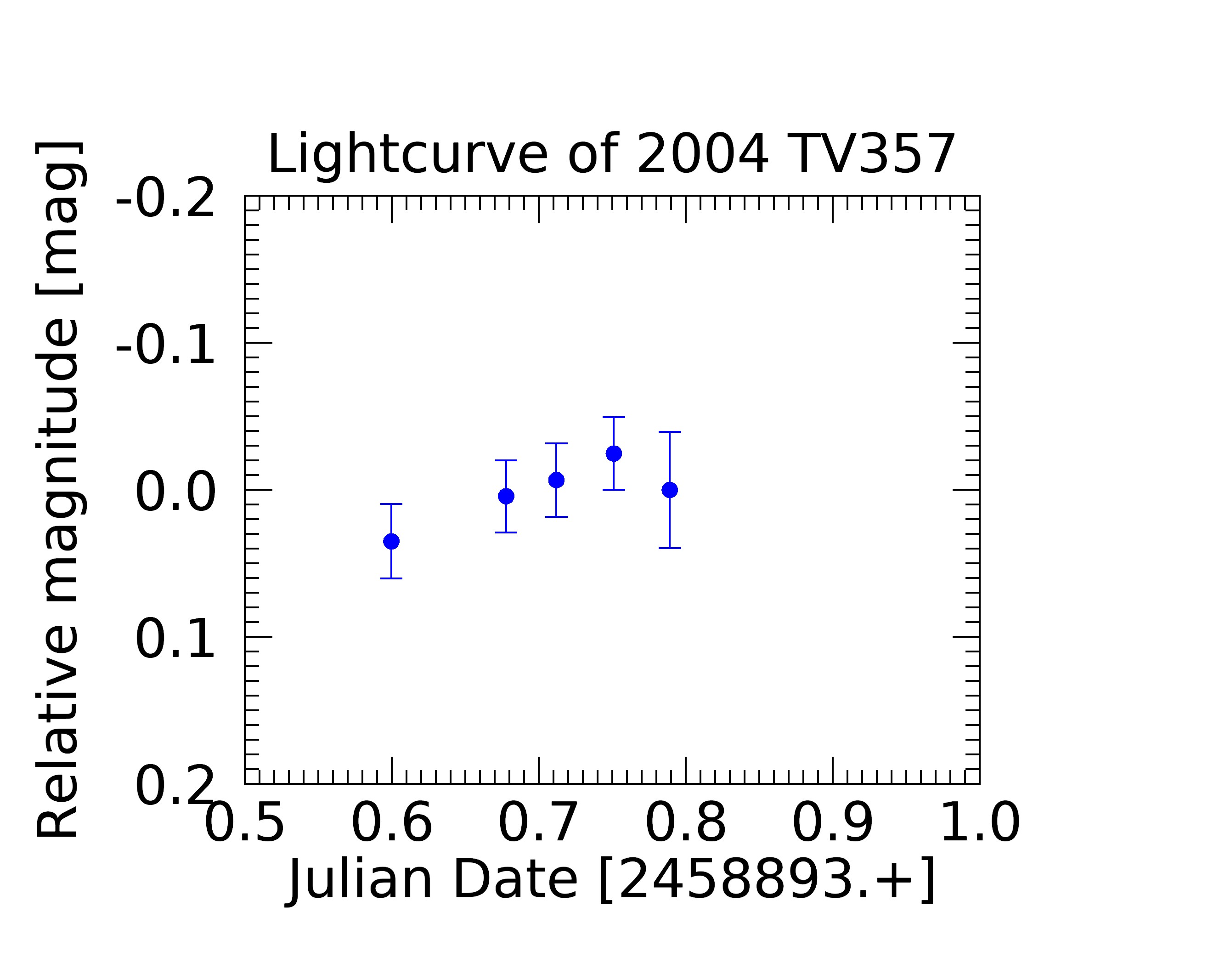}
 \includegraphics[width=9.5cm, angle=0]{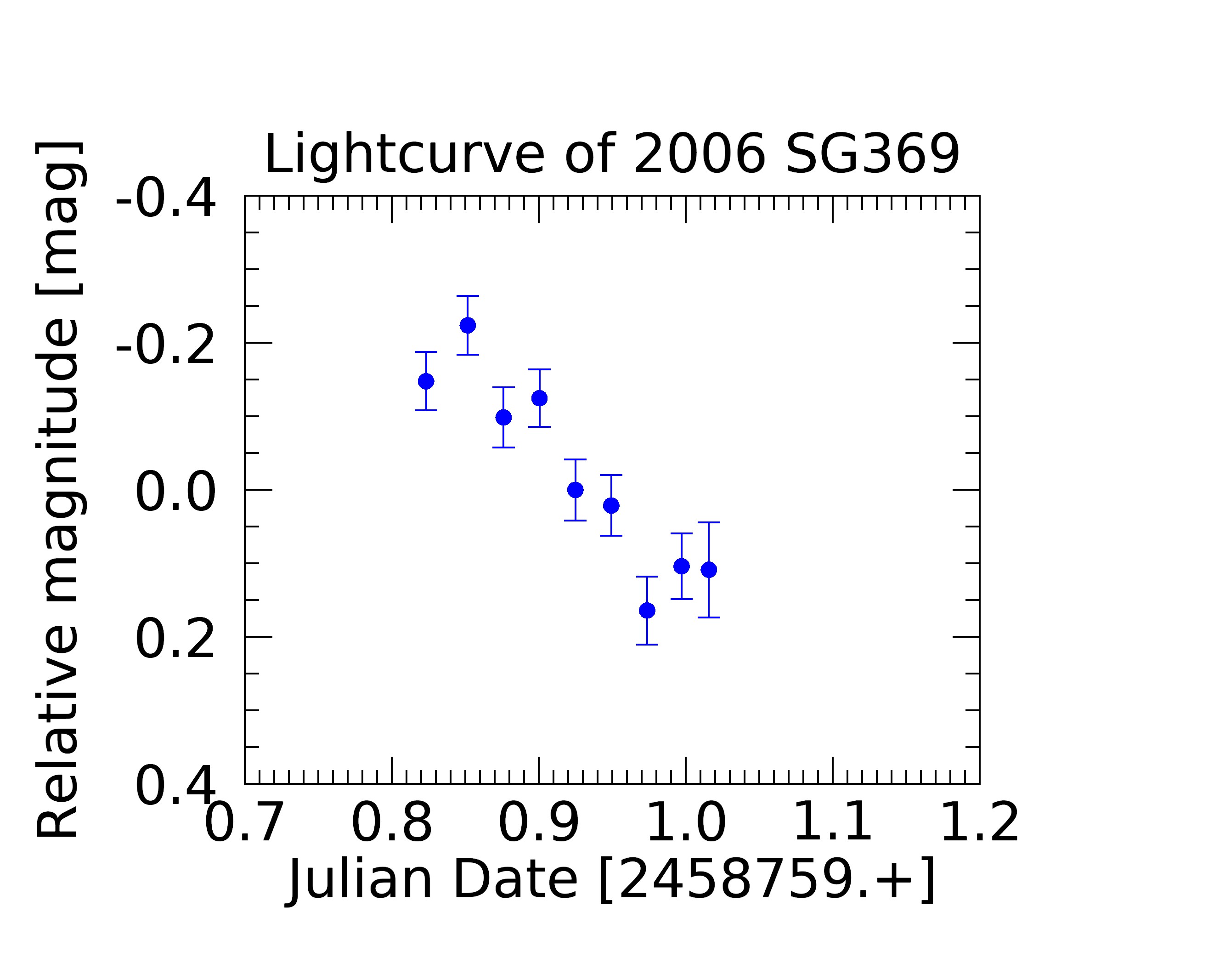}
 \includegraphics[width=9.5cm, angle=0]{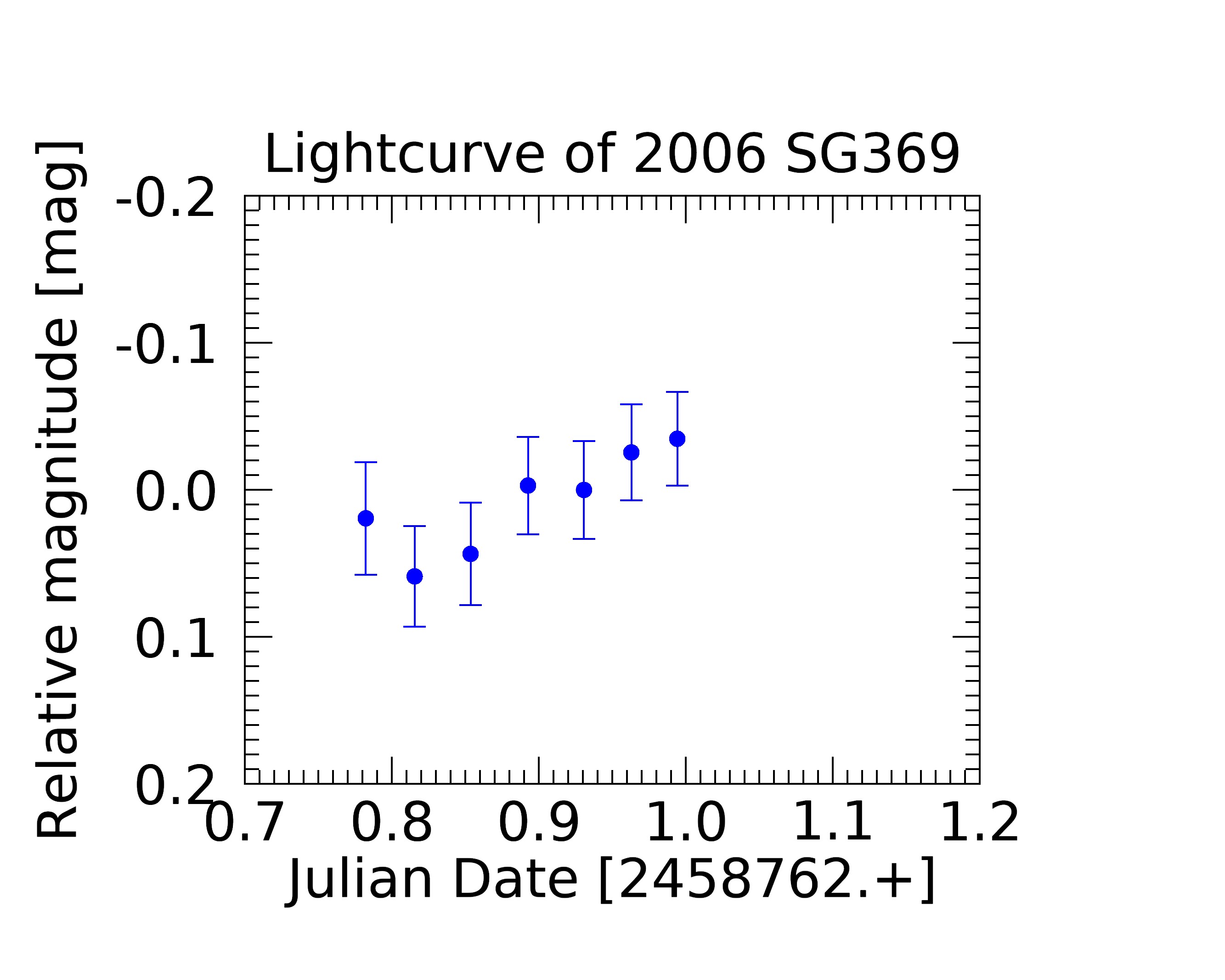}
        \includegraphics[width=9.5cm, angle=0]{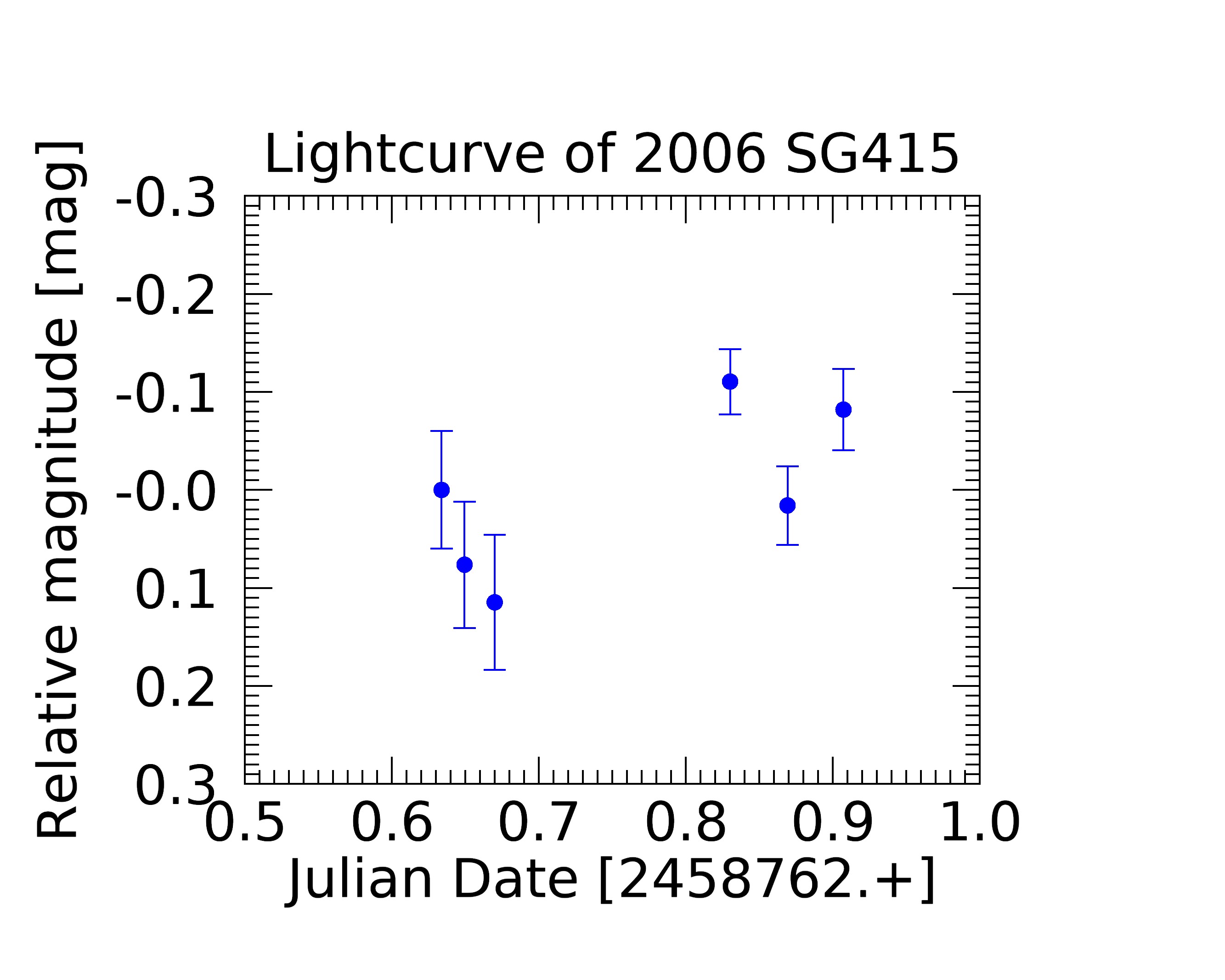} 
 \includegraphics[width=9.5cm, angle=0]{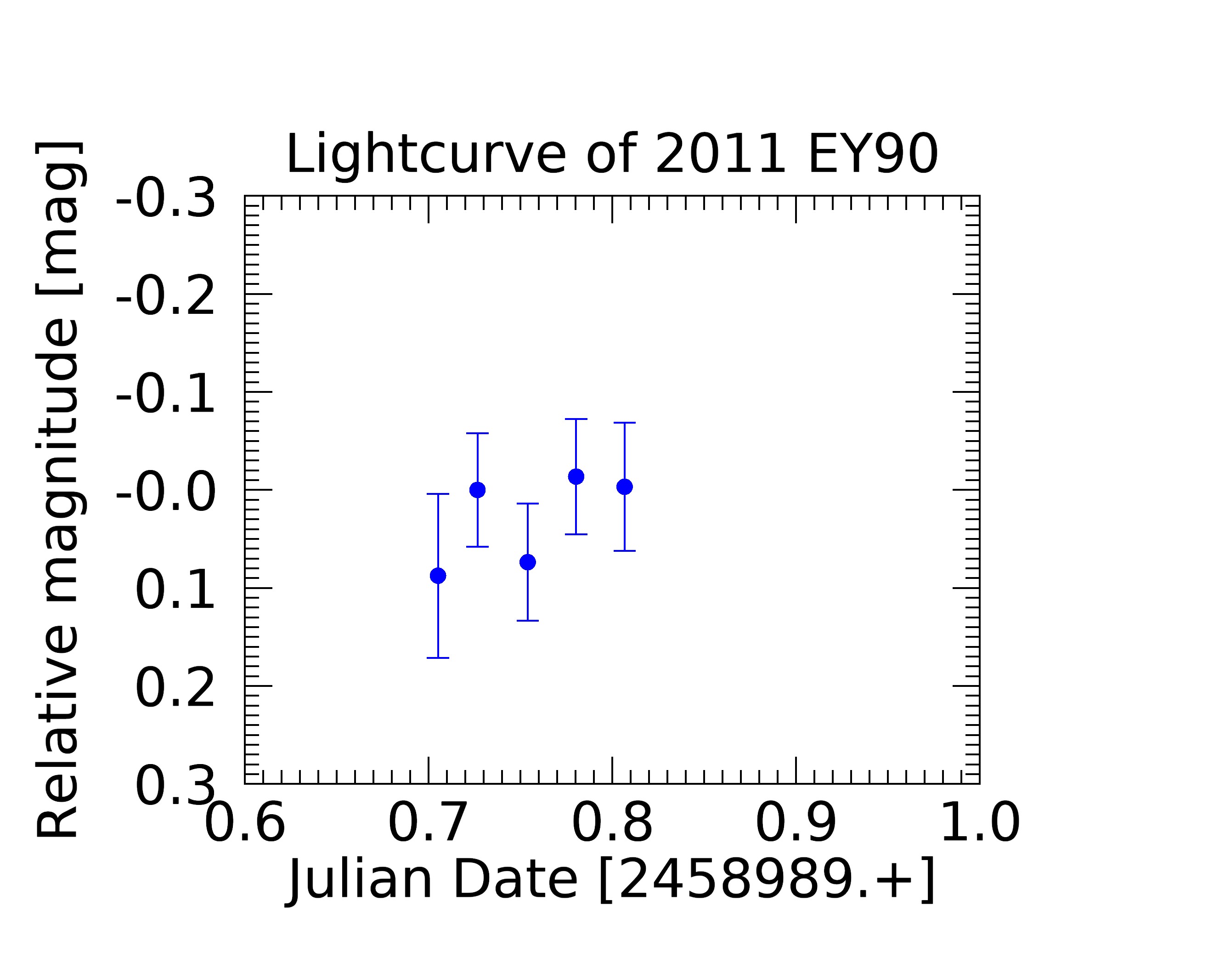}
        \includegraphics[width=9.5cm, angle=0]{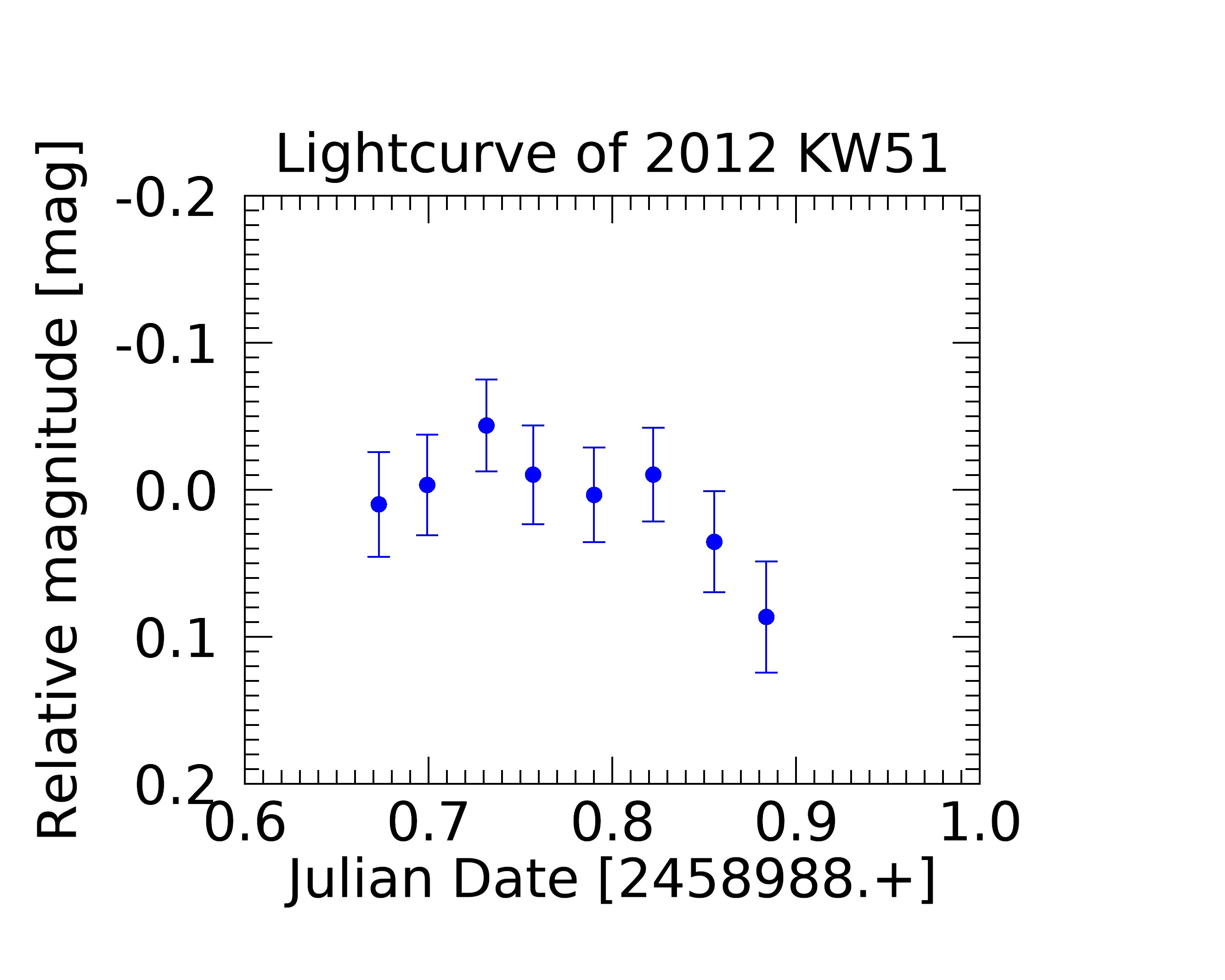} 
\caption{\textit{Objects in the 2:1 mean motion resonance with Neptune }   }
\label{fig:LC21}
\end{figure*}
 
\begin{figure*}
            \includegraphics[width=9.5cm, angle=0]{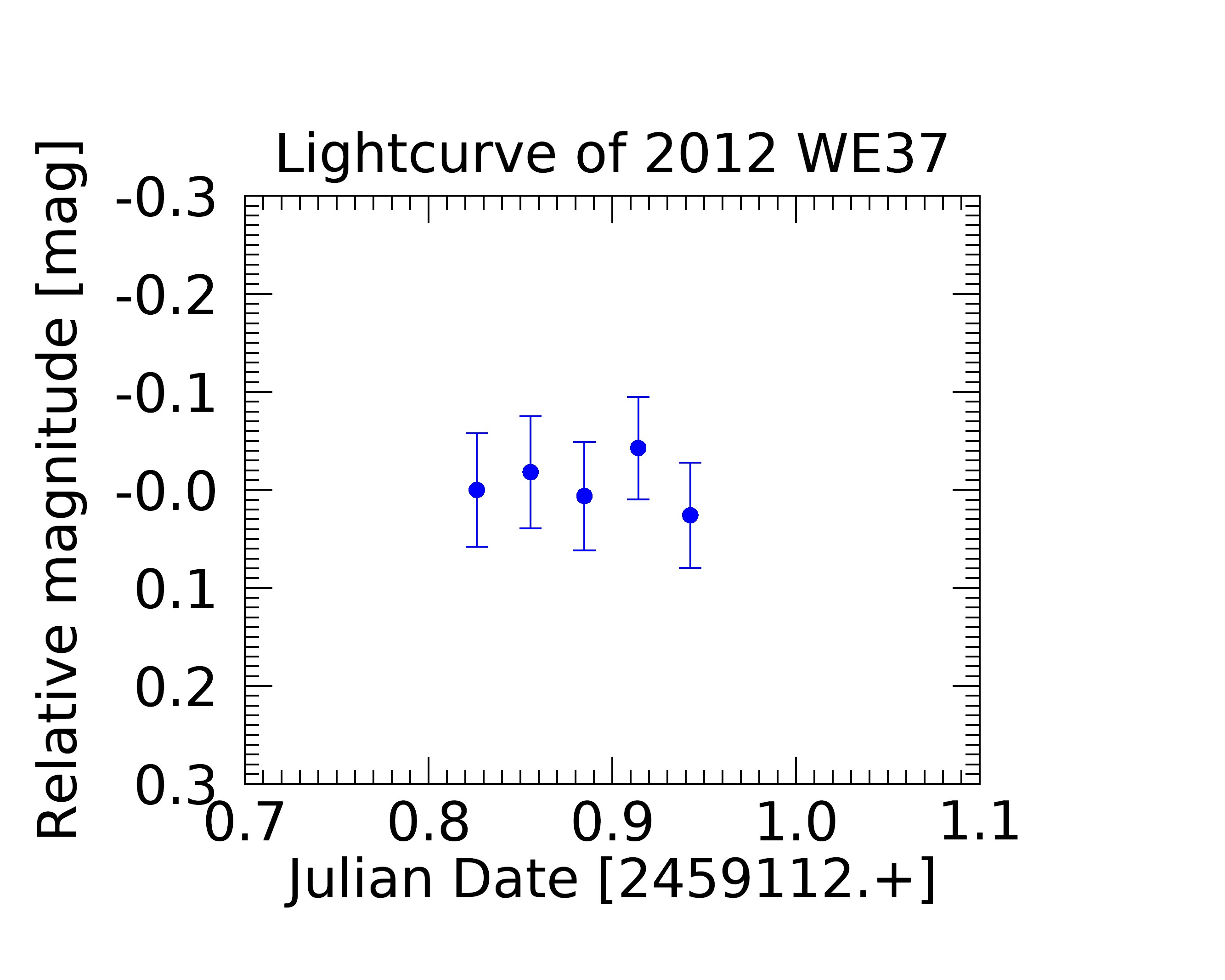}
        \includegraphics[width=9.5cm, angle=0]{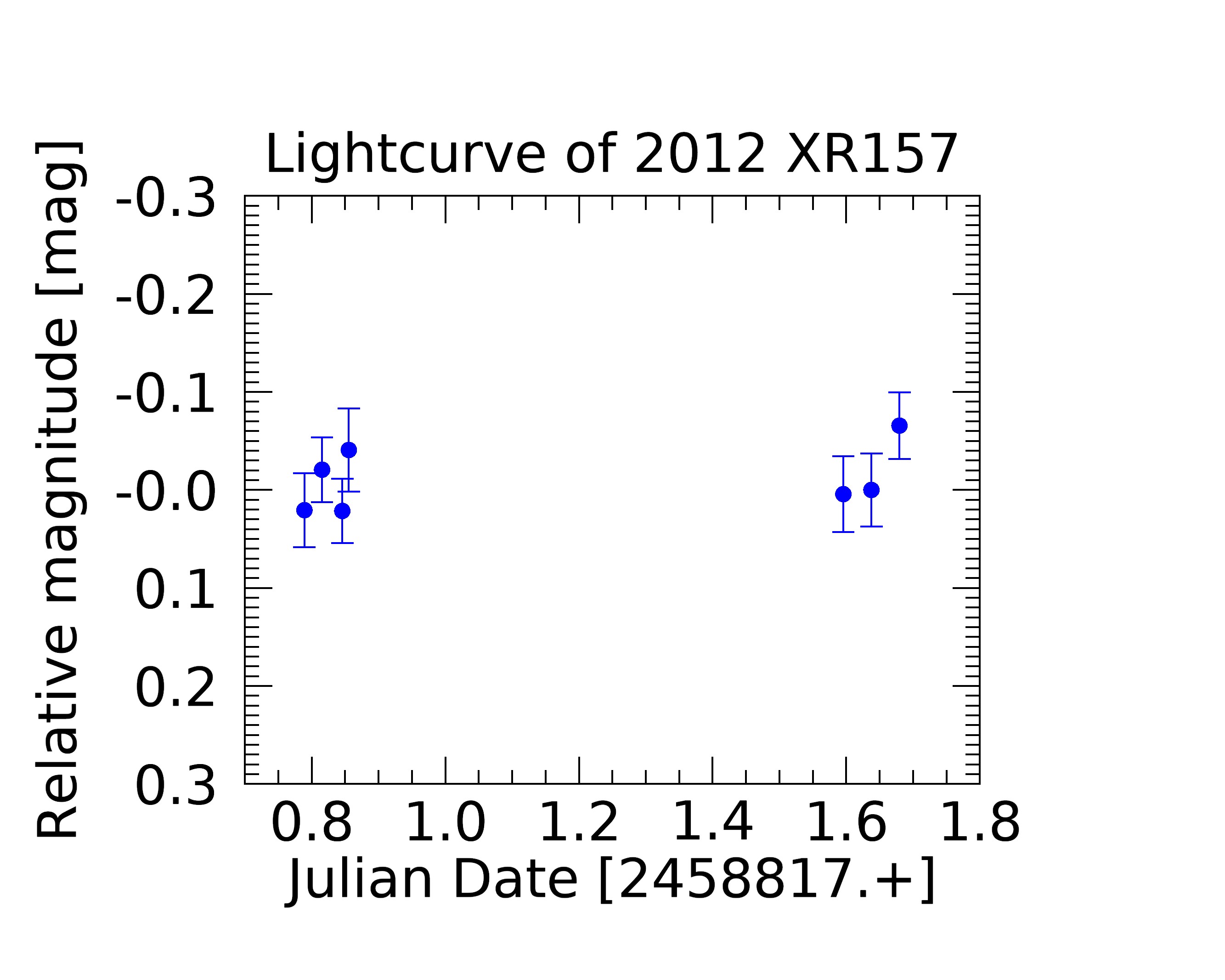}
        \includegraphics[width=9.5cm, angle=0]{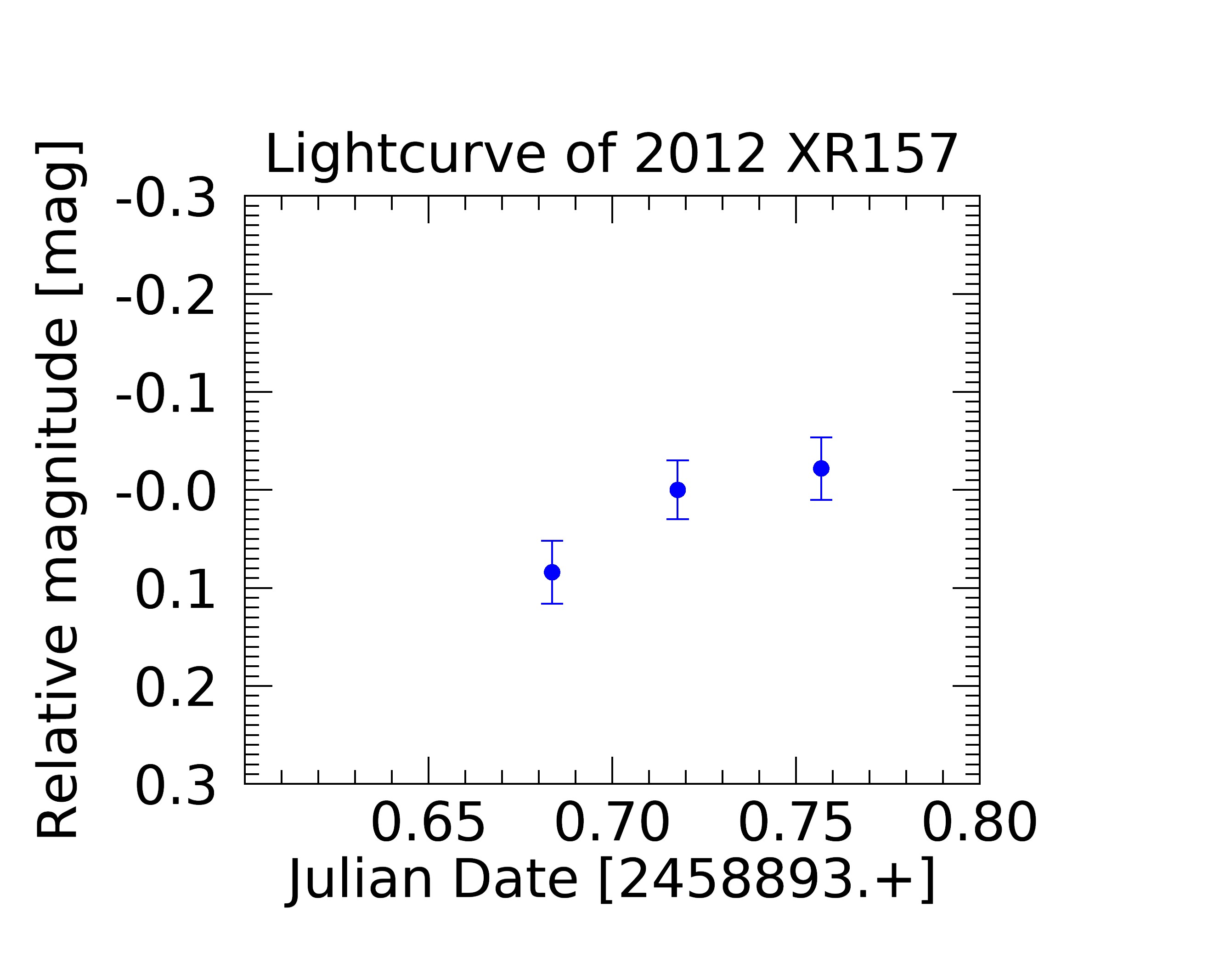}
     \includegraphics[width=9.5cm, angle=0]{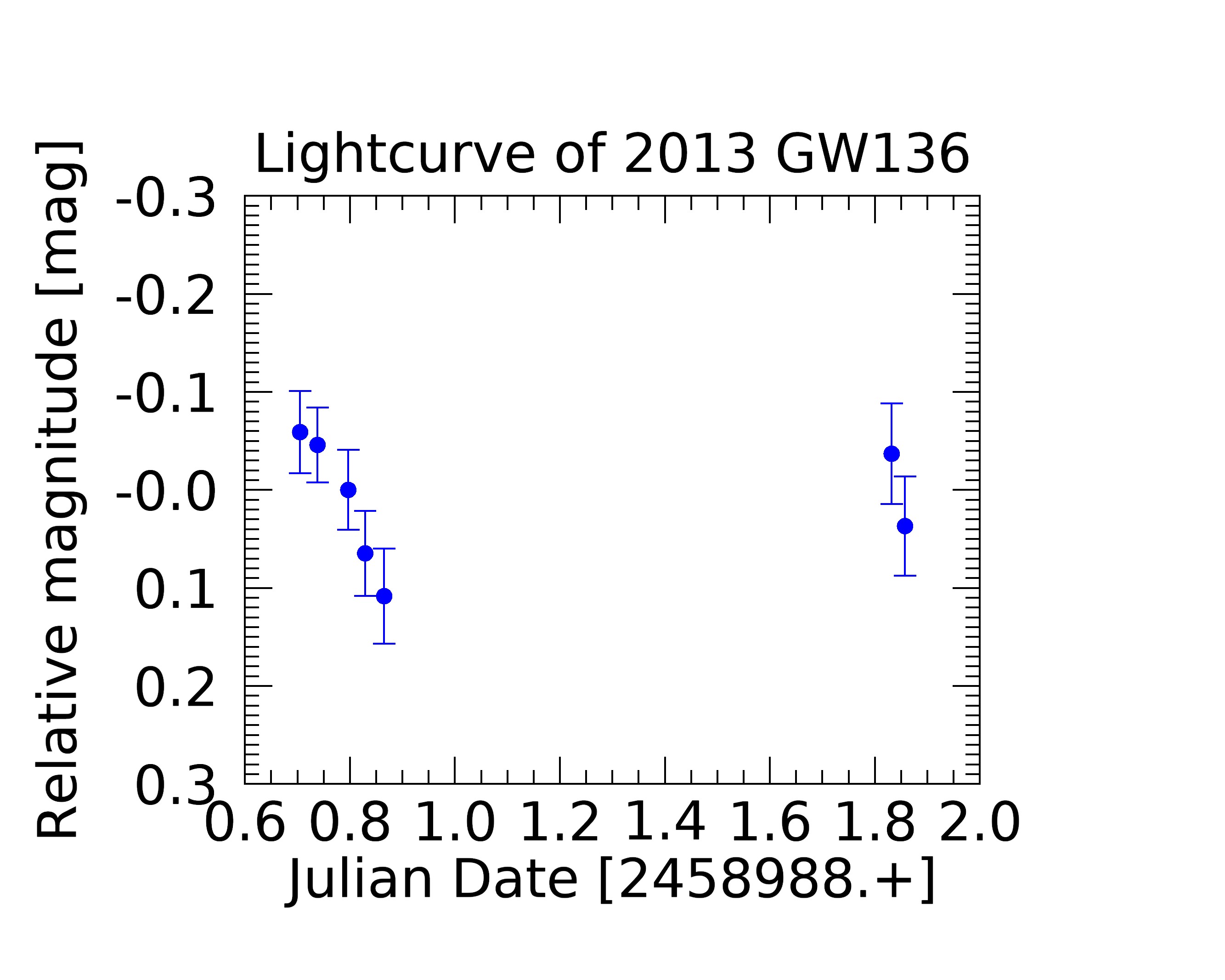}
     \includegraphics[width=9.5cm, angle=0]{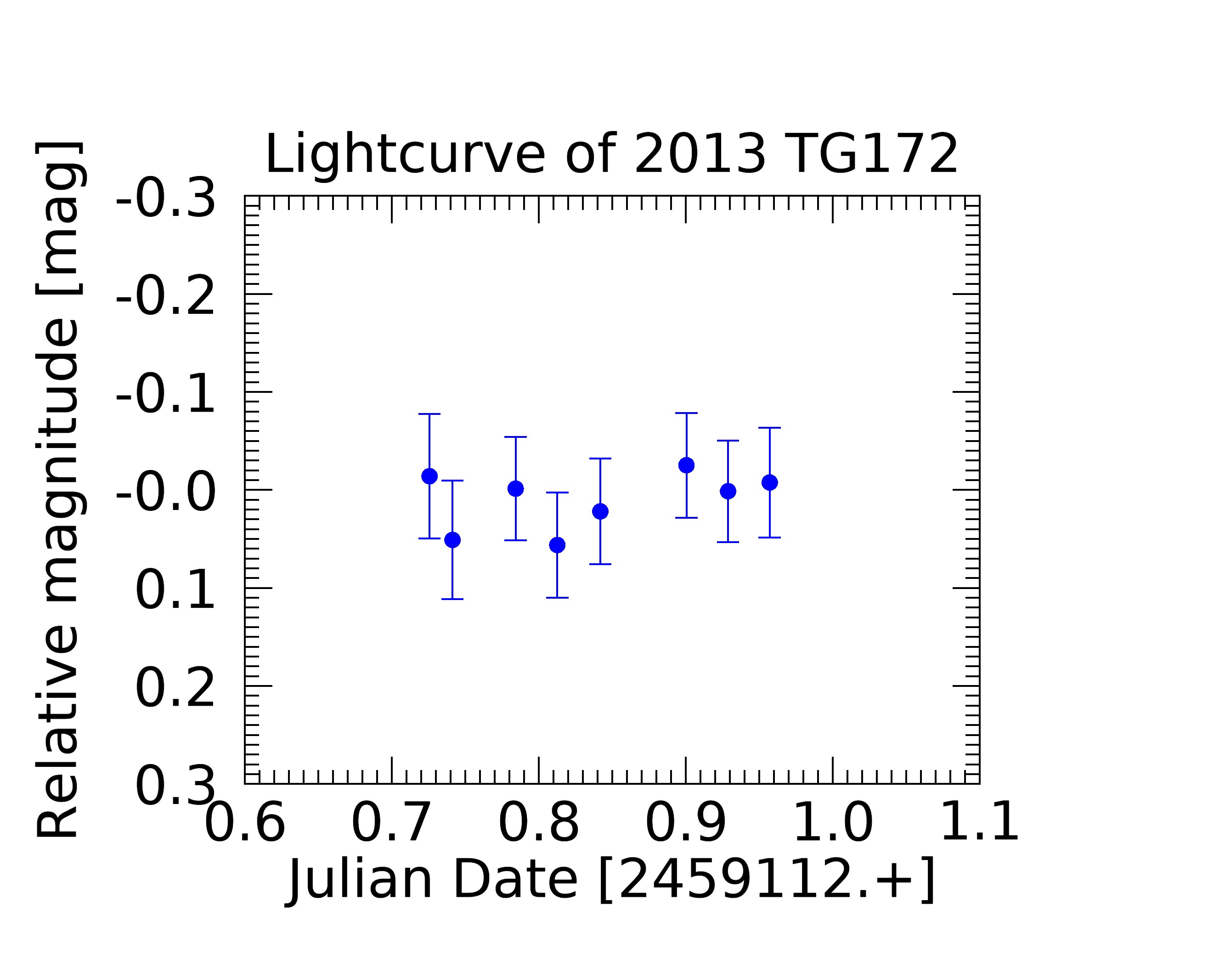}
        \includegraphics[width=9.5cm, angle=0]{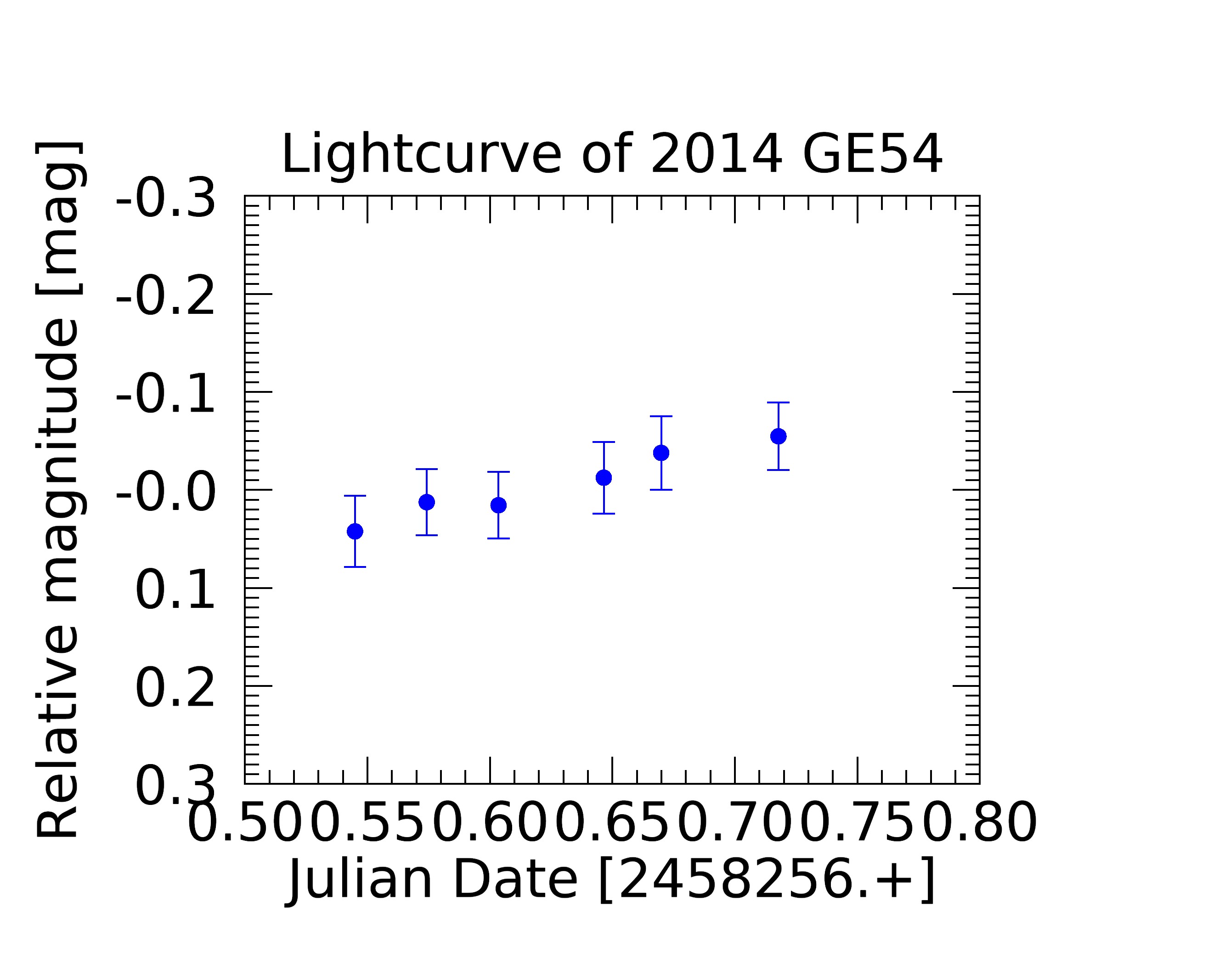}
  \caption{\textit{Objects in the 2:1 mean motion resonance with Neptune }   }
\label{fig:LC21}
\end{figure*}

\begin{figure*}
        \includegraphics[width=9.5cm, angle=0]{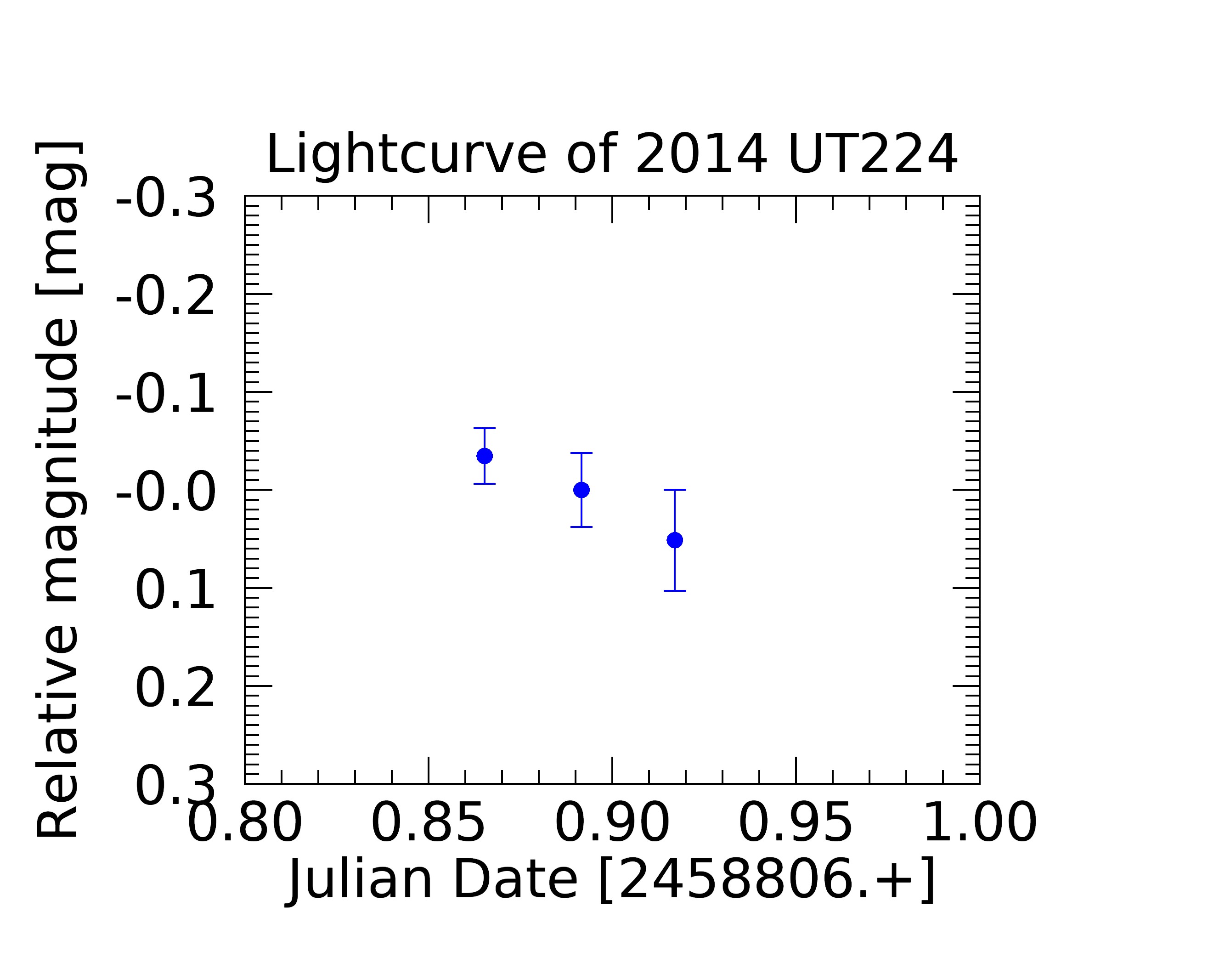}
    \includegraphics[width=9.5cm, angle=0]{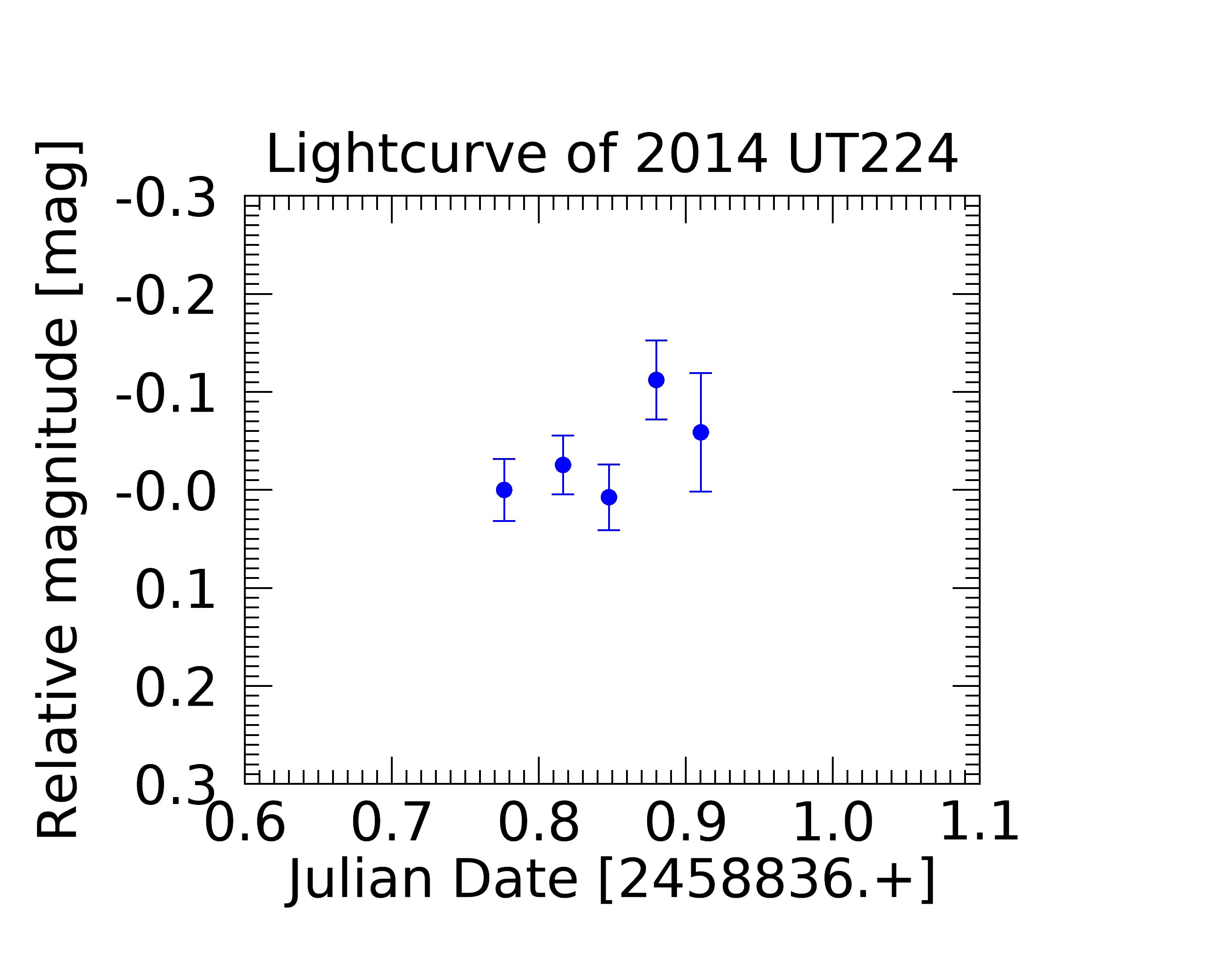}
    \includegraphics[width=9.5cm, angle=0]{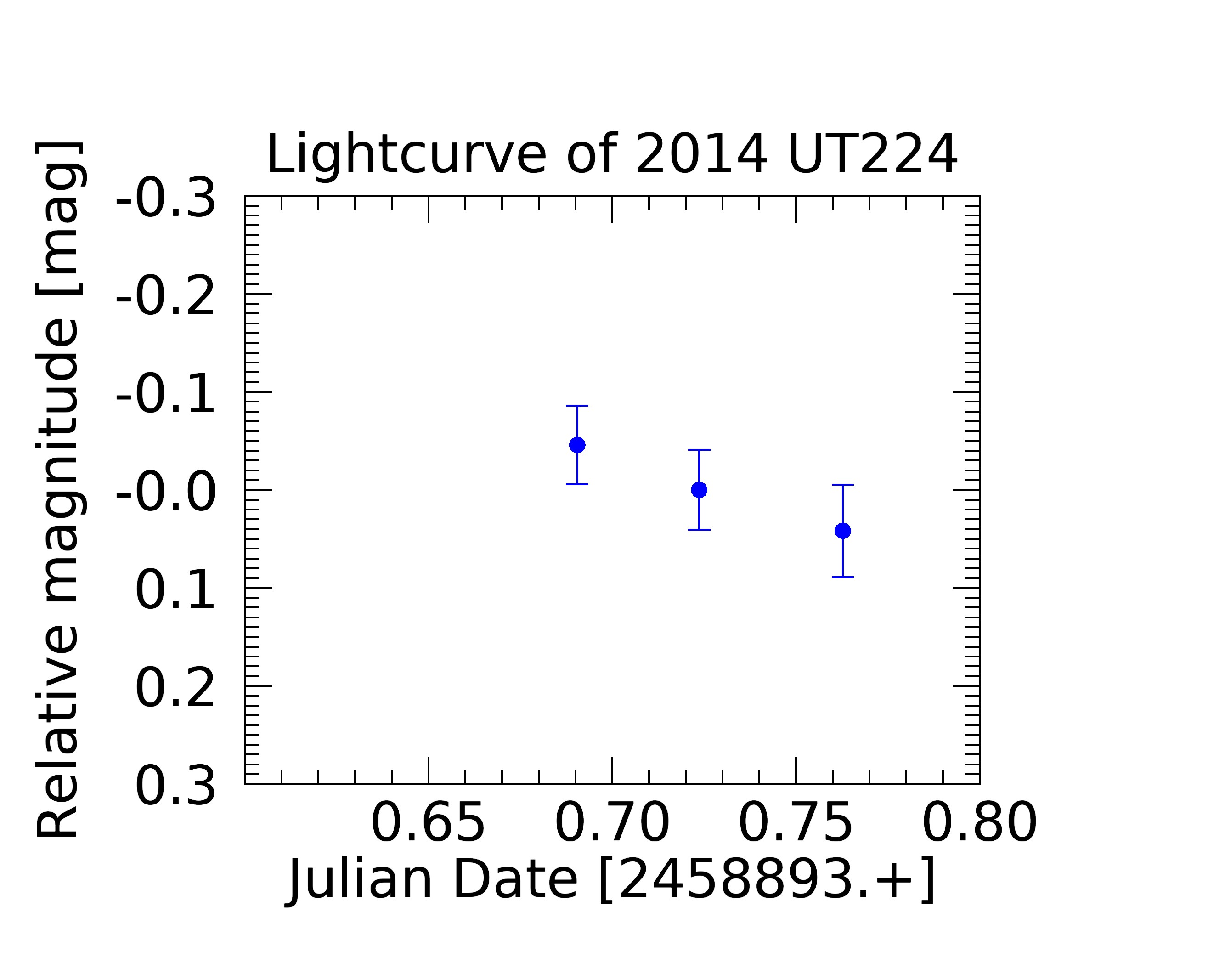}
    \includegraphics[width=9.5cm, angle=0]{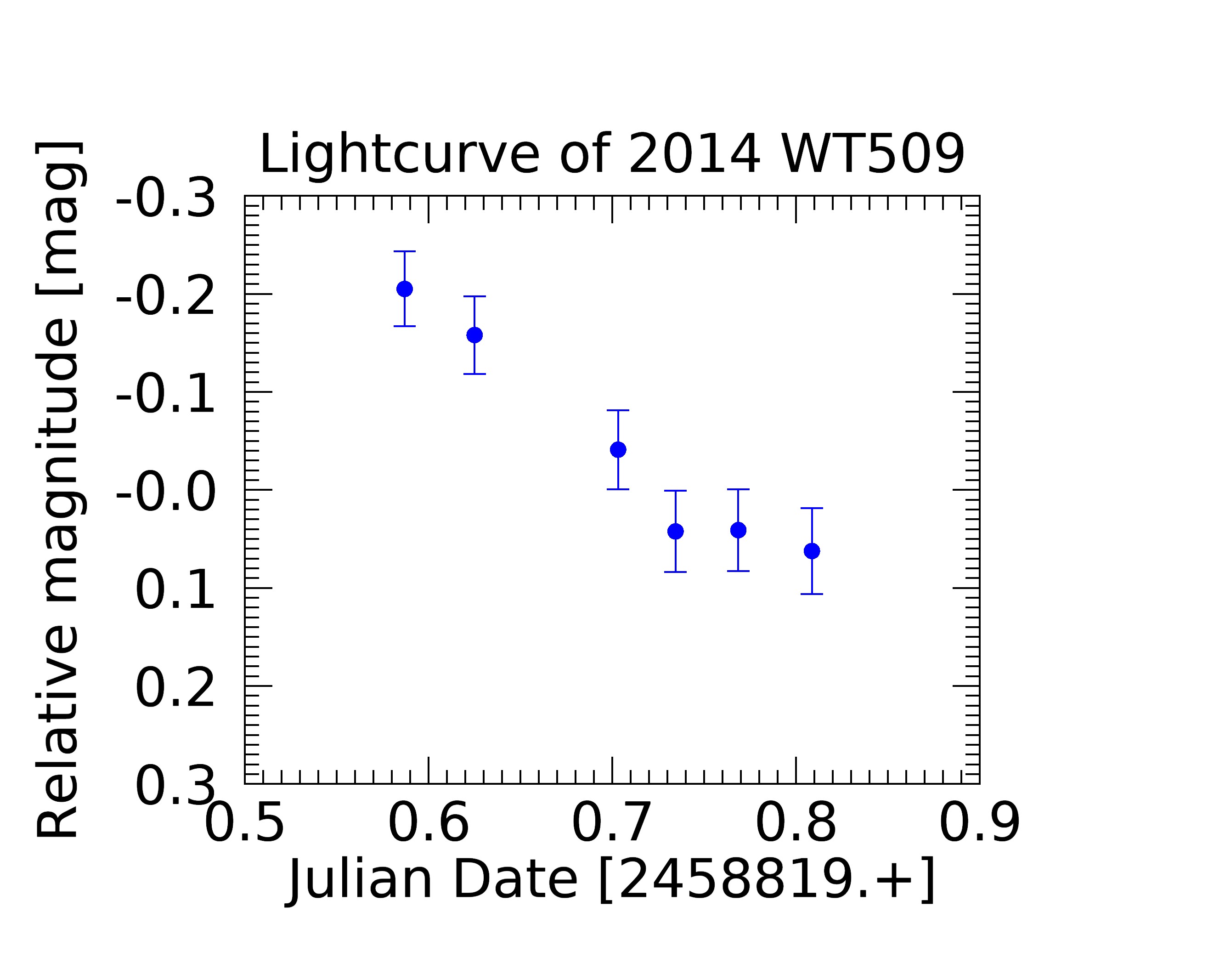}
    \includegraphics[width=9.5cm, angle=0]{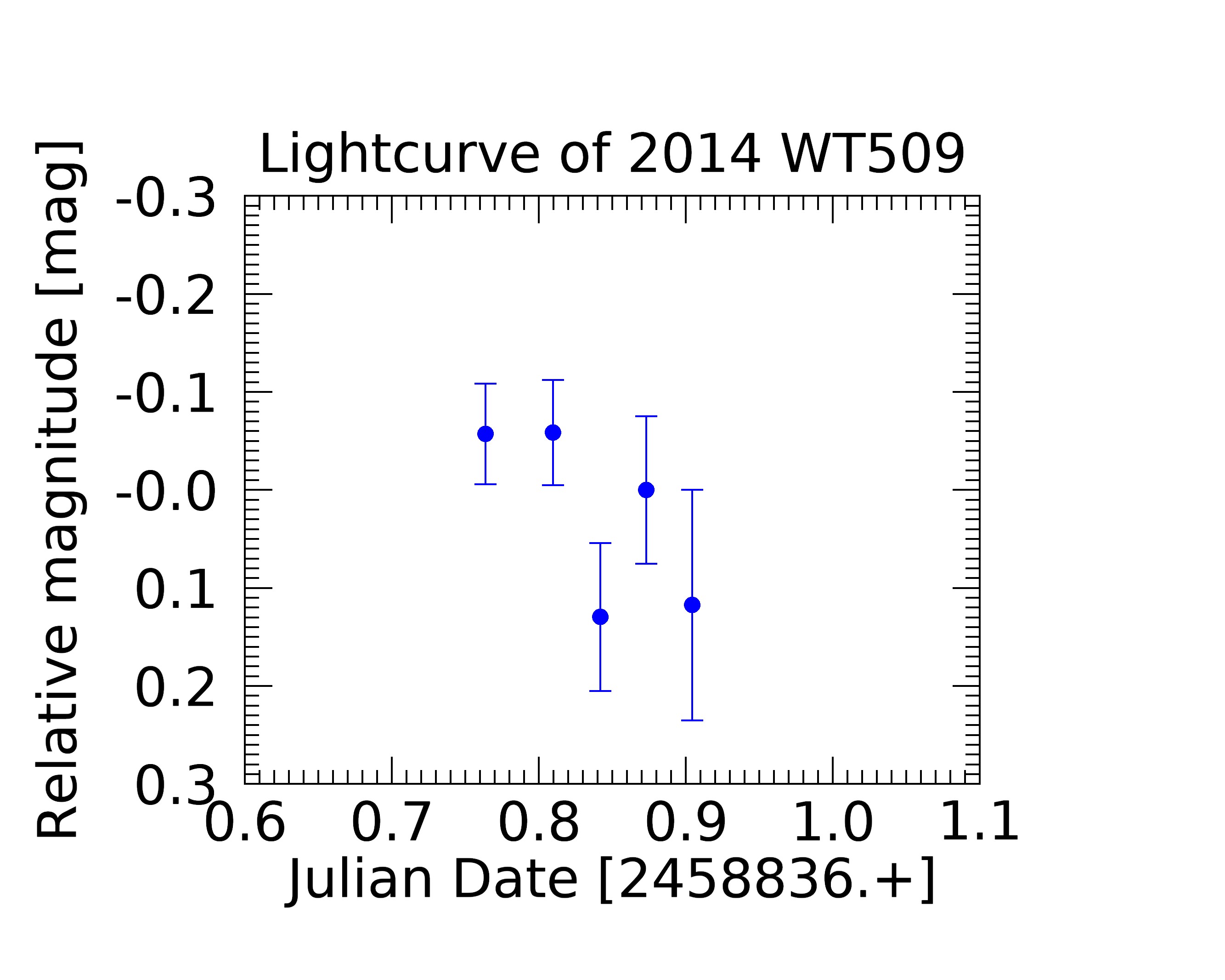}
        \includegraphics[width=9.5cm, angle=0]{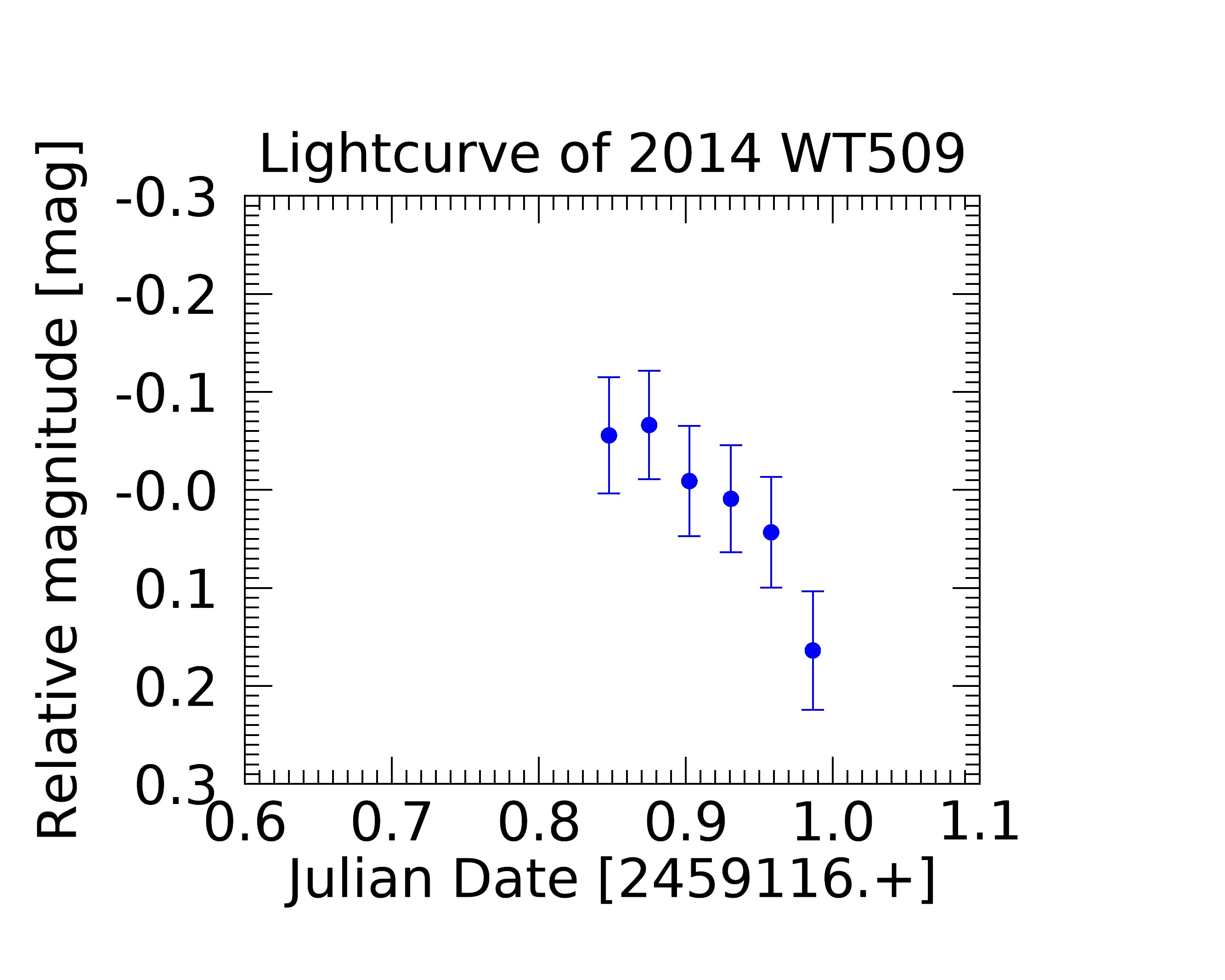}
\caption{\textit{Objects in the 2:1 mean motion resonance with Neptune }   }
\label{fig:LC21}
\end{figure*}

\begin{figure*}
    \includegraphics[width=9.5cm, angle=0]{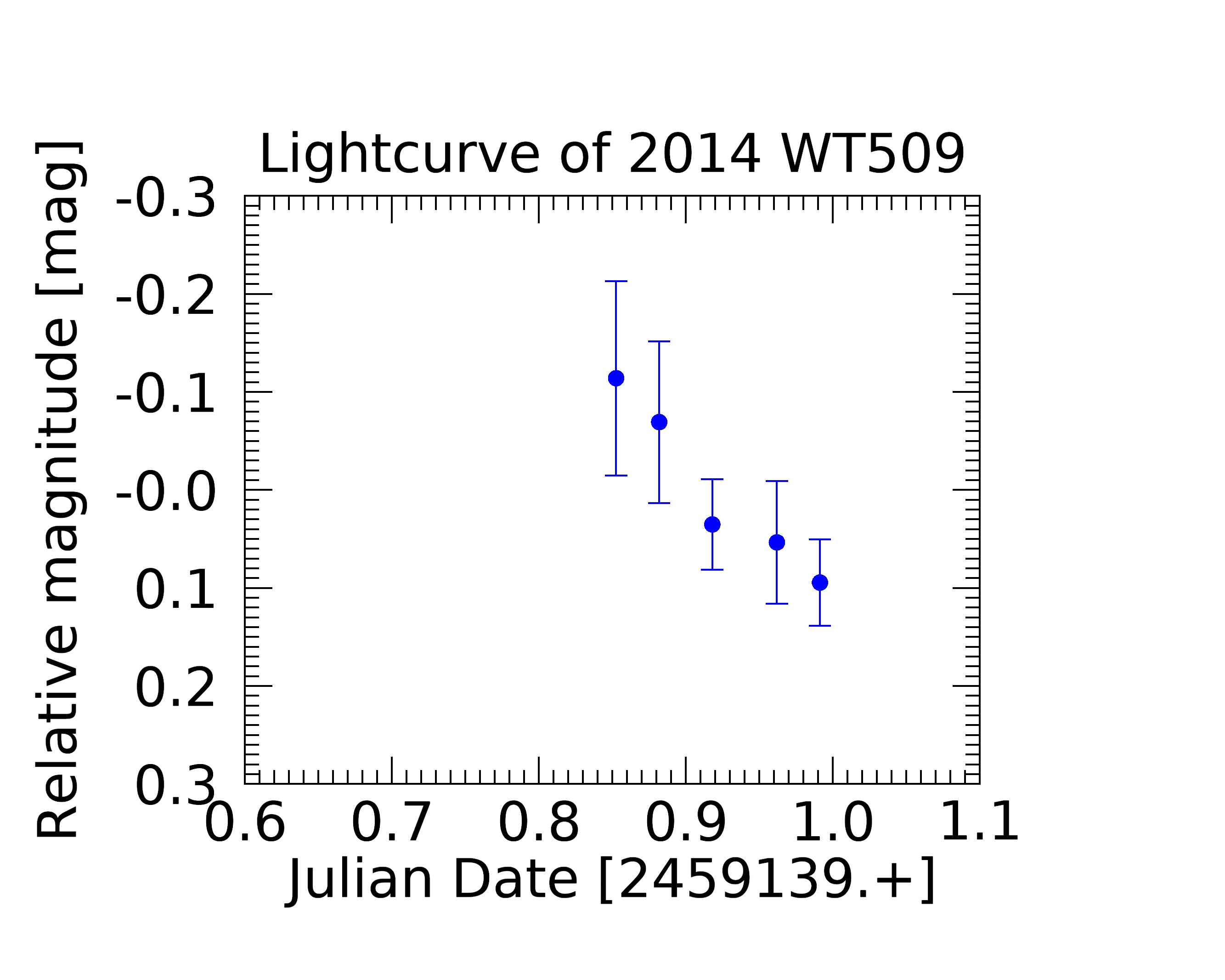}
    \includegraphics[width=9.5cm, angle=0]{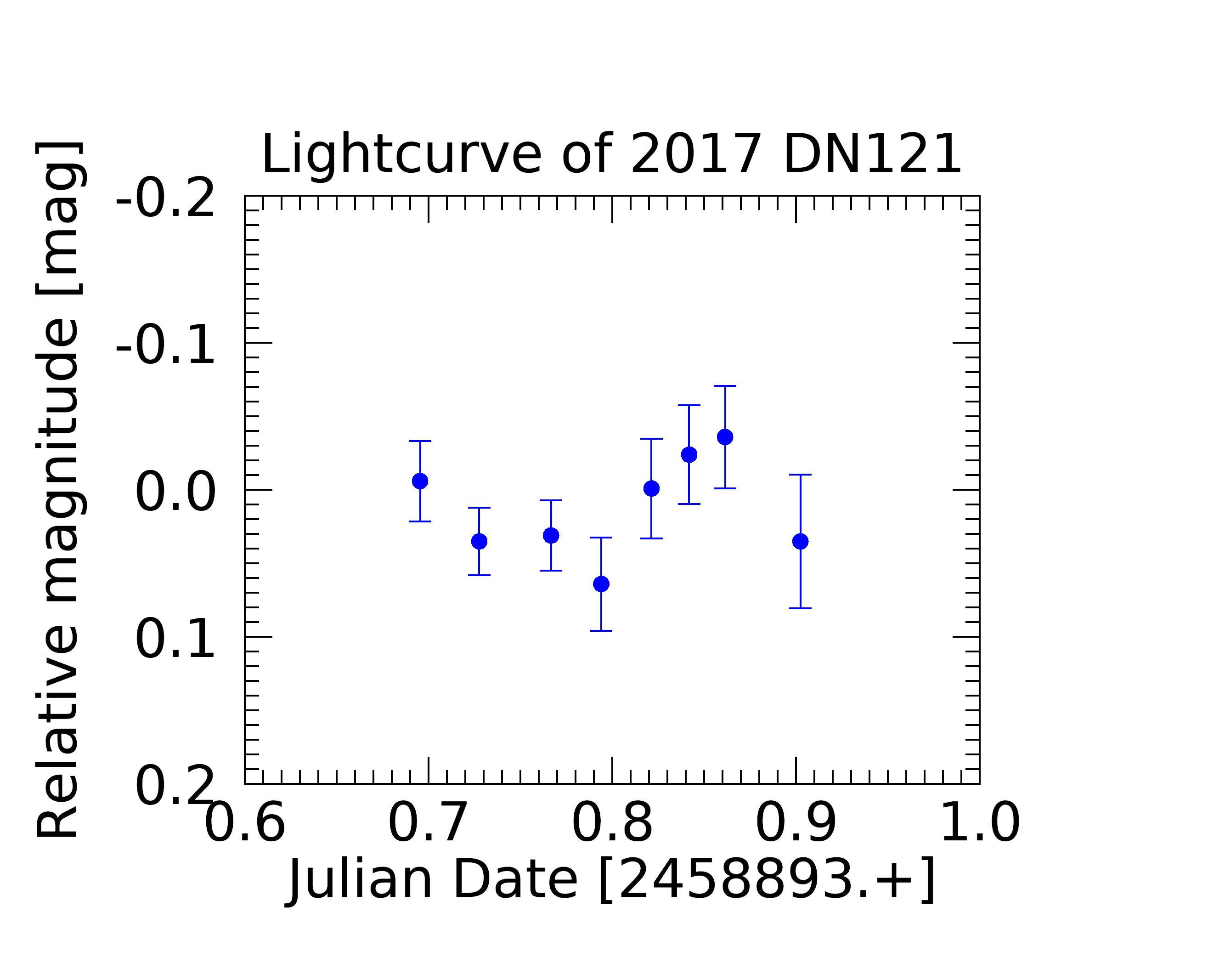}
    \caption{\textit{Objects in the 2:1 mean motion resonance with Neptune }   }
\label{fig:LC21}
\end{figure*}

\end{document}